\documentclass[11pt]{article}
\usepackage{latexsym,amsmath,amssymb,amsthm,amsfonts,graphicx}
\numberwithin{equation}{section}
\numberwithin{figure}{section}

\newcommand{\nc}{\newcommand} 
\nc{\D}{\partial}
\nc{\diff}[2]{\frac{d #1}{d #2}}
\nc{\abs}[1] {\lvert #1 \rvert} 
\nc{\Norm}[1] {\lVert #1 \rVert} 
\nc{\norm}[2] {{\lVert #1 \rVert}_{#2}} 
\nc{\cA}{{\mathcal A}} 
\nc{\cD}{{\mathcal D}} 
\nc{\cE}{{\mathcal E}} 
\nc{\cH}{{\mathcal H}} 
\nc{\cN}{{\mathcal N}} 
\nc{\cO}{{\mathcal O}} 
\nc{\cV}{{\mathcal V}} 
\nc{\cW}{{\mathcal W}}
\nc{\Eplus}{E_+} 
\nc{\Eminus}{E_-} 
\nc{\Epm}{E_\pm}
\nc{\tEplus}{\tilde E_+} 
\nc{\tEminus}{\tilde E_-} 
\nc{\tEpm}{\tilde E_\pm}
\nc{\CEplus}{{\mathcal E}_+} 
\nc{\CEminus}{{\mathcal E}_-} 
\nc{\vE}{\Vec {\mathcal E}} 
\nc{\eplus}{e_+} 
\nc{\eminus}{e_-} 
\nc{\epm}{e_\pm}
\renewcommand{\d}{\delta}
\nc{\w}{\omega} 
\nc{\eps}{\epsilon} 
\nc{\g}{\gamma} 
\nc{\z}{\zeta}
\nc{\G}{\Gamma} 
\renewcommand{\k}{\kappa}
\nc{\kinf}{\kappa_\infty}
\nc{\pZ}{\partial_Z} 
\nc{\pT}{\partial_T} 
\nc{\veps}{\varepsilon}
\nc{\infint}{\int_{-\infty}^{\infty}}
\nc{\bbC}{\mathbb{C}}
\nc{\vetap}{\boldsymbol{\eta}^+}
\nc{\vetam}{\boldsymbol{\eta}^-}
\nc{\ve}{\vec{e}}
\nc{\vf}{\vec{f}}
\nc{\vv}{\vec{v}}
\nc{\vy}{\vec{y}}
\nc{\cAA}{\cA^{(2)}}
\nc{\tcAA}{\tilde\cA^{(2)}}
\nc{\tpsi}{\tilde\psi}
\nc{\nn}{\nonumber}
\nc{\spec}{{\rm spec}}
\nc{\fsign}{\mathfrak{s}}
\nc{\XYZ}[1]{\textbf{XYZ: {#1}}}

\newtheorem{prop}{Proposition}
\DeclareMathOperator{\sech}{sech}
\DeclareMathOperator{\sign}{sign}

\title{Stability and instability of nonlinear defect states in the coupled mode equations---analytical and numerical study}
\author{Roy H. Goodman%
\thanks{Department of Mathematical Sciences, New Jersey Institute of 
Technology, Newark, NJ 01710}\hspace{2cm}%
Michael I. Weinstein%
\thanks{Department of Applied Physics and Applied Mathematics,
Columbia University, New York, NY 10027}
}
\begin{document}
\maketitle
\abstract{Coupled backward and forward wave amplitudes of an electromagnetic field propagating in a periodic and nonlinear medium at Bragg resonance are governed by the nonlinear coupled mode equations (NLCME). This system of PDEs, similar in structure to the Dirac equations, has gap soliton solutions that travel at any speed between $0$ and the speed of light.
A recently considered strategy for spatial trapping or capture of gap optical soliton light pulses is based on the appropriate design of localized defects in the periodic structure. Localized defects in the periodic structure give rise to defect modes, which persist as {\it nonlinear defect modes} as the amplitude is increased. Soliton trapping is the transfer of incoming soliton energy to  {\it nonlinear} defect modes. To serve as targets for such energy transfer, nonlinear defect modes must be stable. We therefore investigate the stability of nonlinear defect modes. Resonance among discrete localized modes and radiation modes plays a role in the mechanism for stability and instability, in a manner analogous to the nonlinear Schr\"odinger / Gross-Pitaevskii (NLS/GP) equation. However, the nature of instabilities and how energy is exchanged among modes is considerably more complicated than for NLS/GP due, in part, to a continuous spectrum of radiation modes which is unbounded above and below.  In this paper we (a) establish the  instability of branches of nonlinear defect states which, for vanishing amplitude, have a linearization with eigenvalue embedded within the continuous spectrum, (b) numerically compute, using Evans function,  the linearized spectrum of nonlinear defect states  of an interesting multiparameter family of defects, and (c) perform direct time-dependent numerical simulations in which we observe the exchange of energy among discrete and continuum modes.
}

\tableofcontents
\section{Introduction}
\label{sec:intro}
In optical fiber communication systems, data is carried as pulses of light.  Expensive and rate-limiting steps in these systems come in processing the data at so-called optical / electrical interfaces. Processing is typically done electronically; signals are converted to electronic form, read, processed, and retransmitted.  A major goal, therefore, is to bypass the optical / electrical interface and implement ``all-optical'' processing by using the nonlinear optical properties of the medium.  Hence, there has been great interest in finding novel materials and optical structures (arrangements of materials) which effect light in different ways.  Among these are Bragg gratings and Bragg gratings with defects--one dimensional arrangements, as well as higher-dimensional structures such as photonic crytal fibers in which the grating structure is transverse to the direction of propagation~\cite{KniBirRus:96}, and other two and three-dimensional structures~\cite{ChrLedSil:03}.

\begin{figure}
\begin{center}
\includegraphics[width=2in]{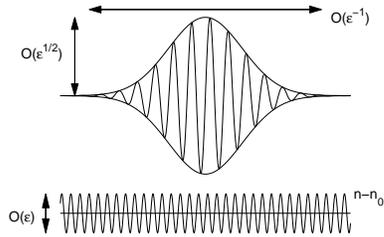}
\caption{Schematic of Bragg resonance condition. Periodic refractive index (lower figure) and electric field envelope with carrier wave in Bragg resonance ($\lambda=2\times{\rm period}$).}
\label{fig:schematic}
\end{center}
\end{figure}

In an optical fiber Bragg grating, the refractive index of the glass varies
periodically, with period resonant with the carrier wavelength of the propagating light; see figure \ref{fig:schematic}.  Light propagation through such fibers has several interesting
properties that makes them useful in technological applications.   The grating
structure couples forward-moving light at the resonant wavelength to
backward-moving light of the same wavelength.  In the low-amplitude (linear) limit, this makes the
fiber opaque to light in a certain range of wavelengths, the so-called photonic band-gap.  When the amplitude
of the light is increased, the band-gap is shifted due to the nonlinear  dependence of refractive index on intensity. Thus, a range of wavelengths that are non-propagating  at low intensity  are shifted into  the range of propagating wavelengths (the pass-band) at higher intensities.
 This is the mechanism by which  localized pulses known as gap
solitons exist; see subsection \ref{sec:gapsolitons}.  In the regime of weak nonlinearity and Bragg resonance, Maxwell's equations can be reduced, via multiple scale asymptotic methods, to the nonlinear coupled mode equations (NLCME), reviewed in section~\ref{sec:NLCME}~\cite{AW,CJ,dSS,GWH}.%
\footnote{A special case of NLCME (equation~\eqref{eq:NLCME} with the self-phase modulation terms omitted) is the massive Thirring system~\cite{FanGir:76}, a completely integrable PDE modeling the interaction of massive fermions.  The methods described in this article apply equally well to the massive Thirring system.}.

Gap solitons may, in theory, propagate (in the stationary reference
frame) at any speed between 0 and $c/n$. Here, $c$ denotes the vacuum speed of light, $n$ the refractive index of the optical fiber core and $c/n$ is the speed of light in the fiber 
without the grating.  In~\cite{GSW}, we showed  via modeling, analysis and numerical simulation, that  propagating optical gap solitons, could be trapped at specially-constructed defects in the grating structure. Trapping and interaction of gap solitons in a number of related structures is considered in~\cite{MakMalChu:03a,MakMalChu:03b,CheMalChu:05}.  Recent experimental advances have made possible the slowing of propagating gap soliton light pulses from $0.5\times c$~\cite{ESdSKS:96} to $0.16\times c$~\cite{MokEtAl:06}.

The focus of the present paper is on the stability and dynamics of such trapped light.
Localized defects in a grating appear as spatially localized {\it potentials} in the NLCME model. In the linear (low light intensity) limit, the resulting {\it linear coupled mode equations} with potentials have spatially localized {\it linear defect eigenstates}; see subsection \ref{sec:linear}.  As the intensity is increased from zero, {\it nonlinear defect modes}, ``pinned'' at the defect location,  bifurcate from the zero state at the linear eigenfrequencies; see subsection \ref{sec:nlmodes}. 

Trapping of a gap soliton by a defect can be understood as the resonant transfer of energy from an incoming soliton\ to a pinned nonlinear defect mode. In \cite{GSW} we demonstrated through numerical experiments such resonant energy transfer / trapping, for sufficiently slow soliton pulses. The analogous question has also been studied for the nonlinear Schr\"odinger / Gross-Pitaevskii  (NLS/GP) equations; see   
\cite{GooHolWei:04,HolMarZwo:07,HolZwo:07,Zworski:06a} and references cited therein.
For the purpose of comparison, we review the stability of and interactions between nonlinear defect modes of NLS/GP in section~\ref{sec:NLS}.

 In order for the energy localized in a defect to remain spatially confined in a nonlinear defect mode, it is necessary that the mode be stable. Thus, in this paper we consider the stability of nonlinear defect modes for a large class of defects introduced in~\cite{GSW} by
 \begin{itemize}
 \item studying the linearized spectral problem about different families (branches in the global bifurcation diagram) of nonlinear defect modes; see sections~\ref{sec:linearization} and~\ref{sec:evans} for analytical perturbation theory and numerics, and 
 \item studying time-dependent simulations of the initial value problem for NLCME. We consider defects which support multiple nonlinear linear defect modes. As the time-evolution proceeds these nonlinear bound state families compete for the energy localized in the defect; see section~\ref{sec:timedependent}.
 \end{itemize} 

The linearized stability of a nonlinear defect mode is governed by a spectral problem of the form:
\begin{equation}
\Sigma_3 H\psi\ =\ \beta\ \psi,\ \ \Sigma_3^*=-\Sigma_3,\ \ H^*=H;
\label{s3hevp}\end{equation} 
see section \ref{sec:linearization}. Eigenvalues are complex values of $\beta$, for which (\ref{s3hevp}) has a nontrivial $L^2$ solution, giving rise to a time-dependent solution of the linearized dynamics $\phi=\psi e^{-i\beta t}$.
The self-adjoint operator, $H$, can be expressed as $H=H_0+W$, where $H_0$ corresponds to the linear (zero intensity) coupled mode equation operator and $W$ tends to zero quadratically in the amplitude ($L^2$-norm) of the nonlinear defect mode about which we have linearized. 

The continuous spectrum of $\Sigma_3H$ typically consists of two symmetric semi-infinite real intervals with an open gap centered at $\beta=0$. A priori, since $\Sigma_3H$ is not self-adjoint,  discrete eigenvalues may lie anywhere in the complex plane, subject to constraints inherited by (\ref{s3hevp}) from NLCME, a Hamiltonian system. In particular, if $\beta$ is an eigenvalue, then so are $-\beta, \beta^*$ and $-\beta^*$. Thus, a necessary condition for linearized stability of the flow $i\D_t\phi=\Sigma_3H\phi$ is that the spectrum of $\Sigma_3H$ be real.

Now $\Sigma_3H_0$ has stable spectrum and may have real discrete eigenvalues in the spectral gap or real {\it embedded eigenvalues} within the continuous spectrum. Numerous different instability-onset scenarios corresponding to different types of bifurcations arise as $\Sigma_3H_0$ deforms to $\Sigma_3H=\Sigma_3(H_0+W)$, as the intensity ($L^2$ norm) of a nonlinear defect mode is increased along the bifurcation branch.  We focus on one such scenario here and in more detail in section~\ref{sec:embedded}.  The other scenarios are discussed briefly in section~\ref{sec:scenarios}.

The first scenario occurs when $\Sigma_3H_0$ has a symmetric pair of real {\it embedded} eigenvalues within the continuous spectrum. Generically these perturb, for $\Norm{W}$ positive and arbitrary, to two pairs of complex conjugate eigenvalues and therefore yield instability.
The mechanism is related to the notion of {\it Krein signature}. A detailed perturbation calculation demonstrating this instability is presented in section \ref{sec:embedded}, showing that the instability is of order $\Norm{W}^2$, or equivalently, of order equal to the {\it fourth power} of the defect mode amplitude. In the particular examples we study numerically, these instability rates are observed, but are seen in some cases to be quite small.
\begin{figure}
\begin{center}
\includegraphics[width=4in]{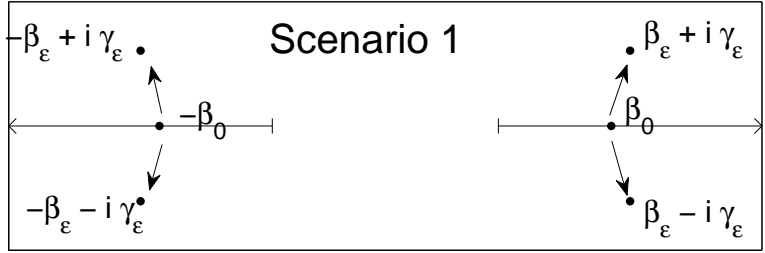}
\caption{Instability onset scenario 1: as the amplitude of the nonlinear defect mode is increased from $0$ to $\eps$ the embedded frequencies $\pm \beta_0$ split into complex frequencies $\pm \beta_\eps \pm i \g_\eps$.}
\label{fig:scenario1}
\end{center}
\end{figure}

In section \ref{sec:timedependent} we explore the temporal dynamics of the initial value problem for NLCME via direct numerical simulation.  We present simulations with various defects supporting one, two and three modes which illustrate both the spectral instability scenarios explored in section \ref{sec:linearization} and the temporal dynamics of energy exchange among defect modes and radiation modes. The mechanisms are similar to, but more complicated than, those studied by Soffer and Weinstein \cite{SofWei:03,SofWei-PRL:05} for the nonlinear Schr\"odinger / Gross-Pitaevskii (NLS / GP) equation,  who consider NLS / GP with a ``defect potential'' supporting two bound states and thus, two branches of nonlinear defect modes (ground and excited), which compete for energy confined by the potential; see also \cite{Tsai-Yau:02}. In the NLS/GP problem, resonance of an embedded eigenvalue, associated with the linear excited bound state,  with the continuous spectrum leads to energy transfer from the excited state to the ground state and to radiation modes. The asymptotic distribution of excited state energy is also studied; see \cite{SofWei:03,SofWei-PRL:05,GW:07}. 

We recap the structure of the paper. In  section~\ref{sec:prelim}, we introduce NLCME, its gap solitons, and two families of defects and their linear and nonlinear bound states.  We also summarize the results of a previous paper in which these defects are used to trap gap solitons.  In section~\ref{sec:NLS} we review some analogous results for the NLS equation with a localized defect.  Section~\ref{sec:linearization} discusses the properties of the linearized operator, outlining several types of instability with particular attention to the bifurcations of \emph{embedded} frequencies. Section~\ref{sec:evans} provides a brief introduction to the Evans function, a useful analytical tool for exploring the discrete spectrum of the linearized operator, as well numerical results using the Evans function to study the spectral stability of nonlinear defect modes.  Section~\ref{sec:timedependent} shows time-dependent simulations confirming the stability predictions of the previous section. Section~\ref{sec:conclusion} contains a short summary and discussion. The appendices contain a list of symbols, a derivation of NLCME with potentials from the problem of wave propagation in a Bragg grating with defects, and a detailed description the numerical measures that are taken to ensure accurate computations of both the nonlinear defect modes and the discrete spectrum of the linearization.

\section{NLCME and NLCME with defects}
\label{sec:prelim}
\subsection{The Nonlinear Coupled Mode Equations}
\label{sec:NLCME}
Consider an electromagnetic wave-packet, a pulse which is spectrally concentrated about a {\it carrier} wavelength, propagating in a waveguide whose refractive index varies periodically in the direction of propagation. If carrier wavelength is in resonance with the waveguide periodicity (Bragg resonance), then backward and forward waves couple strongly.

Envelope equations for these forward and backward wave amplitudes satisfy the nonlinear coupled mode equations. In appendix~\ref{sec:nlcme-sketch} we show that localized defects in the periodic structure give rise to the following modification of the NLCME, which we call by the same name, as derived in~\cite{GSW,We-Bell:99}:
\begin{equation}
\begin{split}
i\pT\Eplus + i \pZ\Eplus
+ \k(Z) \Eminus + V(Z)\Eplus
+ (\abs{\Eplus}^2 + 2\abs{\Eminus}^2)\Eplus
&=0 \\
i\pT\Eminus - i \pZ\Eminus
+ \k(Z) \Eplus + V(Z) \Eminus
+ (\abs{\Eminus}^2 + 2\abs{\Eplus}^2)\Eminus
&=0.
\end{split}
\label{eq:NLCME}
\end{equation}
The coefficient functions (``potentials'') $\k$ and $V$ are determined from the refractive index profile. 
Here, $Z$ is the ``slow'' coordinate along the direction of propagation,  $T$ is a slow time variable and the full electric field is given by
$$
E= \Eplus(Z,T)e^{i(z-t)} + \Eminus(Z,T)e^{-i(z+t)} +{\rm c.c.},
$$
i.e. $\Eplus$ and $\Eminus$ are complex envelopes of rapidly varying electromagnetic fields, assumed to be small amplitude and linearly polarized. The length and time variables $(z,t)$ are chosen such that the wavenumber and frequency are both one, i.e.\ if in dimensional variables $(\tilde z,\tilde t)$, the carrier wave has wavelength $2\pi/\lambda$ and period $2\pi c/ (n \lambda)$, then we choose $z=2\pi\tilde z/\lambda $ and $t=2\pi c\tilde t/ (n \lambda)$.  The condition that the grating is uniform (exactly periodic) away from the defect region implies that $\k(Z) \to \kinf$, constant, and $V(Z) \to 0$ as $\abs{Z} \to \infty$. 
Defining $\vec E = \binom{\Eplus}{\Eminus}$, equation~\eqref{eq:NLCME} may be rewritten as
\begin{equation}
\left( i\D_T +  i\sigma_3\D_Z + V(Z) + \k(Z)\sigma_1 \right) \vec E +\cN(\vec E,{\Vec E}^*)\vec E=0,
\label{eq:nlcme1}
\end{equation}
Where $\sigma_1$ and $\sigma_3$ are the Pauli matrices
$\sigma_1 = \left( \begin{smallmatrix} 0 & 1 \\ 1 & 0\end{smallmatrix}\right)$ and 
$\sigma_3 =  \left( \begin{smallmatrix}1 & 0\\ 0 & -1\end{smallmatrix}\right)$,
the superscript asterisk represents complex conjugation, 
and $\cN$ represents the nonlinear term of~\eqref{eq:NLCME},
\begin{equation}
\cN(\vec E,{\vec E}^*) = 
\begin{pmatrix}
\abs{\Eplus}^2+2\abs{\Eminus}^2 & 0\\  0 & \abs{\Eminus}^2 + 2 \abs{\Eplus}^2
\end{pmatrix}.
\label{eq:nonlinearity}
\end{equation}
A detailed derivation, including careful accounting of the nondimensionalization and scalings, is given in~\cite{GSW,GWH,We-Bell:99}.   

The NLCME system~\eqref{eq:NLCME} has two conserved quantity, the total intensity $I(\vec E)$ and the Hamiltonian (energy) $H(\vec E)$. The total intensity, $I(\vec E)$,  is defined as
\begin{equation}
\label{eq:intensity}
I(\vec E) = \infint \left( \abs{E_+}^2 + \abs{E_-}^2 \right) dZ,
\end{equation}
i.e.\ the square of the $L^2$ norm of the solution. The Hamiltonian is given by
\begin{multline}
H(\vec E) = \infint \Big(
iE_+ \pZ E_+ - i E_- \pZ E_- 
+  \k(Z) \left( E_+ E_-^* + E_+^* E_- \right)\\
+ V(Z) \left(\abs{E_+}^2 + \abs{E_-}^2\right) 
+ 2\abs{E_+}^2  \abs{E_-}^2
+ \frac{1}{2} \abs{E_+}^4 + \frac{1}{2} \abs{E_-}^4 
\Big) \ dZ
\label{eq:H}
\end{multline}
In the absence of a defect ($\k(Z)$ constant, $V(Z)\equiv 0)$, NLCME also conserves momentum, as a consequence of Noether's theorem.

The primary focus of this paper is the standing wave solutions of~\eqref{eq:nlcme1} of the form 
$$\vec E(Z,T) = \vec \cE(Z) e^{-i \w T},$$ 
for which~\eqref{eq:nlcme1} reduces to  the \emph{nonlinear eigenvalue problem}
\begin{equation}
\label{eq:stationary}
\left( \w +\  i\sigma_3\D_Z  + V(Z) + \k(Z)\sigma_1 \right)
\vec \cE +\cN(\vec \cE,\vec \cE^*)\vec \cE=0, \, \vec\cE \in L^2 \times L^2.
\end{equation}
.
\subsection{Constant coefficient NLCME---The spectral gap and ``gap solitons"}
\label{sec:gapsolitons}
NLCME with a uniform grating has $\k(Z) \equiv \kinf$, constant, and
$V(Z)\equiv 0$, and has been studied extensively. In the linearized system, setting the cubic terms to zero, one may look for solutions of the form
$$
\binom{\Eplus}{\Eminus}(Z,T) = e^{i (k Z -\w T) }\binom{e_+}{e_-}
$$
and find the dispersion relation $\w^2 = \kinf^2 + k^2$, so that for frequencies $\abs{\w}<\kinf$,  $k$ is purely imaginary, i.e.\ there exists a \emph{spectral gap} $(-\kinf,\kinf)$.  This results in the reflection of light at the resonant frequencies in the gap.

The fully nonlinear coupled mode system was shown in~\cite{AW,CJ} to support uniformly propagating bound states called gap solitons, parameterized by a velocity $v$ and a detuning parameter $\d$  satisfying $\abs{v}<1$ and $0 \le \d \le \pi$: 
\begin{equation}
\Epm = \pm \Delta^{\mp 1} 
\alpha \sqrt{\frac{ \kinf}{2}}\ (\sin{\d}) \
e^{i(\eta+ \sigma) } \sech{(\theta \mp i \d/2)}\enspace;
\label{eq:gapsoliton}
\end{equation}
where:
\begin{gather*}
\Delta = \left(\frac{1-v}{1+v}\right)^{\frac{1}{4}}\enspace ; \enspace
\alpha = \sqrt{\frac{2(1-v^2)}{3-v^2}}\enspace ;  \enspace
e^{i\eta}= \left( - \frac
{e^{2\theta} + e^{- i \d}}
{e^{2\theta} + e^{i \d}} \right)^{\frac{2v}{3-v^2}};\\
\theta =\frac{ \kinf}{\sqrt{1-v^2}}(\sin{\d})(z - v t)\enspace ; \enspace
\sigma =  \frac{ \kinf}{\sqrt{1-v^2}}  (\cos{\d})(v z - t).
\end{gather*}

In the Lorenz-shifted reference frame, the term $e^{i\sigma}$ is responsible for an internal oscillation with frequency $\kinf \cos \d$ which is \emph{inside} the spectral gap, the nonlinearity having served to effectively shift the gap. This is also known as self-induced transparency. A comprehensive introduction to gap solitons is given in the review paper by de Sterke and Sipe~\cite{dSS}. Gap solitons may propagate in the laboratory reference frame at any velocity $v$ below the speed of light (here $c=1$ by the 
 leading to~\ref{eq:scales}) and thus are intriguing as possible components of all-optical communications systems.  Zero-speed waves, for example, are candidates for bits in an optical buffer or memory device.

\subsection{Defect modes of the linear coupled mode equations}
\label{sec:linear}
In the zero-intensity limit, we ignore the nonlinear term in~\eqref{eq:stationary}, giving a linear eigenvalue problem
\begin{align}
\label{eq:linear}
&\left(\ \w_* I\ +\ h_{V,\k}\ \right)\ \vec \cE_*\ \equiv\ \left( \w_*I +\  i\sigma_3\D_Z  + V(Z) + \k(Z)\sigma_1 \right)
\vec \cE_* =0,  \nn\\
&\ \ \ \ \  \vec\cE_* \in L^2 \times L^2.
\end{align} 
\emph{Hereafter, frequencies and vectors with subscript  asterisks will refer to solutions the linear eigenproblem~\eqref{eq:linear}, and those without will refer to solutions of~\eqref{eq:stationary}}.

  Note that $h_{V,\k} $ is self-adjoint on $L^2 \times L^2$ and therefore has real spectrum.
Moreover, we assume  $V(Z) \to 0$ and $\k(Z) \to \kinf$ rapidly  as $\abs{Z}\to\infty$, and therefore this spectral problem has two branches of continuous spectrum and a finite number of discrete frequencies given by 
\begin{equation}
\spec\left( \w_* I+h_{V,\k}\right)\ =\  \spec_{\rm continuous} \cup \spec_{\rm discrete} = \{\w_* \in {\mathbb R} : \abs{\w_*} \ge \kinf\} \cup 
\{ \w_{j*}: j = 1,\ldots, N\},
\nn\end{equation}
where the discrete frequencies satisfy $-\kinf < \w_{j*} < \kinf$.

We consider two families of defects in the current study which we refer to as ``even defects'' and  ``odd defects''.

\subsubsection*{ (a) Even defects}
If we specify $\k(Z)$ to be even and $V(Z)=0$, then it is simple to show that if $\w_*$ is a real eigenvalue of~\eqref{eq:linear}, then so is $-\w_*$.  A simple argument eliminates the possiblity of nontrivial null eigenstates. Thus, system~\eqref{eq:linear} must have an even number of discrete eigenvalues with corresponding $L^2$ eigenfunctions.  As we do not know of any $\k(Z)$ for which the eigenvalue problem~\eqref{eq:linear} may be solved in closed form, we choose a relatively simple form of $\k(\cdot)$,
\begin{equation}
\label{eq:kappa}
\k(Z) = 1 - b \sech(k Z).
\end{equation}
Thus, $\kinf = 1 $ and the linear operator $\w_*I+h_{0,\k}$ has continuous spectrum $\{\w \in {\mathbb R}:  \abs{\w} \ge 1\}$.  We have found via numerical simulation that as $b$ is increased with $k$ held fixed, the existing frequencies migrate toward the origin, and new discrete frequencies bifurcate in pairs from opposite endpoints of the band gap at $\w=\pm 1$.  Such emergence from the edges has been studied for related problems, for example, in~\cite{KivPelCre:98,Rau:80}.
\subsubsection*{ (b) Odd defects}

In~\cite{GSW}, we construct a three-parameter family
\footnote{By a rescaling of the space and time coordinates, we may set $k=1$, so in reality this is a two-parameter family, but the above form makes it easier to create different defects with the same limiting grating strength $\kinf$.}
of defects of the form
\begin{equation}
\k(Z) = \sqrt{\w^2 + n^2 k^2 \tanh^2 {(kZ)}}; \qquad
V(Z) =  \frac{\w n k^2 \sech^2{(kZ)}}
               {2(\w^2 + n^2 k^2 \tanh^2{(kZ)})}. 
\label{eq:potential}
\end{equation}
with $k>0$, $n>0$, and $\w$ nonzero.
\footnote{In the special case $\w=0$, then $V(Z)=0$ and $\k(Z)=n k \tanh{(kZ)}$.  The expressions for the eigenmodes are also altered.}
Standing wave solutions exist of the form:
\begin{align}
&\vec \cE_{(0)*} = \binom{e^{ i\Theta}}{-\fsign e^{-i\Theta}} \sech^n{(kZ)} e^{-i\w_{0*} T} ,\ \ \w_{0*}= \w,\  \fsign = \sign{\w}
\label{eq:Epm} \\
&\Theta=\int_0^Z V(\z)\ d\z=\frac{1}{2} \arctan{\frac{n k\tanh{(kZ)}}{\w}}. \label{eq:Theta}
\end{align}
For $n>1$, the defect supports a total of $2\lfloor n\rfloor -1$ distinct eigenvalues, where
$\lfloor n \rfloor$ is the greatest integer less than or equal to $n$.  Its eigenvalues are 
\begin{equation} 
\w_{\pm j*}=\pm \sqrt{\w^2 + (2nj-j^2)k^2}; \;  
j=1,\dots, n-1.
\label{eq:higher_eigs} 
\end{equation}
If $\w$ and $k$
are held constant, while $n$ is increased, the ``ground state'' eigenvalue
$\w_{0*}=\w$ remains constant, while the other eigenvalues move from the edge of
the spectral gap toward $\pm \w_{0*}$.  Each time $n$ passes through an integer
value, two new eigenvalues are created in \emph{edge bifurcations} from the two ends of the continuous spectrum. 

Reference~\cite{GSWK} contains the general formula for the eigenfunctions. The
solutions can ultimately be expressed in terms of hypergeometric functions of
$\tanh{kZ}$, which simplify to Legendre functions of $\tanh{kZ}$ when $n \in
\mathbb{Z}$, and can be expressed as  algebraic
combinations of $\tanh{kZ}$ and $\sech{kZ}$.  The general formula is
quite complicated. 

In much of what follows we specialize to the case $n=2$, in which case there are three linear defect modes.  The first eigenfunction, with frequency
$\w_0=\w$, is given  by~\eqref{eq:Epm}.  The remaining two, with
frequencies $\w_{\pm 1*}= \pm \sqrt{\w^2+3k^2}$, are given (in a non-normalized form) by 
\begin{equation}
\label{eq:nextmodes}
\vec \cE_{(\pm 1)*} = 
\binom{(k+i(\w_{\pm 1*}+\w)\tanh{kZ})e^{i\Theta(Z)}}
{\fsign\left(k-i(\w_{\pm 1*}+\w)\tanh{kZ}\right)e^{-i\Theta(Z)}} \sech{kZ}.
\end{equation}

\subsection{Defect modes of \emph{nonlinear} coupled mode equation}
\label{sec:nlmodes}
  
Consider equation~\eqref{eq:stationary}, the nonlinear eigenvalue problem solved by the nonlnear defect modes.  In the small amplitude (linear) limit, we expect solutions to be well-approximated by those of the linear eigenvalue problem~\eqref{eq:linear}, whose explicit eigenstates are displayed in section~\eqref{sec:linear}.

In~\cite{GSW}, we used perturbation analysis to show that in analogy with the NLS/GP case~\cite{RosWei:88}, there exist nonlinear defect mode solutions of~\eqref{eq:stationary}
\begin{align}
\vec \cE_{(j)}(Z)\ &=\ \alpha\left(\  \vec \cE_{(j)*} (Z) + O(\abs{\alpha}^2)\ \right) \text{ and } \\
\w &=  \w_{j*} + q \abs{\alpha}^2 +  O(\abs{\alpha}^4), \, \alpha \in {\mathbb C}, \, \alpha \to 0\\
q &= - \frac{\langle \cE_{j*} , \cN(\cE_{j*},\cE_{j*}^*) \cE_{j*}\rangle}
{\langle \cE_{j*} , \cE_{j*} \rangle} 
\end{align}
Notice that $q<0$, due to the focusing character of the nonlinearity. Thus, as the amplitude is increased, the nonlinear frequency is shifted toward the left edge of the spectral gap. 

We aim to compute the nonlinear modes and their associated frequencies as accurately as
possible. This is important so that errors in the numerical solutions contribute as little as possible to errors in their numerically calculated linearized spectrum and determination of stability.  Details of the numerical calculation are provided in appendix~\ref{sec:nl_numerics}.  Briefly, the derivatives are discretized using Fourier transforms, while the infinite interval is truncated using a nonuniform-grid method and the resulting algebraic equations derived are solved using MINPACK routines~\cite{minpack}.  
 
\begin{figure}
\begin{center}
\includegraphics[width=3in]{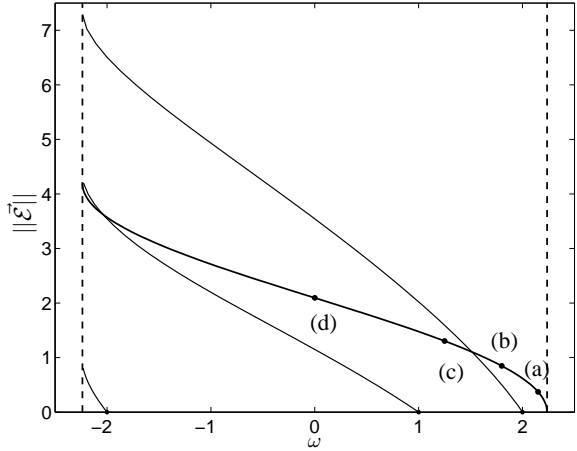}
\caption{The intensity as a function of the frequency of the numerically calculated nonlinear defect modes, with parameter values $\w=1$, $k=1$, $n=2$.  The band edges are at $\pm
  \sqrt{5}$ and the linear eigenvalues are $\w_0=1$ and $\w_{\pm 1} = \pm 2$.  The darker curve is plots the amplitude and frequency of a stationary gap soliton~\eqref{eq:gapsoliton}.}
\label{fig:all}
\end{center}
\end{figure}

Figure~\ref{fig:all} shows a typical example of the dependence of the defect mode's frequency
on its amplitude.  The frequency of each of the three defect modes shifts to
the left as the intensity is increased; $q<0$.  At small intensities the frequency
depends linearly on the intensity, and at larger intensities, the curves
steepen near the left band edge. We believe that these branches terminate at the left band edge, but the nonlinear modes decay very slowly as the edge is approached, and thus become difficult to calculate numerically.

\subsection{Summary of previous numerical experiments}
In~\cite{GSW}, we explored the behavior of gap solitons incident on localized
defects of the form~\eqref{eq:potential}.  We found
parameter regimes in which the soliton---or at least the energy it carries---can
be captured.  We hypothesized the existence of a 
nonlinear  resonance mechanism between the soliton, and a \textit{nonlinear} defect
mode.  By~\eqref{eq:gapsoliton}, the gap soliton oscillates with an internal frequency of about~$\kinf \cos \d$ as it propagates.  If there exists a nonlinear defect mode of the same frequency and smaller $L^2$-norm, then the solitary wave may transfer its energy to the defect mode. In particular, as  $\d$ is increased from $0$ to $\pi$ this internal frequency decreases from $\kinf$ to $-\kinf$.  This is the thickest curve in figure~\ref{fig:all}; a family of states of the translation invariant NLCME, which bifurcates from the zero state at the band edge frequency.

 For sufficiently small $\d$ (frequency near the right band edge, marked \textbf{(a)} in the figure), there exists no nonlinear defect mode of the same frequency and lower amplitude to which the gap soliton can transfer its energy.  In this case, the defect behaves as a barrier.  For solitary waves above a critical velocity, the pulse passes by the defect with almost all its amplitude and its original velocity almost unchanged.  Below the critical velocity, the solitary wave is reflected, again almost elastically.    If, by contrast, there exists a nonlinear defect mode resonant and of with the solitary wave and of smaller amplitude, then the pulse may be trapped, for example points~\textbf{(b)} and~\textbf{(d)}.  In this case there exists a critical velocity, below which solitary waves are transmitted (inelastically) and below which they are largely trapped.  Behavior at the point marked \textbf{(c)} is similar to that at \textbf{(a)}--the defect mode that bifurcates from $\w_{1*}=2$  is a larger-amplitude state than the gap soliton, so it cannot be excited, and the defect mode bifurcation from $\w_{0*}$ is not resonant with this frequency.

Similar behavior, namely the dichotomy between the nonresonant and elastic transmit/reflect behavior and the resonant and inelastic capture/transmit behavior, has also been seen for nonlinear Schr\"odinger solitons~\cite{GooHolWei:04,LeeBra:06}. Estimates for $v_{\rm c}$ in several related problems and an explanation for the existence of a critical velocity are given in~\cite{GooHab:07} and references therein.
This trapping phenomenon was subsequently seen by Dohnal and
Aceves for a two-dimensional generalization of equation~\eqref{eq:NLCME}~\cite{AceDoh:06,DohAce:05}.

\begin{figure}
\begin{center}
\includegraphics[width=2in]{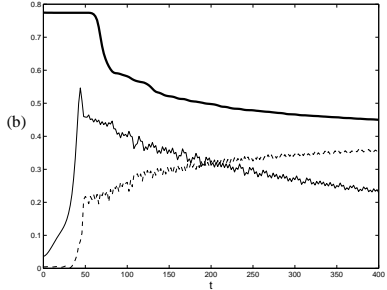}
\caption{The projections onto the $\w_2$ (thin solid line) and $\w_1$ (dashed), showing that energy is originally, and quickly transferred from the moving soliton to the $\w_2$ mode, and then slowly transferred to the $\w_1$ mode.  The thick line shows the total $L^2$ norm of the solution, which is not conserved due to absorbing boundary conditions at the endpoints.}
\label{fig:JOSAB}
\end{center}
\end{figure}

Trapping could be more accurately described as a transfer of energy from the traveling soliton mode to a stationary nonlinear
defect mode.  This is pictured in figure~\ref{fig:JOSAB} reprinted from~\cite{GSW}.  This figure describes the evolution of a pulse captured by a wide defect supporting 5 linear defect modes. Energy initially moves into the mode associated with the frequency $\w_2$ but is subsequently transferred to the mode associated with frequency $\w_1$. A part of the energy is transferred to radiation modes and is dispersed to infinity. It is thus important for us to understand the stability of these nonlinear modes, to assess their suitability as long-time containers for electromagnetic energy, as well as the mechanisms through which energy moves among discrete and continuum modes.

\section{Dynamics and energy transfer for NLS/GP}
\label{sec:NLS}
Many of the phenomena and mechanisms present in the dynamics of NLCME~\eqref{eq:NLCME} are present in the related  and more studied nonlinear Schr\"odinger / Gross-Pitaevskii (NLS/GP) equation:
\begin{equation}
\label{eq:NLS}
i\D_t \Phi=- \Delta \Phi+ V(x) \Phi+ g\abs{\Phi}^2 \Phi,
\end{equation}
where $\Phi=\Phi(x,t)$ is complex-valued, $x\in\mathbb{R}^n$, $t\in\mathbb{R}$. $g>0$ corresponds to a repulsive or defocusing nonlinear potential,
while $g<0$ corresponds to an attractive or focusing nonlinear potential.
Before embarking on a study of NLCME, we briefly review results for NLS/GP.

To fix ideas, we take $V(x)$ to be a smooth potential well ($V\le0$), decaying to zero rapidly  as $|x|\to\infty$. NLS/GP is a Hamiltonian system with conserved Hamiltonian energy 
\begin{equation}
\cH[U]= \int_{\mathbb{R}^n} \left(\abs{\nabla U}^2 + V(x) \abs{U}^2 + \frac{g}{2}\abs{U}^4\right)dx.
\label{nls-gp-energy}
\end{equation}

Solitary standing wave  solutions are solutions of the form: $
 \Phi(x,t) = \Psi(x)e^{-i\Omega t}$, from which we obtain the nonlinear eigenvalue problem for $u(x)$
\begin{equation}
\label{eq:NLSomega}
\Omega \Psi =  - \Delta \Psi + V(x) \Psi+g\abs{\Psi}^2 \Psi; \; \Psi \in L^2
\end{equation}
In the low intensity (linear) limit, (\ref{eq:NLSomega}) reduces to the  the linear eigenvalue problem for the Schr\"odinger operator $-\Delta+V$.
 
The operator $-\Delta+V$ has continuous spectrum covering the non-negative half-line: $\Omega\ge 0$ and, in general,  a finite number of negative discrete eigenvalues 
$\Omega_{0*} < \Omega_{1*} < \ldots < \Omega_{n*} <0$, with corresponding eigenfunctions $\psi_{j*},\ j=0,\dots,n$,\ $\| \psi_{j*}\|_2=1$; recall subscript asterisks denote solutions in the linear limit.
  The eigenfunction $\Psi_{0*}(x)$ corresponding to $\Omega_{0*}$ is the {\it ground state}, a minimizer of the (linearized) energy  
  $\cH_{\rm linear}[U]=
  \int \left(\abs{ \nabla U}^2 + V(x) \abs{U}^2\right)\ dx$, 
  subject to the constraint: $\norm{U}{2} = 1$. Nonlinear defect mode families, $(\Psi_{\alpha_j}(x),\Omega_{\alpha_j}(x))$,  bifurcate from the zero state at the linear eigenfrequencies \cite{RosWei:88}. The  nonlinear ground state, $(\Psi_{\alpha_0},\Omega_{\alpha_0})$ can also be characterized variationally as a minimum of $\cH[U]$ subject to the constraint  $\norm{U}{2} =\alpha$. For small $L^2$-norm we have  for $j=0,1,\dots$ and $\alpha\to0$
  \begin{align}
& \Psi_{\alpha_j}(x) = 
 \alpha \left(\ \psi_{j*}(x)\ +\ \cO(|g|\ |\alpha|^2)\ \right)\nn\\
& \Omega_j(\alpha) = \Omega_{j*}\ +\ gc_{j*}^2 |\alpha|^2\ +\ \cO(g^2|\alpha_j|^4),\ \ \alpha_j\in{\mathbb C}
\nn\end{align}
 where $c_{j*}^2>0$ is a positive constant.

Soffer and Weinstein \cite{SofWei:03,SofWei-PRL:05}
 studied the dynamics of  the initial value problem for NLS/GP 
  (\ref{eq:NLS}) in the case where the potential well, $V(x)$,  supports exactly two  eigenstates, ``linear defect modes'', $(u_{0*}(x),\Omega_{0*})$ and $(u_{1*}(x),\Omega_{1*})$.  By the above discussion, there exist two bifurcating branches of nonlinear bound states $(\Psi_{\alpha_j},\Omega_{\alpha_j}),\ j=0,1$. These are used to parameterize general small amplitude solutions of NLS/GP:
 \begin{equation}
 \Phi(x,t)\ \sim\ \Psi_{\alpha_0(t)}(x)\ +\ \Psi_{\alpha_1(t)}(x)\ +\ \Phi_{\rm dispersive}(x,t)
 \nn\end{equation}
  and it is shown that generic finite energy solutions of the initial value problem converge as $t\to\pm\infty$
  to a nonlinear ground state: $\Phi(x,t)\ \to\ \Psi_{\alpha_0^\pm},\ \alpha_0^\pm\in\mathbb{C}$, locally in $L^2$. This {\it ground state selection} phenomenon has been experimentally observed; see ~\cite{ManLahSil:05}.
  
  The very large time dynamics, which involve energy leaving the excited state and being transferred to the ground state and radiation modes is  governed by {\it nonlinear master equations}
  \begin{align}
 & \frac{dP_0(t)}{dt}\ \sim \G P_1^2(t)P_0(t)\nn\\
&   \frac{dP_1(t)}{dt}\ \sim -2\G P_1^2(t)P_0(t).
   \label{nl-master}
   \end{align}
   Here, $P_j(t)\sim |\alpha_j(t)|^2$ and $\G=\cO(g^2)$ is non-negative and generically strictly positive, provided an arithmetic condition implying
coupling to of the discrete to continuum radiation modes at second order in $g$:
\begin{equation}
\label{eq:NLS_resonance}
2\Omega_{1*}- \Omega_{0*}  >0.
\end{equation}
The expression for $\G$ is a nonlinear analogue of Fermi's golden rule (see also \cite{SW-Inventiones:99}),  which arises in the  calculation of the spontaneous emission rate of an atom to its ground state
 \cite{Cohen-Tannoudji-etal}.

 The resonance condition (\ref{eq:NLS_resonance}) also appears in a natural way in the linearized  (spectral) stability analysis of nonlinear defect modes. In particular, consider the linearization about a nonlinear excited state. This  yields a linear spectral problem of the form: $\sigma_3H Y= \beta Y$, where $H$ is a two by two self-adjoint matrix operator and $\sigma_3$ is the standard Pauli matrix. Now $H=H_0+W$, where $W$ tends to zero quadratically in the nonlinear excited state amplitude. Since $W$ is spatially localized, the continuous spectrum of $\sigma_3H$ is  given by two semi-infinite intervals, the complement of an open symmetric interval about the origin. 
 Now {\it under the resonance condition (\ref{eq:NLS_resonance}), 
   $\sigma_3H_0$ has an embedded eigenvalue within the continuous spectrum}. The perturbation theory of embedded eigenvalues is a fundamental problem in mathematical physics; see, for example, \cite{RS4:78,SW-TDRT:98}. These embedded eigenvalues can be shown generically to perturb for $W$ arbitrarily small to complex eigenvalues, corresponding to instabilities \cite{CPV}.
 
 In subsequent sections we explore the analogous picture for NLCME.
 In particular, in section~\ref{sec:linearization}, we obtain an analogous (more complicated)    resonance  condition, yielding embedded eigenvalues for a spectral problem, perturbing to instabilities and 
 analogous dynamics of energy transfer among modes.
 \medskip
 
 We conclude this section by considering numerical simulations for  NLS/GP, with a simple family of potentials that support two linear defect modes is 
 \begin{equation}
 V_L(x) = -2 \sech^2{(x-L)} - 2 \sech^2{(x+L)}.
 \nn\end{equation}
 More detailed numerical simulations and their interpretation in light of 
 \cite{SofWei:03,SofWei-PRL:05} is given in \cite{ShlWei:07}.
  
A potential well $V=-2 \sech^2{x}$ has a single localized eigenfunction $u = \sech{x}$ with frequency $\Omega = -1$.  For all values of $L$, the potential $V_L(x)$ supports exactly two stationary solutions.  For $L$ sufficiently large, the normalized ground state is given by $\psi_{0*} \approx  2^{-\frac{1}{2}}\ \left(\sech{(x-L)} +\sech{(x+L)}\right)$, with frequency $\Omega_{0*} \approx -1 - \eps(L)$ where $\eps(L)$ is positive and exponentially small in $L$. The excited state is  $\psi_{1*} \approx  2^{-\frac{1}{2}}\left(\sech{(x-L)} -\sech{(x+L)}\right)$, with frequency $\Omega_{1*} \approx -1 +\eps(L)$ .  As $L$ is decreased, $\Omega_{0*}$ increases and $\Omega_{1*}$ increases toward $-1$.  For $L < 1.13$, condition~\eqref{eq:NLS_resonance} is satisfied.  

\begin{figure}
\begin{center}
\includegraphics[width=3in]{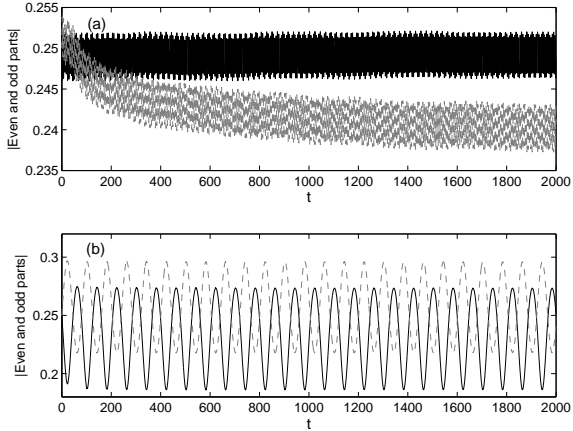}
\caption{The projections on the ground state (black) and excited state (gray, dashed) of numerical solutions to equation~\eqref{eq:NLS}.  In (a), the potential is defined with $L=1$ leading to a resonance.  In (b), the potential is defined with $L=2$, leading to no resonance between the two modes.  Plotted are not the amplitudes of projections onto the two modes, but the $L^2$-norm of the even and odd projections of the solution, which are less noisy.}
\label{fig:nls2well}
\end{center}
\end{figure}

We have simulated the initial-value problem for this problem with initial conditions given by a linear combination of the two eigenfunctions, in both the resonant and non-resonant cases.  The results of these simulations is shown in figure~\ref{fig:nls2well} and the behavior in the two cases is strikingly different.  Subfigure (a) shows the case $L$=1, in which the ground state grows slightly, while the excited state  decays.  In addition to this transfer of energy, there is a fast oscillation in the amplitudes of the two modes.  Subfigure (b) shows the nonresonant case $L=2$.  Here there is a much larger amplitude oscillation between the two modes, but none of the one-way energy transfer. In section~\ref{sec:timedependent}, we perform analogous numerical experiments, which confirm analytical / numerical predictions of section~\ref{sec:evans}.

\section{Linearized stability analysis of nonlinear defect modes}
\label{sec:linearization}
We now study the linearization of the variable-coefficient NLCME~\eqref{eq:NLCME} about a nonlinear defect mode. In the introduction, we reviewed one type of instability that may arise. The analysis of subsection~\ref{sec:embedded} is focused on this scenario---instabilities arising from perturbations of eigenvalues, embedded within the continuous spectrum.
We first discuss the conditions under which, in the limit of vanishing-amplitude defect modes, the linearization contains discrete eigenvalues embedded in the continuous spectrum, and then develop a time-dependent perturbation theory that shows when these embedded modes may lead to instability.  In section~\ref{sec:scenarios}, we discuss other types of instability that may arise.

\subsection{Conditions for embedded eigenvalues in the linearization}
\label{sec:embed_cond}
Letting $\vec E  =\binom {\cE_+}{\cE_-} e^{-i\w t}$ be a solution to the
nonlinear eigenvalue problem~\eqref{eq:stationary} constrained to have
intensity $I$, we linearize about $\vec E$ by letting 
\begin{align*}
\Eplus &= ( {\cE_+} + y_1(Z,T))e^{-i\w T} &  E^*_+ &= ( \cE^*_+ + y_3(Z,T)) e^{i\w T} \\
\Eminus &= ( {\cE_-} + y_2(Z,T)) e^{-i\w T} &  E^*_- &= ( \cE^*_- + y_4(Z,T)) e^{i\w T} .
\end{align*}
Letting $y_i(Z,T)=y_i(Z) e^{-i\beta T}$ yields the eigenvalue problem for $\beta$ and an  $L^2$ function $\vec y = (y_1,y_2,y_3,y_4)^{\rm T}:{\mathbb R}\mapsto{\mathbb C}^4$
\begin{equation}
\label{eq:eigen}
\beta \vec y(Z)=-\Sigma_3 \cdot \left(V(Z) + \w + i
\begin{pmatrix}
\sigma_3 & 0 \\ 0 & -\sigma_3
\end{pmatrix}
\pZ 
+\begin{pmatrix}
\sigma_1 & 0 \\ 0 & \sigma_1
\end{pmatrix}\k(Z) + W(Z) \right)
\vec{y}
\end{equation}
where $\Sigma_3=\left(\begin{smallmatrix} I & 0 \\ 0 & -I\end{smallmatrix}\right)$ is a $4\times4$ Pauli matrix. The matrix multiplication operator $W(Z)$ is given by
\begin{equation}
W = \begin{pmatrix}
2 \abs{\vec \cE}^2 & 
2 \cE_+ \cE^*_- &
\cE_+^2 &
2\cE_+ \cE_- \\
2\cE^*_+ \cE_- &
2 \abs{\vec \cE}^2 & 
2 \cE_+ \cE_- &
\cE_-^2  \\
{\cE^*_+}^2 &
2 \cE^*_+ \cE^*_- &
2 \abs{\vec \cE}^2 &
2\cE^*_+ \cE_- \\
2 \cE^*_+ \cE^*_- &
{\cE^*_-}^2 &
2 \cE_+ \cE^*_- &
2 \abs{\vec \cE}^2.
\end{pmatrix}
\label{Wdef}
\end{equation}
An eigenfunction is a square-integrable solution of~\eqref{eq:eigen},
 with corresponding discrete eigenvalue $\beta$.
Thus if any non-trivial solutions exist with $\Im \beta > 0$, the nonlinear defect mode is
unstable.   

\begin{figure}
\begin{center}
\includegraphics[width=4in]{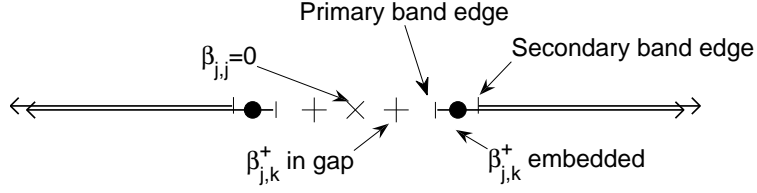}
\caption{A schematic depicting the spectrum of the linearization about the defect mode at zero amplitude.}
\label{fig:spec_linearized}
\end{center}
\end{figure}

As $\Norm{\vec \cE_{(j)}} \searrow 0$, the spectrum of the linearization about a given defect mode $\vec{\cE}_{(j)}$ approaches
$$
\spec_{\rm linearized} = 
\{\pm( \w_{j*} - \w_*) : \w_* \in \spec_{\rm linear}\}
$$
In particular, from the form of equation~\eqref{eq:eigen} in the limit of $\abs{Z}\to\infty$,  the branches of continuous spectrum must be given by 
$$ 
\spec_1=\{\beta : \abs{\beta} \ge \kinf - \abs{\w_{j*}}\}
$$
and 
$$ 
\spec_2=\{\beta : \abs{\beta} \ge \kinf + \abs{\w_{j*}}\}
$$
so that spectrum has multiplicity two for $\abs{\beta} \ge\kinf + \abs{\w_{j*}} $.   In addition the discrete spectrum is given by
$$
\spec_{\rm discrete} = \{\beta_{j,k}^{\pm} =\pm(\w_{j*} - \w_{k*}) : 1 \le k \le N\}.
$$
This implies that the spectral gap is the interval
\begin{equation}
{\rm gap}=(-\kinf + \abs{\w_{j*}}, \kinf-\abs{\w_{j*}})
\label{eq:gap}
\end{equation}
This is summarized in figure~\ref{fig:spec_linearized}.
If for some $k\in\{1,\dots,N\}$,   
\begin{equation}
\label{eq:embed_cond}
\abs{\beta_{j,k}^\pm} \ge  \abs{\kinf} - \abs{\w_{j*}}.
\end{equation} 
then the discrete eigenvalue $\beta_{j,k}^{\pm}$ is embedded in the continuous spectrum.
We will refer to the boundary between the band gap and the multiplicity-one continuous spectrum as the ``primary band edge'' and the boundary between multiplicity-one and multiplicity-two continuous spectrum as the ``secondary band edge.''

\noindent \textbf{Example 1} \emph{For ``even'' defects of the form~\eqref{eq:kappa} with exactly two discrete eigenvalues $\w_{-1*}=-\w_{1*}$, therefore there will exist an embedded frequency if $\w_{1*}-\w_{-1*} = 2\w_{1*}< \kinf-\w_{1*}$, i.e. if $3 \w_{1*}>\kinf$.  Since $\kinf=1$ in this example, the condition for an embedded eigenvalue is $\w_{1*}>1/3$.}

A similar calculation shows that for ``odd'' defects of the form~\eqref{eq:potential}, the linearization about the ``excited states,'' $\cE_{(\pm 1)*}$ always produces embedded frequencies, while the linearization about the ``ground state'' $\cE_{(0)*}$ has an embedded frequency if $\w/k \ge 1/\sqrt{56}$.  Under perturbation, the frequencies in the gap will generically remain real at small amplitude, while the embedded frequencies will be shown to move off into the complex plane, giving growing modes.

\subsection{Perturbation theory of embedded eigenvalues---time dependent approach}
\label{sec:embedded}

In example 1, we observe that there are cases when the linear spectral problem~\eqref{eq:eigen}, in the limit of vanishing nonlinear defect mode amplitude, has eigenvalues embedded in the continuous spectrum.  The perturbation theory of embedded eigenvalues in continuous spectrum is an important fundamental question in mathematical physics~\cite{RS4:78,SW-TDRT:98}.  In the self-adjoint case, such embedded eigenvalues perturb to \emph{resonances}, time-decaying states; self-adjointness precludes instability.  In the present non-self-adjoint setting, instability cannot be precluded.  

We sketch a time-dependent approach to the theory of embedded eigenvalues~\cite{SW-TDRT:98,SofWei:03,SofWei-PRL:05} to study instabilities which arise. A time-independent approach can also be applied; see~\cite{CPV}.  In particular, we sketch a proof of the following 

\begin{prop}
\label{prop1}
Consider the linear spectral problem (\ref{eq:eigen}), associated with a branch of nonlinear bound states , which bifurcates from the zero solution at frequency $\w_j$. If the linearized operator corresponding to the zero amplitude limit (the operator on the right hand side of (\ref{eq:eigen}) with $W\equiv0$) has an embedded eigenvalue in its continuous spectrum (see condition 
(\ref{eq:embed_cond}) ), then for arbitrary sufficiently small nonlinear bound states of order $\|\vec\psi\|$, the linearized evolution equation has an exponentially growing solution, with exponential rate $e^{\G t}$, where $\G\ge0$ (generically $>0$) is of order $\|\vec\psi\|^4$ and is given by the expression (\ref{Gamma-def}).
\end{prop}

Recall that the  nonlinear coupled mode equations are a system of semilinear evolution equations for complex wave amplitudes $E_+$ and $E_-$. 
Using~\eqref{eq:Theta}, the changes of variables 
$\Epm = e^{\pm i \Theta(Z)} \tilde{E}\pm$ and
$\tilde \k(Z) = \k(Z) e^{-2i\Theta} = \w - i n k \tanh{kZ}$
simplify system~\eqref{eq:nlcme1} to
\begin{equation}
\left( i\D_T +  i\sigma_3\D_Z + 
\begin{pmatrix}
0 & \tilde \k(Z) \\ \tilde\k^*(Z) & 0
\end{pmatrix}
\right) \binom{{\tilde E}_+}{{\tilde E}_-}
+\cN({\tilde E}_+,{\tilde E}_- ,{\tilde E}_+^*,{\tilde E}_-^* )
 \binom{{\tilde E}_+}{{\tilde E}_-}=0,
\label{eq:nlcme2}
\end{equation}

The linearized evolution equation, governing the perturbation:
\begin{equation}
\phi\ =\ \vec y\ =\ (y_+,y_-,y^*_+,y^*_-)e^{-i\w T}
\label{phidef}
\end{equation}
 about a fixed nonlinear defect mode has the form
\begin{equation}
i\D_t\phi\ =\ \Sigma_3\ H\ \phi\label{phi-eqn}
\end{equation}
Here, $H=H_0+W$, with $H_0$ and $W$ self-adjoint operators.
\begin{equation}
H_0\ =\ 
\begin{pmatrix}
h_0 & 0\\
0 & h^*_0
\end{pmatrix}, 
\label{Hdef}
\end{equation}
where the $4\times4$ matrix $H_0$ is naturally expressed in terms of  $2\times2$ blocks, where 
\begin{equation}
h_0\ =\ \begin{pmatrix} 
\w_j - i \pZ & -\tilde \k(Z)  \\
-\tilde \k^*(Z) & \w_j + i \pZ
\end{pmatrix}\label{h0def}   
\end{equation}
and the matrix $W$ is given above in~\eqref{Wdef}.

Note that if $\psi$ solves the spectral problem $\Sigma_3H\psi=\lambda\psi$, then $\Sigma_3\psi$ solves the adjoint spectral problem, $H\Sigma_3\psi=\lambda\psi$. If $\psi\in L^2$ we  introduce, $\tpsi$, the normalized adjoint eigenvector:
$$
\tpsi\ =\ \frac{\Sigma_3\ \psi}{\left(\Sigma_3\psi,\psi\right)}
 ,\nn\\
\ \  \text{ such that } (\tpsi,\psi)\ =\ 1,
$$
where $(\cdot,\cdot)$ denotes the inner product on ${\mathbb C}^4$-valued $L^2$ vector functions.
 
We assume that  there is a spectral decomposition associated with the operator $\Sigma_3H$. If $f\in \cD(\Sigma_3H)$, then $f$ can be decomposed in terms of its bound state and continuous spectral parts:
\begin{equation}
f\ =\ P_b\ f\ +\ P_c\ f,\nonumber
\end{equation}
where 
\begin{align}
P_b\ f\ &=\ \sum_k\ \left(\  \tpsi_k, f\ \right)\ \psi_k\nonumber\\
P_c\ f\ &=\ \int_{\sigma_c(\Sigma_3H)}\ \left(\ \tpsi_\lambda,f\ \right)\ \psi_\lambda\  d\lambda,
\label{PbPc}
\end{align}
where
\begin{equation}
\tpsi_k\ =\ \frac{\Sigma_3\ \psi_k}{\left(\ \Sigma_3\psi_k\ ,\ \psi_k\ \right)}
 ,\nn\\
 \text{ such that} \left(\ \tpsi_k\ ,\ \psi_k\ \right)\ =\ 1
 \end{equation}
 An analogous spectral decomposition holds for the operator $\Sigma_3H_0$.

Consider the situation, in which $\Sigma_3H_0$ has a simple eigenvalue $\lambda_0$ embedded in the continuous spectrum and corresponding eigenvector, $\psi_0$.  

\noindent \emph{What are the dynamics for the perturbed evolution equation $i\D_t\phi\ =\ \Sigma_3\ H \ \phi$?}

Following the analysis of ~\cite{SW-TDRT:98}, we consider the initial value problem with data given by the unperturbed state, $\psi_0$:
\begin{equation}
H\psi_0\ =\ \lambda\psi_0,\ \ \lambda_0\in\sigma_c(\Sigma_3H_0)
\label{psi0emb}
\end{equation}
For $W$ small (small amplitude nonlinear defect states), we seek a solution of the perturbed evolution equation
\begin{equation}
i\D_t\phi\ =\ \Sigma_3\left(\ H_0\ +\ W\ \right)\ \phi.
\label{phi-evolution}
\end{equation}
We show the existence of an exponential instability with growth proportional to $e^{\G t}$ where
\begin{equation}
\G = -\frac{ \pi{\abs{\left(\psi_{\tilde\lambda_0},W\psi_0\right)}^2}}
{ \left(\Sigma_3\psi_0,\psi_0\right)}.
\label{Gamma-def}
\end{equation}
Here $\psi_{\tilde\lambda_0}$ is a generalized eigenfunction corresponding to the shifted frequency ~\eqref{tlam}.  If $\lambda_0$ is an embedded eigenvalue of $\Sigma_3 H_0$ with corresponding eigenfunction $\psi_0$ having negative ``Krein signature'', $\left(\Sigma_3 \psi_0,\psi_0\right)$, then $\G>0$.
 
We seek
\begin{equation}
\phi(t)\ =\ a(t)\psi_0\ +\ \phi_1(t),\ \ \ \left(\ \tpsi_0,\phi_1(t)\ \right)\ =\ 0
\label{phiexpand}
\end{equation}
Let $P_c^\#$ denote the spectral projection (with respect to the unperturbed  operator $\Sigma_3H_0$ ) onto the 
part of the spectrum which is (i) continuous, (ii) bounded away from infinity and (iii) from {\it thresholds} (endpoints of branches of continuous spectra). In~\cite{SW-TDRT:98} it is shown that the full dynamics are subordinate to the coupled dynamics of $a(t)$ and $\phi_d(t)=P_c^\#\phi_1(t)$, which are approximately governed by the system
\begin{align}
i\D_ta(t)\ &=\ \tilde\lambda_0a(t)\ +\ \frac{\left(\ \psi_0,W\phi_d(t)\ \right)}{\left(\Sigma_3\psi_0,\psi_0\right)}\nn\\
i\D_t\phi_d(t)\ &=\ \Sigma_3H_0\phi_d(t)\ +\ a(t)P_c^\#\Sigma_3W\psi_0,
\label{aphi}
\end{align}
where 
\begin{equation}
\tilde\lambda_0\ =\ \lambda_0\ +\ \frac{\left(\psi_0,W\psi_0\right)}{\left(\Sigma_3\psi_0,\psi_0\ \right)}.
\label{tlam}
\end{equation}
Terms neglected in arriving at~\eqref{aphi} can be treated perturbatively
~\cite{SW-TDRT:98}.
We consider the system~\eqref{aphi} with initial data
\begin{equation} a(0)=1,\ \ \ \phi_d(0)=0
\nn\end{equation}
corresponding to the unperturbed \emph{embedded} state.

We determine the asymptotic (in time) dynamics of~\eqref{aphi} by solving for $\phi_d(t)$ as a functional of $a(t)$ and then substituting the result into the equation for  $a(t)$. To facilitate this, we first extract
the rapidly varying time dependence of $a(t)$ by setting
\begin{equation}
A(t)\ =\ e^{i\tilde\lambda_0 t} a(t)
\label{Adef}
\end{equation}
Then, the system~\eqref{aphi} can be equivalently rewritten as 
\begin{align}
\D_tA(t)\ &=\ -i\ e^{i\tilde\lambda_0 t}\ 
\frac{\left(\ \psi_0,W\phi_d(t)\ \right) }{\left(\Sigma_3\psi_0,\psi_0\ \right) }\label{A-Aphi}\\
i\D_t\phi_d(t)\ &=\ \Sigma_3H_0\phi_d(t)\ +\  e^{-i\tilde\lambda_0 t} A(t)\ P_c^\#\Sigma_3W\psi_0,
\label{phid-Aphi}
\end{align}

Now solving for $\phi_d(t)$ with the initial condition $\phi_d(0)=0$, gives by duHamel's principle
\begin{equation}
\phi_d(t)\ =\ -ie^{-i\tilde\lambda_0 t}\ \int_0^t\ e^{-i\left(\Sigma_3H_0-\tilde\lambda_0I\right)(t-s)}\ A(s)\ P_c^\#\Sigma_3W\psi_0\ ds
\label{phid-solve}
\end{equation}
Substitution of~\eqref{phid-solve} into~\eqref{A-Aphi} yields the closed nonlocal equation for $A(t)$:
\begin{equation}
\left(\Sigma_3\psi_0,\psi_0\right)\ \D_tA(t)\ =\ 
-\left(\ W\psi_0\ ,\ \int_0^t\ e^{-i\left(\Sigma_3H_0-\tilde\lambda_0I\right)(t-s)}\ A(s)\ P_c^\#\Sigma_3 W\psi_0\ ds\ \right)
\label{A-nonlocal}
\end{equation}
Denote by $\sigma_c^\#\subset\sigma_c(\Sigma_3H_0)$ the spectral subset onto which to the operator $P_c^\#$ projects.  We now use the spectral representation of $\Sigma_3H_0$ to compute the \emph{dominant and local} contribution of~\eqref{A-nonlocal}.
\begin{equation}\begin{split}
\left(\Sigma_3\psi_0,\psi_0\right)\ \D_tA(t)\ 
=&\ 
-\left(\ W\psi_0\ ,\ \int_0^t\   \int_{\sigma_c^\#}\ \left(\
\Sigma_3\psi_\lambda, e^{-i\left(\Sigma_3H_0-\tilde\lambda_0I\right)(t-s)} \ \Sigma_3 W\psi_0\ \right) \psi_\lambda\ d\lambda\ \ A(s)\ ds\ \right)\\
=&\ 
-\left(\ W\psi_0\ ,\ \int_0^t\   \int_{\sigma_c^\#}\ \left(\
e^{i\left(\Sigma_3H_0-\tilde\lambda_0I\right)(t-s)}\Sigma_3\psi_\lambda,  \ \Sigma_3 W\psi_0\ \right)\psi_\lambda\ d\lambda\ \ A(s)\ ds\ \right)\\
=&\ 
-\left(\ W\psi_0\ ,\ \int_0^t\   \int_{\sigma_c^\#}\ \left(\
e^{i(\lambda-\tilde\lambda_0)(t-s)}\Sigma_3\psi_\lambda,  \ \Sigma_3 W\psi_0\ \right) \psi_\lambda\ d\lambda\ \ A(s)\ ds\ \right)\\
=&\ 
-\left(\ W\psi_0\ ,\ \int_0^t\    \int_{\sigma_c^\#}\ \left(\
e^{i(\lambda-\tilde\lambda_0)(t-s)}\psi_\lambda,  \ W\psi_0\ \right) \psi_\lambda\ d\lambda\ \ A(s)\ ds\ \right)\\
=&\ 
-\left(\ W\psi_0\ ,\  \int_{\sigma_c^\#}\ e^{i(\lambda-\tilde\lambda_0)t}\  \int_0^t\  \left(\
e^{-i(\lambda-\tilde\lambda_0)s}\psi_\lambda,  \ W\psi_0\ \right) \psi_\lambda\ A(s)\ ds\ d\lambda\ \ \right)\\
=&\ 
-\lim_{\veps\to0}\left(\ W\psi_0\ ,\  \int_{\sigma_c^\#}\ e^{i(\lambda-\tilde\lambda_0)t}\  \int_0^t\ \left(\
e^{-i(\lambda-\tilde\lambda_0+i\veps)s}\psi_\lambda,  \ W\psi_0\ \right) \psi_\lambda\ A(s)\ ds\ d\lambda\ \  \right)\\
\sim&\ 
-\lim_{\veps\to0}\left(\ W\psi_0\ ,\  \int_{\sigma_c^\#}\   \left(\
\frac{1}{-i(\lambda-\tilde\lambda_0+i\veps)}\psi_\lambda,  \ W\psi_0\ \right) \psi_\lambda\ d\lambda\ \right)\ A(t)\\
=&\ 
-\lim_{\veps\to0}\left(\ W\psi_0\ ,\  \int_{\sigma_c^\#}\   \left(\
\psi_\lambda\ ,\ \frac{1}{i(\lambda-\tilde\lambda_0-i\veps)} W\psi_0\ \right) \psi_\lambda\ d\lambda\ \right)\ A(t)\\
=&\ 
-\lim_{\veps\to0}\left(\ W\psi_0\ ,\  \int_{\sigma_c^\#}\   \left(\
\psi_\lambda\ ,\ \frac{1}{i}\frac{\lambda-\tilde\lambda_0+i\veps}{(\lambda-\tilde\lambda_0)^2+\veps^2} W\psi_0\ \right) \psi_\lambda\ d\lambda\ \right)\ A(t)\\
=&\ i\ \    {\rm P.V.}\ \int_{\sigma_c^\#}\   \left(\
\psi_\lambda\ ,\ \frac{1}{\lambda-\tilde\lambda_0} W\psi_0\ \right)\ \left(\ W\psi_0\ ,\ \psi_\lambda\ \right)\
d\lambda\ A(t)\\
&-\ \pi\   \int_{\sigma_c^\#}\   \left(\
\psi_\lambda\ ,\ \d(\lambda-\tilde\lambda_0)\  W\psi_0\ \right)\ \left(\ W\psi_0\ ,\ \psi_\lambda\  \right)\ d\lambda\ A(t)\\
=&\   i\ \    {\rm P.V.}\  \int_{\sigma_c^\#}\ \frac{1}{\lambda-\tilde\lambda_0}\ \left|\left(\psi_\lambda,W\psi_0\right)\right|^2\ d\lambda\ A(t)\ -\
 \pi\ \left|\left(\psi_{\tilde\lambda_0},W\psi_0\right)\right|^2\ A(t).
 \label{long}
\end{split}
\end{equation}
This last line,
\begin{equation}
\boxed{ \left(\Sigma_3\psi_0,\psi_0\right)\ \D_tA(t)\ = 
  i\ \    {\rm P.V.}\  \int_{\sigma_c^\#}\ \frac{1}{\lambda-\tilde\lambda_0}\ \left|\left(\psi_\lambda,W\psi_0\right)\right|^2\ d\lambda\ A(t)
   -\ \pi\ \left|\left(\psi_{\tilde\lambda_0},W\psi_0\right)\right|^2\ A(t),}
 \label{A-final}
 \end{equation}
 is precisely the conclusion of Proposition~\ref{prop1}.
 The first term on the right hand side of~\eqref{A-final} contributes an $\cO(W^2)=\cO (\lVert\vec\cE\rVert_2^4)$  phase correction, while the second term determines the stability or instability of the dynamics.  An explanation of the $\veps-$ regularization  in~\eqref{long} is given in  appendix~\ref{sec:details}.
 
Define the  {\it Krein signature} of the unperturbed bound state $\psi_0$, to be the sign of  $\left(\Sigma_3\psi_0,\psi_0\right)$. If the signature is positive, then the dynamics the solution to~\eqref{A-final} decay exponentially for $t>0$ and the dynamics are stable. If the signature is negative, then the solutions to~\eqref{A-final} grow exponentially with $\cO(W^2)=\cO (\lVert\vec\cE\rVert_2^4)$ growth rate and the dynamics are unstable. These two cases correspond to the time-independent perturbation theory, which gives that positive signature embedded states perturb to {\it resonances} and negative signature embedded states perturb to genuine finite energy unstable eigenstates of $\sigma_3H$; see ~\cite{CPV}.

The scenario discussed here is generic; if $\G\equiv0$, then for typical small perturbations of the potentials $\kappa$ and $V$ it will be non-zero.  Since the continuum eigenmode $\psi_{{\bar \lambda}_0}$ is not easily computed, it is difficult to estimate the constant $C$ in the formula
$$\G \sim C\norm{\vec\psi_0}{2}^4.$$ Depending on the magnitude of this growth rate and the appearance of other instability mechanisms, discussed next, this may or may not be the most important instability of a nonlinear defect state.

\subsection{Other scenarios for instability onset}
\label{sec:scenarios}
We here discuss four additional scenarios under which a defect mode may lose stability.  Unlike scenario one, these transitions to instability arise at nonzero amplitude, and cannot be determined immediately from a condition such as~\eqref{eq:embed_cond} derived entirely from quantities known exactly in the zero-amplitude limit. In the first two scenarios, instability arises due to bifurcations involving the continuous spectrum while in the last two, instability arises from bifurcations involving only frequencies in the discrete spectrum.

\begin{figure}
\begin{center}
\includegraphics[width=.4\textwidth]{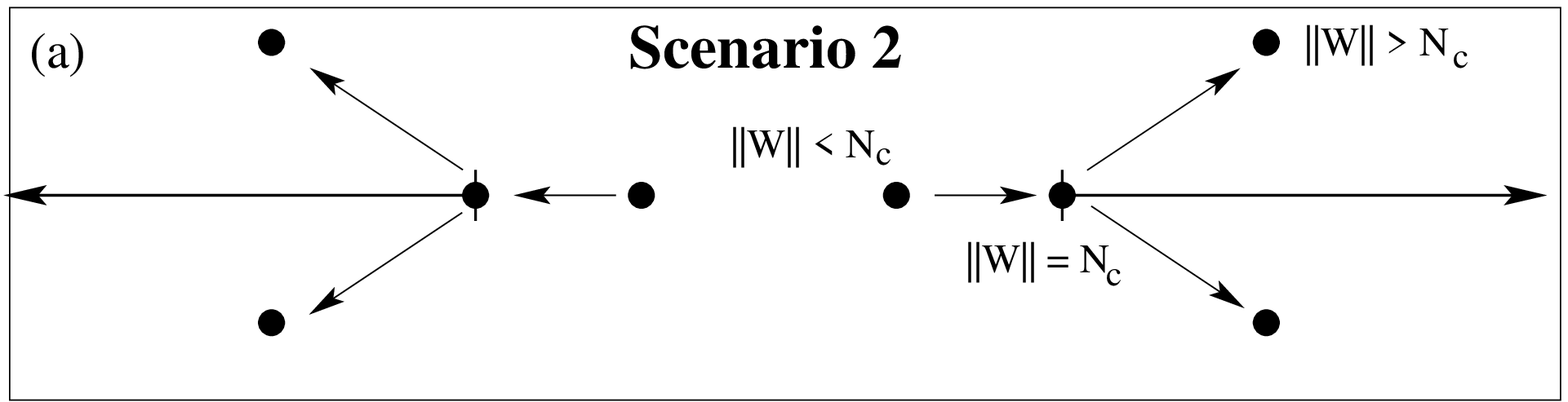}
\includegraphics[width=.4\textwidth]{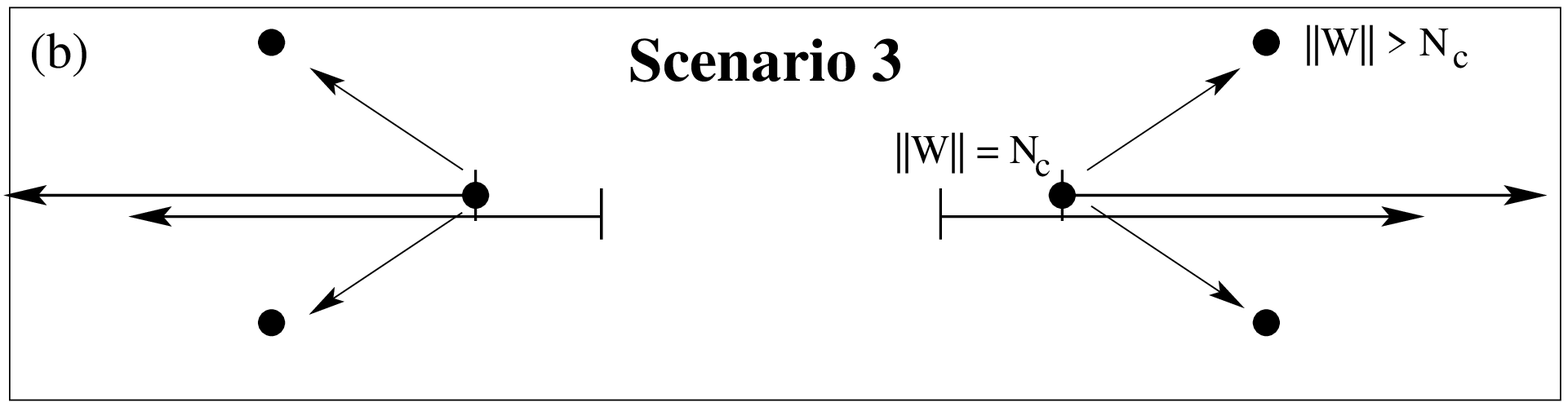}
\includegraphics[width=.4\textwidth]{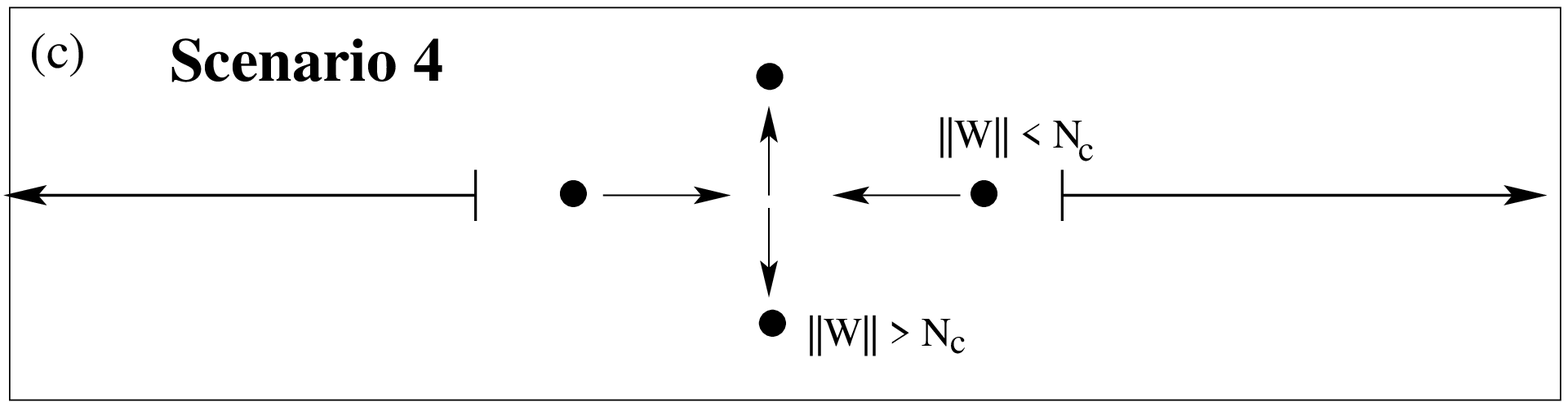}
\includegraphics[width=.4\textwidth]{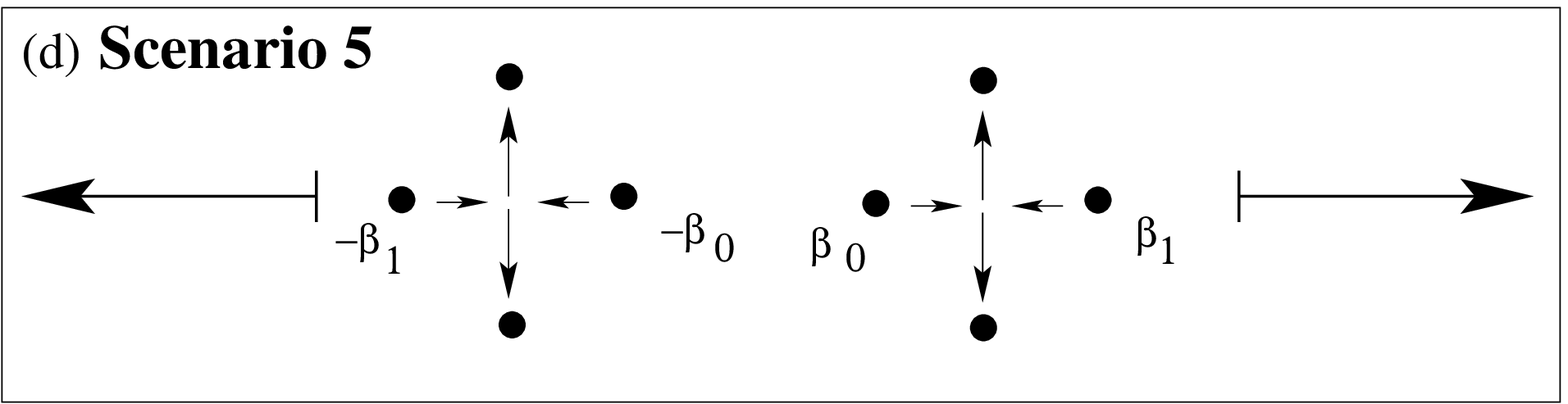}
\caption{(a) Instability scenario 2. (b) Instability scenario 3. (c) Instability scenario 4. (d) Instability scenario 5.}
\label{fig:scenarios}
\end{center}
\end{figure}

In a second scenario, see figure~\ref{fig:scenarios}(a), $\Sigma_3H_0$ has a symmetric pair of eigenvalues in the gap, and as $\Norm{W}$ is increased the  corresponding pair of eigenvalues of $\Sigma_3(H_0+W)$ collide, for some critical positive amplitude,  with the symmetrically located spectral band/gap edges. As $\Norm{W}$ is further increased, each collision gives rise to a pair of complex conjugate eigenvalues. Those in the upper half plane correspond to solutions which grow exponentially as time increases, and have a growth rate given by the imaginary part of $\beta$.

In a third scenario, see figure~\ref{fig:scenarios}(b) as $\Norm{W}$ is increased, a quartet of complex eigenvalues (the four values$\pm \beta$ and $\pm \beta^*$) bifurcates from the secondary band edges on both sides of the spectral gap, resulting in two growing modes and two decaying modes with growth rate given by the imaginary part of $\beta$.
\footnote{Closely related to this is an instability scenario described by Kapitula and Sandstede for a different system of coupled-mode equations. An edge bifurcation may occur when the primary and secondary band-edges coincide, i.e at the value of $\Norm{\vec\cE_{(i)}}$ for which $\w_i=0$; see equation~\eqref{eq:gap}~\cite{KapSan:02}. This was not observed in our numerical simulations.}

In a fourth scenario, see figure~\ref{fig:scenarios}(c), $\Sigma_3H_0$ has a symmetric pair of eigenvalues within the gap, and as $\Norm{W}$ increases, the associated discrete eigenvalues of $\Sigma_3H=\Sigma_3(H_0+W)$ collide at zero at some critical positive amplitude and then symmetrically ascend/descend the imaginary axis as $\Norm{W} $ is further increased.  This scenario is usually associated with Hamiltonian pitchfork bifurcations in finite dimensions.

A fifth scenario---see figure~\ref{fig:scenarios}(d)---is related to Hamiltonian Hopf bifurcations. Suppose that at small values of $\Norm{W}$, the linearized operator~$\Sigma_3(H_0+W)$ has two pairs of real frequencies $\pm\beta_0$ and $\pm\beta_1$ in the spectral gap, such that they collide pairwise at some critical amplitude.  Then above this critical amplitude, the two pairs of real frequencies are replaced by a quartet of complex frequencies, two of which lead to exponential growth.  As we did not investigate any defects that have  more than three linear modes, this scenario was not observed.

\section{Numerical computation of discrete spectrum of $\Sigma_3 H$ via the Evans function}
\label{sec:evans}
We now turn to numerical computation of the eigenvalues of equation~\eqref{eq:eigen} to confirm the above analysis and estimate the growth rates predicted by equation~\eqref{A-final}. The simplest approach to numerically solving the eigenvalue problem~\eqref{eq:eigen} would be to discretize the
derivative and to approximate the solution~\eqref{eq:eigen} by the solution to
the large system of linear equations so derived.  This requires truncating to
a finite domain and applying artificial boundary conditions, usually Dirichlet
or periodic.   

Barashenkov and Zemlyanaya performed such a calculation using a Fourier representation in considering the closely-related question of stability of the gap solitons of~\eqref{eq:gapsoliton} without defect~\cite{BPZ:98,BZ:00}.  This produced a significant number of spurious unstable eigenvalues with relatively large
($\cO(10^{-2})$) imaginary parts, and these errors were found to decay rather slowly as
the number of modes used in their calculation was increased, possibly due to the artificial boundary conditions.  Sandstede and Scheel have analyzed the effect that such boundary conditions has on the spectrum of linear operators~\cite{SS:00}.Another approach was taken by Malomed and Tasgal, who studied the problem using averaged Lagrangian techniques to derive ordinary differential equations governing the evolution of perturbations to the gap solitons~\cite{MalTas:94}, although they, too,  found that their method could yield spurious growing modes. Stability of nonlinear defect modes is studied by~\cite{MakMalChu:03a}, who report simulations of the initial-value problem, showing that certain nonlinear defect modes localized at a point (Dirac delta) defect are ``semi-stable.''

A better way to test for instability is to use the Evans function, the analog of the
characteristic polynomial for the linear operator~\eqref{eq:eigen}.  It was introduced by Evans~\cite{Evans:75} and generalized by Alexander, Gardner, and Jones~\cite{AGJ}.  Evans functions have since
become a standard tool in both the analytic~\cite{GarZum:98,KapSan:02,LiPro:00,PelSche:03} 
and numerical~\cite{BD:99,BD:03,BDG:02,Bri:01,LiPro:00,PSW:93,PW:92}
study of wave stability.  Derks and Gottwald have applied the Evans function numerically to the problem of gap solitons and were able to avoid finding spurious unstable eigenvalues~\cite{DerGot:05}.

\subsection{Definition of the Evans Function}
The Evans function is the analog of a characteristic polynomial for a class of infinite-dimensional operators.  It is an analytic function whose zeroes and their multiplicity correspond to eigenvalues of the operator and their multiplicities.  To construct the Evans function, it is convenient to rewrite the eigenvalue problem~\eqref{eq:eigen} as:
\begin{equation}
\label{eq:evansform}
\diff{}{Z} \vec y(Z)=
\begin{bmatrix}
i\left( \cV(Z) + \beta\right) & 
i \left(\k(Z) + 2 \cE_+ \cE^*_- \right) & 
i \cE_+^2 & 
2i \cE_+ \cE_- \\
-i \left(\k(Z) + 2 \cE^*_+ \cE_- \right) & 
-i\left( \cV(Z) + \beta\right) &
-2i \cE_+ \cE_- &   
-i \cE_-^2\\
-i {\cE^*_+}^2 & 
-2 i \cE^*_+\cE^*_- & 
-i\left( \cV(Z) - \beta\right) &
-i \left(\k(Z) + 2 \cE^*_+ \cE_- \right) \\
2 i \cE^*_+\cE^*_- &
i {\cE^*_-}^2 & 
i \left(\k(Z) + 2 \cE_+ \cE^*_- \right) & 
i\left( \cV(Z) - \beta\right) 
\end{bmatrix}
\vec{y}
\end{equation}
where $\cV(Z) = V(Z) + \w + 2(\abs{\cE_+(Z)}^2 + \abs{\cE_-(Z)}^2)$.

The Evans function $D(\beta)$ is constructed from the stable and unstable manifolds of the trivial solution
$\vec y=0$ to equation~\eqref{eq:evansform} for fixed $\beta$ and $\abs{Z}\to \infty$.
First, rewrite~\eqref{eq:evansform} in the general form   
\begin{equation}
\label{eq:eigen_general}
\diff{}{Z} \vec y = A(Z,\beta) \vec y
\end{equation}
for $y\in \bbC^n$ and $A(Z,\beta)$ an analytic $n\times n$ matrix-valued analytic function of $Z$ and $\beta$ which approaches a constant value $A_\infty(\beta)$ as $\abs{Z} \to \infty$.  
Each discrete eigenvalue of~\eqref{eq:eigen} corresponds to a solution of~\eqref{eq:eigen_general} that decays as $\abs{Z}\to\infty$.  Solutions that decay as $Z \to -\infty$
must approach zero in a direction tangent to the unstable subspace of $A_\infty(\beta)$ (the span of the eigenvectors whose eigenvalues have positive real part), which
we may assume has dimension $k$, while solutions that decay as $Z \to \infty$
approach zero tangent to the $(n-k)$-dimensional stable subspace of
$A_\infty(\beta)$ (corresponding to eigenvalues with negative real part).  An
eigenfunction of~\eqref{eq:eigen_general} exists if the stable and unstable 
manifolds have nontrivial intersection.

The Evans function $D(\beta)$ is defined as follows.  Assume the matrix
$A_\infty{(\beta)}$ has eigenvalues  satisfying
\begin{equation}
\Re \lambda_1\ge \Re \lambda_2 \ge \ldots \ge \Re\lambda_k >0 > \Re\lambda_{k+1} \ge \ldots
\Re \lambda_n 
\label{eq:eigen_order}
\end{equation}
with corresponding (generalized) eigenvectors $\vec{v}_i$. 
For $j=1,\ldots,k$, define $\vetap_j (Z)$ to be a solution
to~\eqref{eq:eigen_general} satisfying the asymptotic condition $\vetap_j(Z)
\sim \vec{v}_j \exp{ \lambda_j Z}$ as $Z\to -\infty$.  For $j=k+1,\ldots,n$,
define $\vetam_j (Z)$ to be a solution satisfying the asymptotic condition
$\vetam_j(Z) \sim \vec{v}_j \exp{\lambda_j Z}$ as $Z\to\infty$.   Note that
$\lambda_j$, and ${\mathbf \eta}^{\pm}_j$ all depend on the frequency parameter
$\beta$. 

If a discrete eigenmode exists with complex frequency $\beta$, then it must
approach the space spanned by $\{\vetap_j(Z): j=1,\ldots,k\}$ as $Z \to -\infty$
and analogously to $\{\vetam_j(Z): j=k+1,\ldots,n\}$ as $Z \to \infty$.  Thus
these two subspaces must be linearly dependent, which will cause the Wronskian
determinant 
\begin{equation}
\label{eq:wronskian}
W(Z;\beta) = \det\begin{bmatrix}
      \vetap_1(Z) & \ldots & \vetap_k(Z) & \vetam_{k+1}(Z) &\ldots&  \vetam_n(Z) 
\end{bmatrix}
\end{equation}
to vanish for all $Z$.  The eigenvectors $\vec{v}_j$ of $A_\infty$ depend
analytically on $\beta$, up to a constant multiple.  We define the Evans
function by the Wronskian evaluated at some point $Z_0$ which may be taken
to be zero: 
\begin{equation}
\label{eq:Evans}
D(\beta) = W(0;\beta).
\end{equation}
The domain of the Evans function consists of the set of complex frequencies $\beta$ for which the asymptotic eigenvectors $\eta_j^\pm$ can be defined analytically and the condition $$\Re\lambda_k>\Re\lambda_{k+1}$$ is maintained. That is, the ordering of the other eigenvalues may change, so long as the $(k,n-k)$ splitting is preserved.
The eigenvectors may be normalized analytically such that 
$$
\lim_{\abs{\beta}\to \infty} D(\beta) = 1
$$
when the limit is taken along any path inside the domain of $D(\beta)$.

\subsection{Numerical experiments}
\label{sec:numer_stability}
The principal way Evans functions are used numerically takes advantage of their analyticity---for a given closed curve $\g$, the winding number of $D(\g)$ equals the number of zeros, counting multiplicity, of $D(\beta)$ inside $\g$.  Since $D(\beta)\to 1$ as $\abs{\beta}\to\infty$ in the domain of $D(\beta)$, the number of zeros of $D(\beta)$, counting multiplicity, above a given line in the half-plane ${\mathbb C}_\beta^+$ is equal to the winding number of its image under $\beta\mapsto D(\beta)$.  

We have used winding number calculations to count eigenvalues, followed by root-finding to trace the evolution of these eigenvalues as the amplitude of a defect mode is increased.  We discuss here some numerical results for defects with one, two, or three bound states in the linear regime.  We looked at a variety of defects beyond those discussed here, with similar results, although no thorough attempt has been made to thoroughly explore the parameter space.

\subsection*{Experiment 1: A defect supporting one bound state}
\begin{figure}
\begin{center}
\includegraphics[width=.3\textwidth]{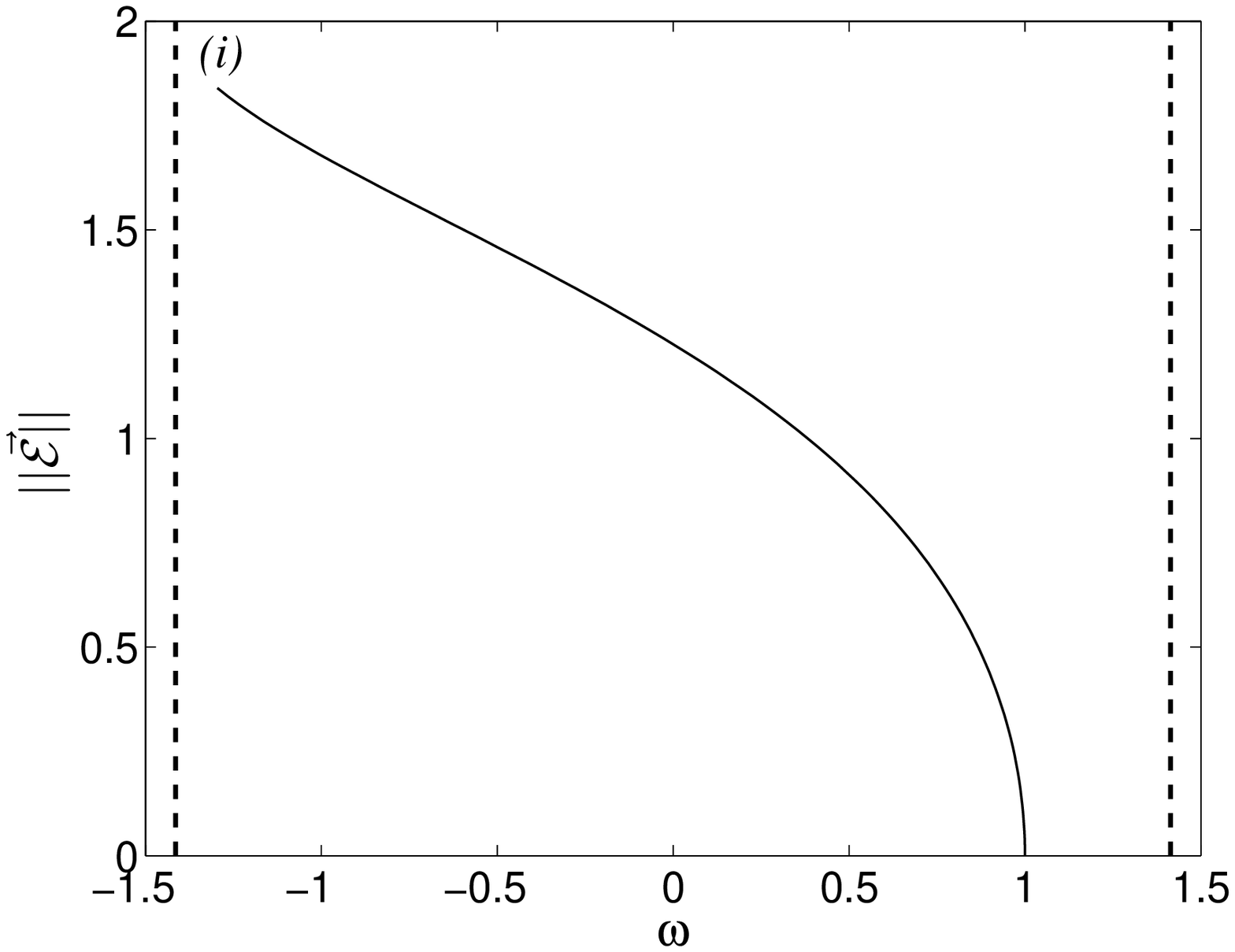}
\includegraphics[width=.3\textwidth]{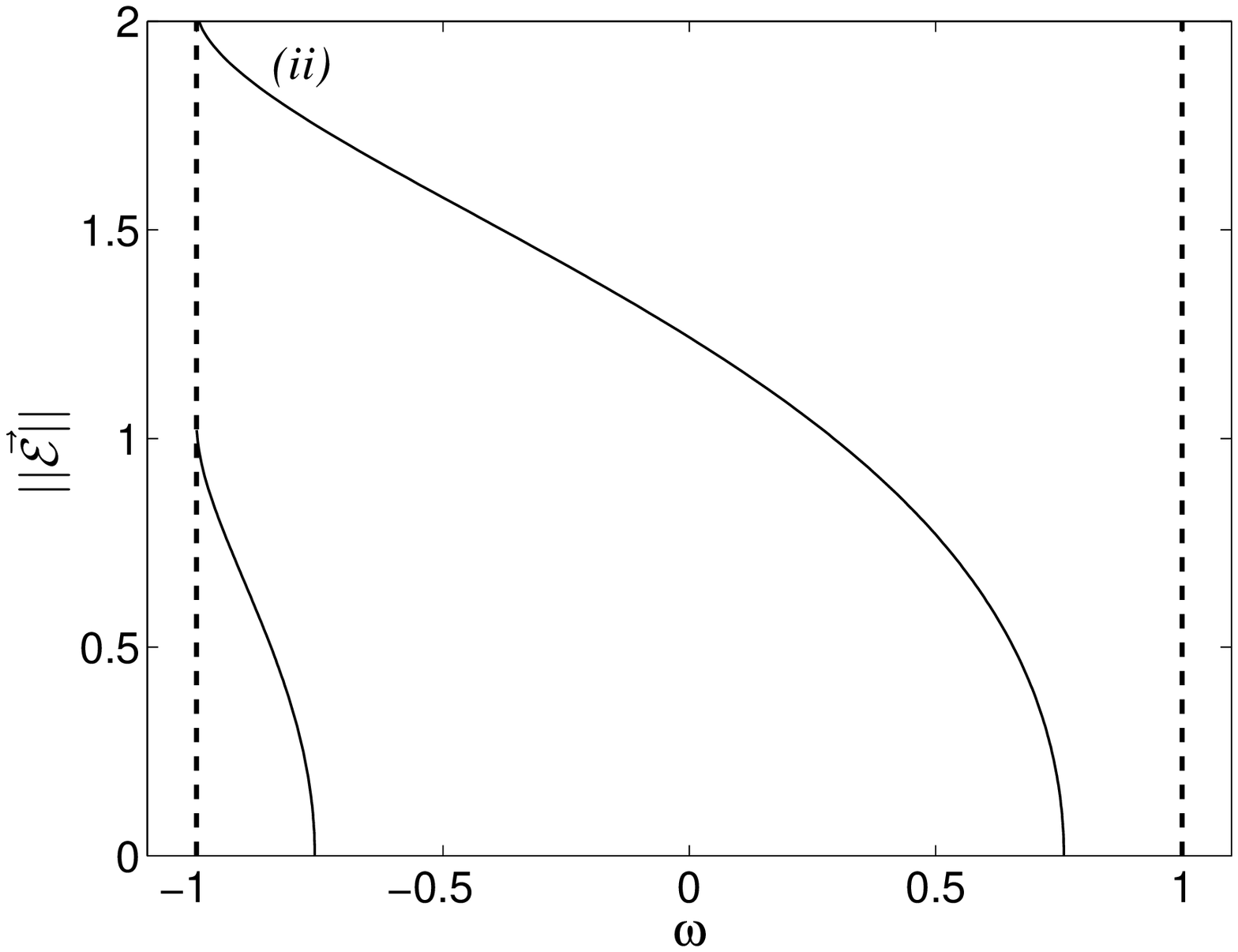}
\includegraphics[width=.3\textwidth]{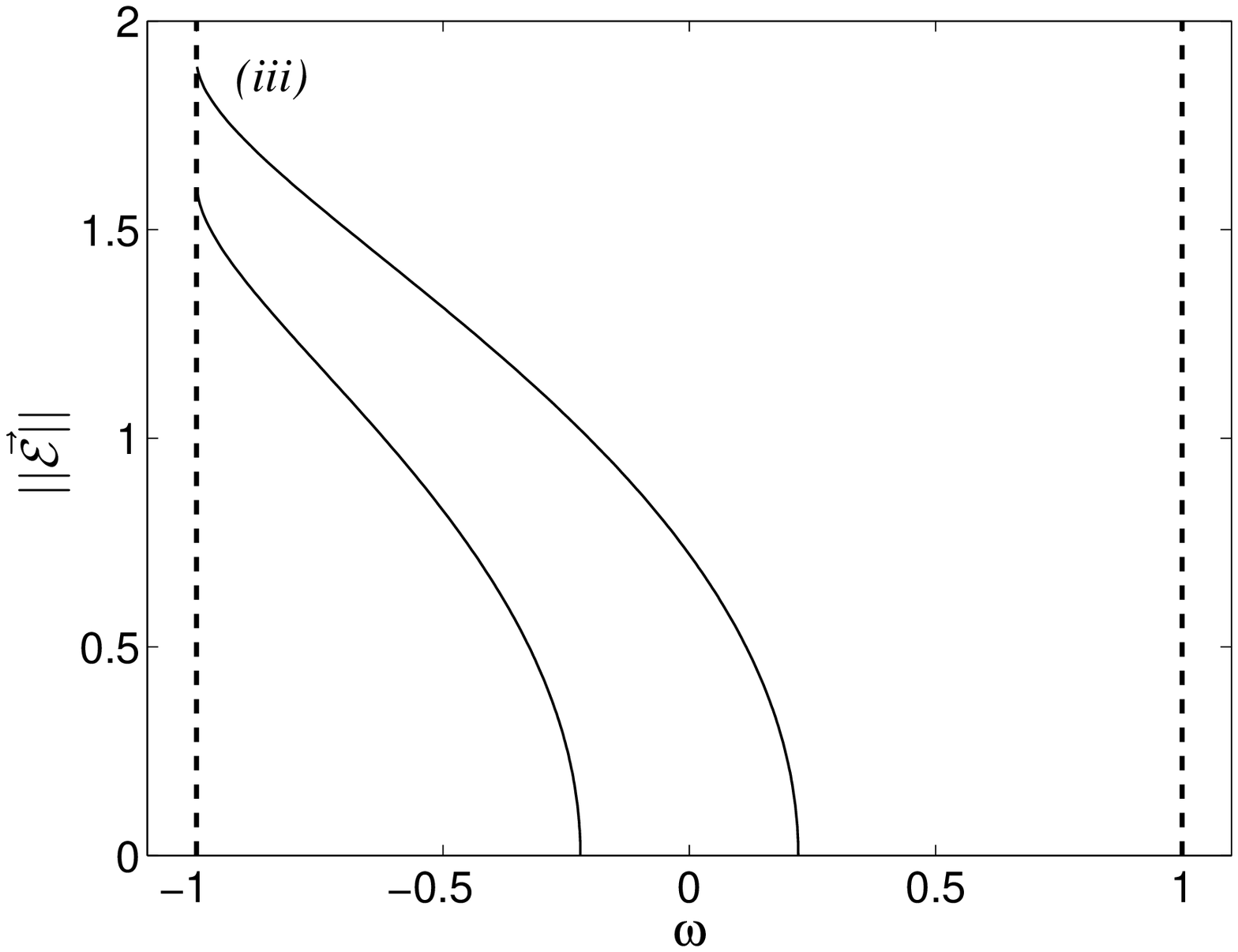}
\caption{\textbf{\emph{(i):}} The amplitude vs. frequency of the single nonlinear defect mode of the ``odd defect'' of equation~\eqref{eq:potential} with $\w=k=n=1$.  The stability of this mode is analyzed in Experiment 1 and figure~\ref{fig:111}. \textbf{\emph{(ii):}} The ``resonant'', ``even'' defect analyzed in Experiment 2.1, with curve \textbf{(a)} analyzed in figures~\ref{fig:Sol5_left} and~\ref{fig:sol5_left_growth} and curve~\textbf{(b)} analyzed in figure~\ref{fig:Sol5_right}. \textbf{\emph{(iii):}} The ``nonresonant'', ``even'' defect analyzed in Experiment 2.2, with curve \textbf{(a)} analyzed in figure~\ref{fig:Sol3_left}  and curve~\textbf{(b)} analyzed in figure~\ref{fig:Sol3_right}. }
\label{fig:amp_freq}
\end{center}
\end{figure}

\begin{figure}
\begin{center}
\includegraphics[width=.4\textwidth]{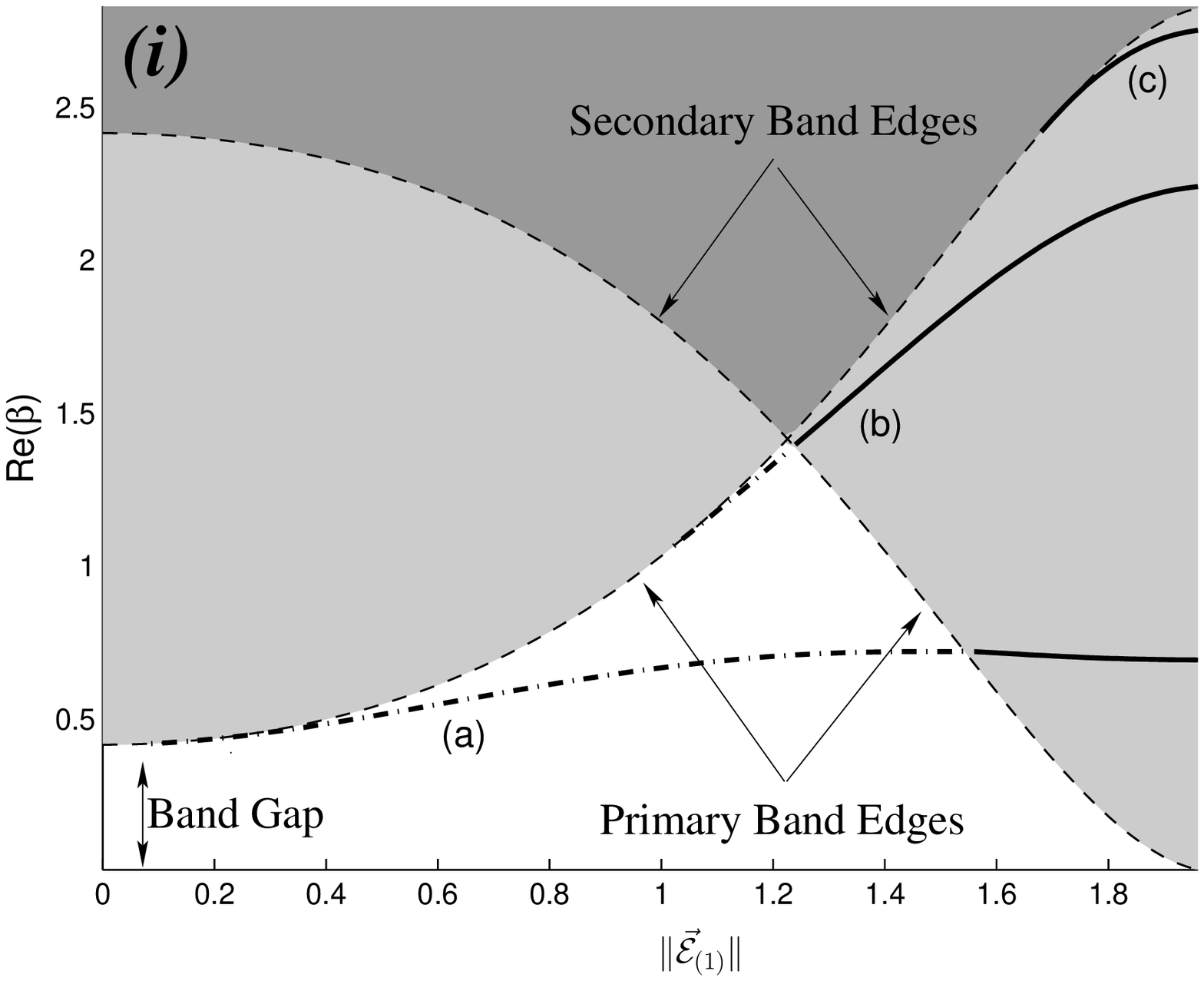}
\includegraphics[width=.4\textwidth]{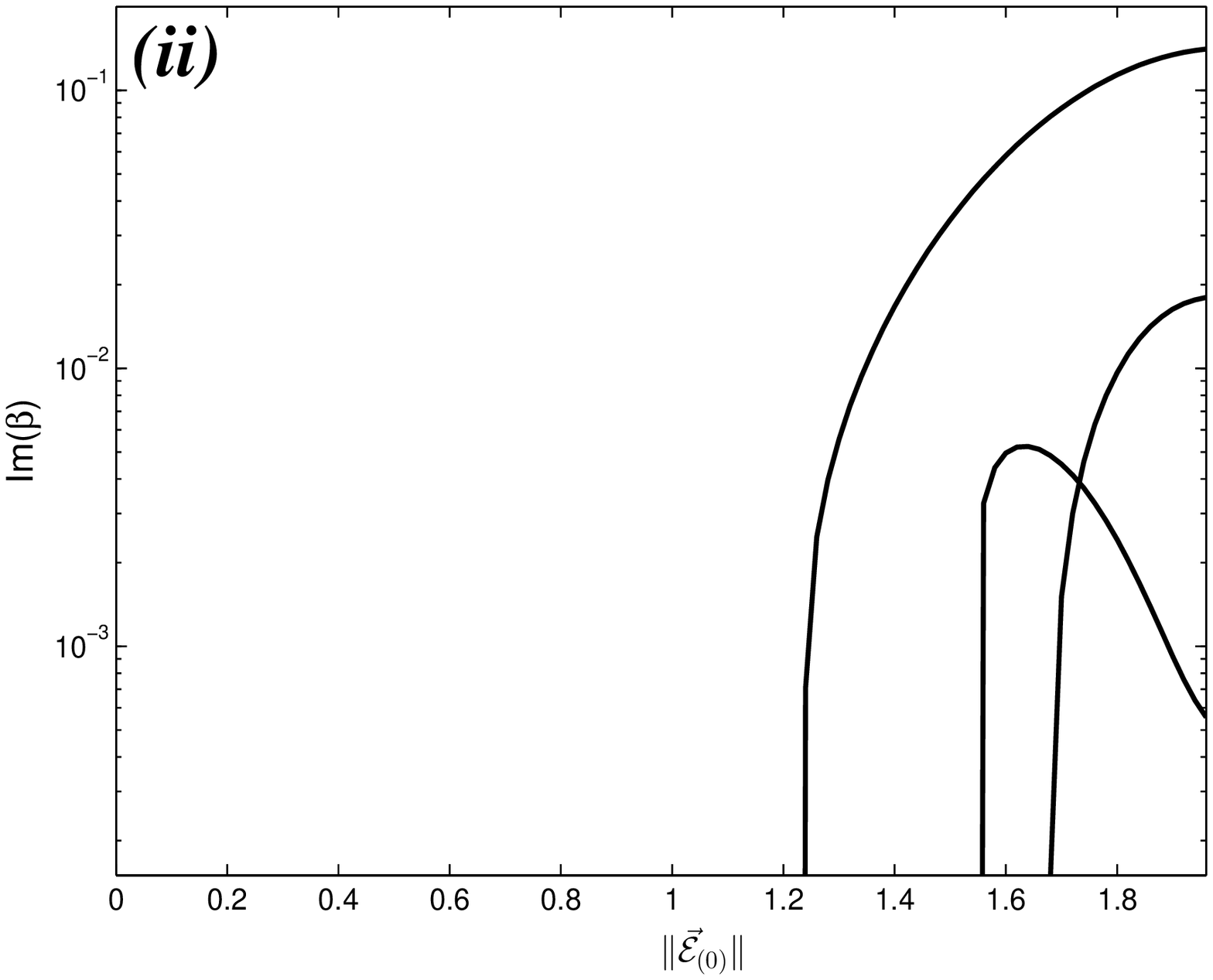}
\caption{Experiment 1: The eigenvalues of the linearization about the nonlinear defect mode of a one-mode defect. \textbf{\emph{(i)}}: The real part of the frequency, showing three modes, marked (a), (b), (c) bifurcating from the edge of the continuous spectrum.  These lines are drawn dash-dot if the frequency is purely real and solid if it has a nonzero imaginary part (growth rate). The dashed lines mark the band edges and the light and dark shaded regions the multiplicity-one and multiplicity-two regions of continuous spectrum.  \textbf{\emph{(ii)}}: The growth rates of the three instabilities, labeled as in {\it (i)}. } 
\label{fig:111}
\end{center}
\end{figure}

The defect defined in equation~\eqref{eq:potential} with $n=1$ supports a single linear bound state.  To discuss the system at zero amplitude, we refer to figure~\ref{fig:spec_linearized}.  The only discrete eigenvalue is given by $\beta=\w_0-\w_0=0$ given by a `$\times$' sign in the figure. Modes like those marked with the `$\bullet$' and `$+$' are absent for this particular defect. As the amplitude is increased, the eigenvalue at zero does not move.  The spectral gap is given by formula~\eqref{eq:gap} with $\w_{0*}$ replaced by $\w_0$, the frequency of the nonlinear bound state.  As seen in figure~\ref{fig:amp_freq}\emph{(i)}, the frequency moves toward the left band edge as the amplitude is increased.  In figure~\ref{fig:111}, we plot the numerically calculated eigenvalues of this defect with $\w=k=1$.  
Figures like this one are repeated in this section, so it is worth spending some time dissecting it.  As the defect mode's amplitude is increased, its frequency decreases from $\w=1$ to $\w = -\sqrt{2}=-\kinf$, the left band edge. Note that equation~\eqref{eq:gap} implies that the width of the gap approaches zero as $\w_0 \searrow - \kinf$. The spectrum is symmetric across the real and imaginary axes, so only the first quadrant and its boundary are plotted.  In the left figure, the mode's amplitude is plotted on the $x$-axis, and the real part of the spectrum on the $y$-axis.  The region of multiplicity-one continuous frequencies is shaded light gray and multiplicity two in dark grey, with dashed lines indicating their boundaries---the band-edges---moving according to~\eqref{eq:gap} and cross when $\w_0=0$.

The linearization has no discrete modes of non-zero frequency at zero amplitude. Since zero remains an eigenvalue for all amplitudes of $\vec\cE_0$, instabilities may only develop from edge bifurcations.  Two discrete modes, labeled (a) and (b), bifurcate from the primary band edge.  At higher amplitude, each collides with the band edge to produce an instability, as described by scenario 2.  A third instability, marked (c) arises due to an edge bifurcation from the secondary band edge, in accordance with scenario 3.  Mode (b) is the first to become unstable and develops the largest growth rate of the three modes. Below the large intensity 1.2, there is no instability. It is worth noting that edge bifurcations only take place on the band edge  that moves away from the origin as $\Norm{\vec \cE_0}$ is increased and not from the other edge.  This same pattern is seen in all the simulations.

\subsection*{Experiment 2: Defects supporting two bound states}

We consider two defects in the family~\eqref{eq:kappa}, with $k=3$ and different values of the parameter $b$.  Because these defects support two linear bound states, the linearization has  discrete consisting of the zero eigenvalue and of another family eigenvalues that bifurcate from $\beta=\pm 2 \w_{0*}$ which we track as $\norm{\vec \cE_{\pm1}}{2}$ from zero.  In experiment 2.1, this eigenvalue is embedded in the continuous spectrum like the mode marked~$\bullet$ in figure~\ref{fig:spec_linearized}.  According to scenario 1, detailed in section~\ref{sec:embedded}, we expect an exponential instability for \emph{any} $\norm{\vec \cE_{\pm1}}{2}>0$. In experiment 2.2, the eigenvalue is in the gap, like the mode marked $+$ in figure~\ref{fig:spec_linearized}.  Surprisingly, a much stronger instability is found in the second case.

\subsubsection*{Experiment 2.1: A defect with an embedded eigenvalue in its linearization}

\begin{figure}
\begin{center}
\includegraphics[width=.45\textwidth]{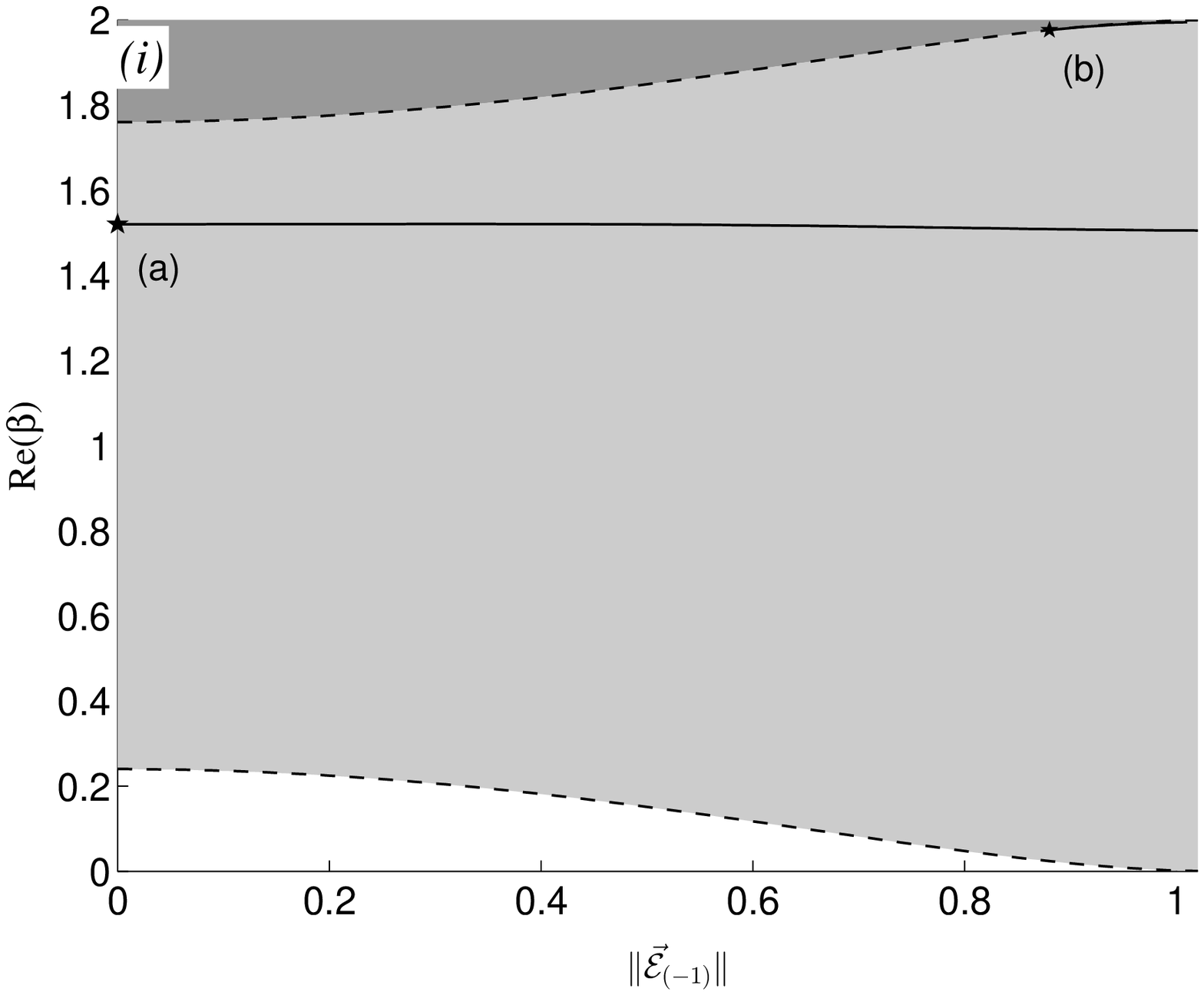}
\includegraphics[width=.45\textwidth]{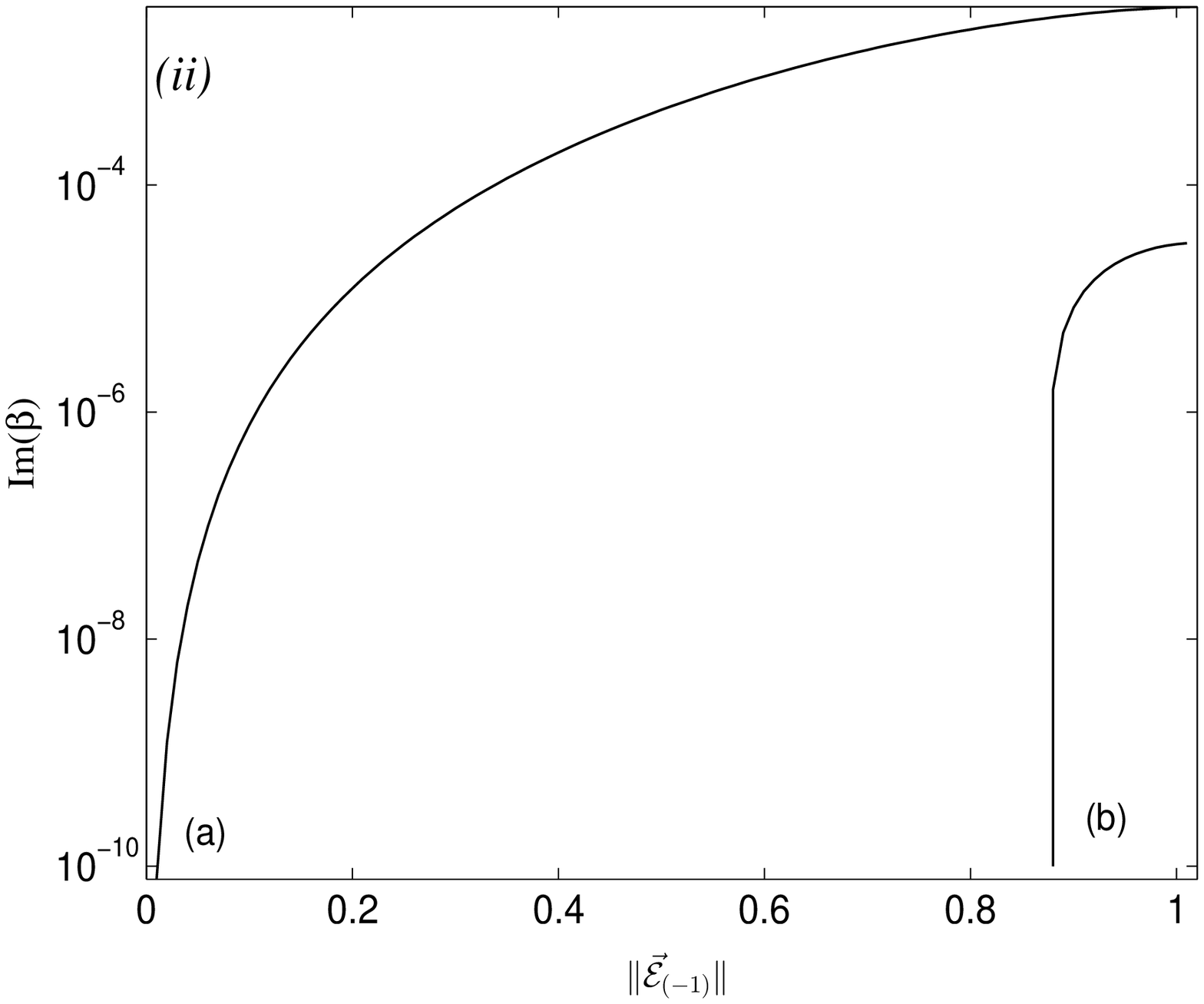}
\caption{Experiment 2.1, annotated as in figure~\ref{fig:111}. The frequency \emph{(i)} and growth rates \emph{(ii)} for the linearization about the left nonlinear defect mode with $b=1$. The embedded frequency develops a small instability as the amplitude is increased.  A second slightly unstable mode (b) bifurcates from the band edge at higher amplitude.}
\label{fig:Sol5_left}
\end{center}
\end{figure}

\begin{figure}
\begin{center}
\includegraphics[width=3in]{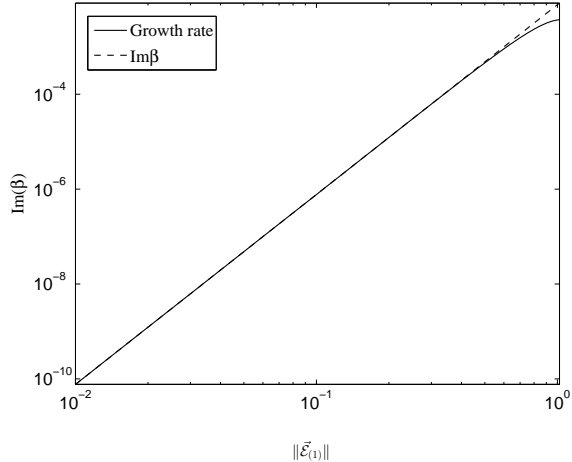}
\caption{Experiment 2.1: Verification that the growth rate scales as $\lVert \vec \cE \lVert_2^4$ as predicted following equation~\eqref{A-final}.}
\label{fig:sol5_left_growth}
\end{center}
\end{figure}

In this case, we set $b=1$ in equation~\eqref{eq:kappa} and find the linear frequencies are $\w_{\pm1*} = \pm 0.76$, embedded in the continuous spectrum.  We look first at the ``left mode,'' for which $\w_{-1*}=-0.76$, the left branch figure~\ref{fig:amp_freq}\emph{(ii)}.   The real and imaginary frequencies of the linearization are given as functions of the amplitude in figure~\ref{fig:Sol5_left}.  As the mode's norm increases, its frequency approaches $-1$, so the band gap closes and the edges of the secondary bands go to $\pm 2$.   The mode labeled (a), which is embedded in the continuous spectrum (like the eigenvalues marked by `$\bullet$' in figure~\ref{fig:spec_linearized}) develops an instability in the manner of scenario 1 (figure~\ref{fig:scenario1}). The real part of the frequency does not change significantly: from 1.52 to 1.505.  A second frequency (b) bifurcates from the secondary band edge at $\norm{\vec\cE_{-1*}}{2}=0.88$ under scenario 3.    Figure~\ref{fig:Sol5_left}\emph{(ii)} shows that $\Im \beta$ for mode (a) is never larger than $.004$ and for (b) is never larger than $1.6 \times 10^{-6}$, so no large instabilities arise.
Figure~\ref{fig:sol5_left_growth} shows via a log-log plot hat the growth rate scales as $\Norm{\vec \cE}_2^4$ for small amplitudes, in agreement with equation~\eqref{A-final} (recall $\Norm{W} = \cO(\Norm{\vec \cE}^2$).

\begin{figure}
\begin{center}
\includegraphics[width=.45\textwidth]{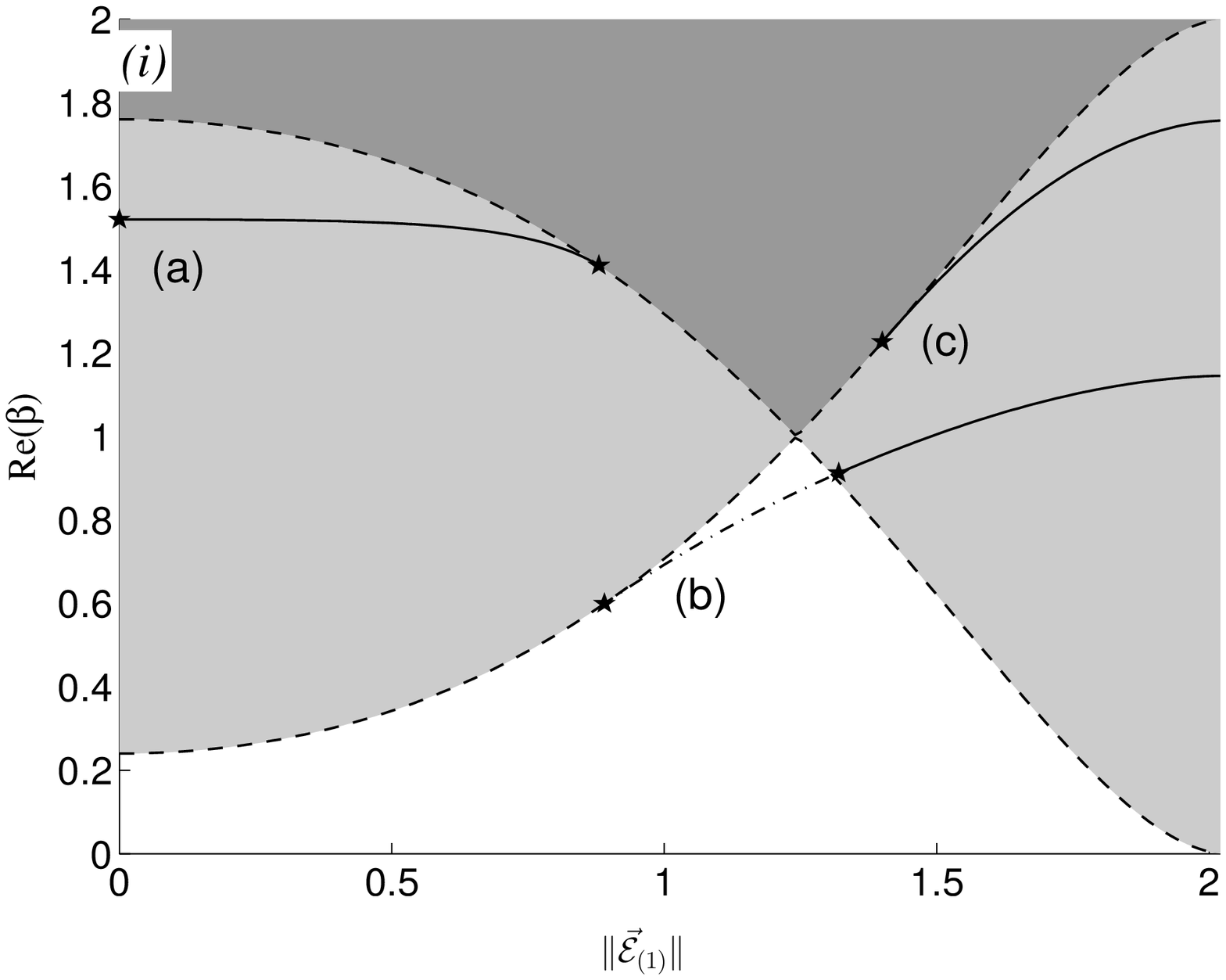}
\includegraphics[width=.45\textwidth]{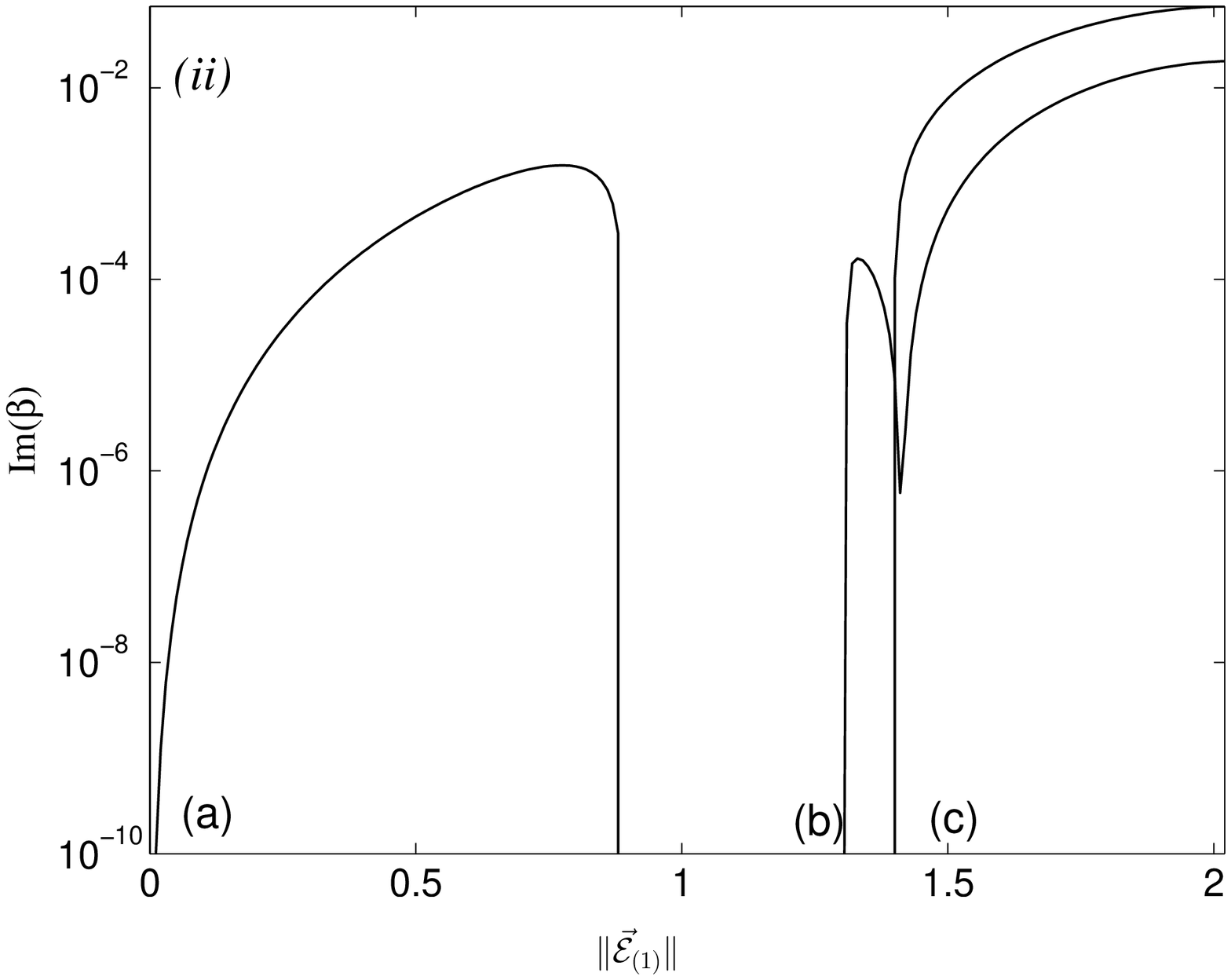}
\caption{Experiment 2.1, annotated as in figure~\ref{fig:111}: The frequencies \emph{(i)} and growth rates \emph{(ii)} for the linearization about the right nonlinear defect mode with $b=1$.}
\label{fig:Sol5_right}
\end{center}
\end{figure}

Figure~\ref{fig:Sol5_right} shows the real and imaginary parts of the frequency for the mode with linear frequency  $\w_{1*} = +0.76$ (the right branch in figure~\ref{fig:amp_freq}).  In this case, the frequency $\w_1$ decreases monotonically from 0.76 to -1.0 and thus the band gap widens until $\w_1=0$ and then contracts.   This allows for some more interesting behavior.  The discrete eigenvalue marked (a) begins in the continuous spectrum at $\norm{\vec \cE_{1}}{2}=0$ and its imaginary part grows in accordance with scenario 1.  At larger amplitude it is subsequently absorbed in the band edge at $\norm{\vec\cE_{-1*}}{2}\approx 0.88$ (essentially scenario 3) in reverse.   A second mode, marked (b) bifurcates from the primary band edge soon after. This frequency must be real unless it collides with another eigenfrequency or band edge.  However it soon collides again with the primary band edge and picks up a nonzero imaginary part, as in scenario 2.  At even higher intensity, a third discrete mode (c) bifurcates from the secondary band edge with nonzero imaginary part (scenario 3).   The growth rates for this defect mode all remain relatively small, with the largest growth rate arising from branch (c) at larger amplitudes. 

\subsubsection*{Experiment 2.2: A defect with no embedded eigenvalue in its linearization}

\begin{figure}
\begin{center}
\includegraphics[width=.45\textwidth]{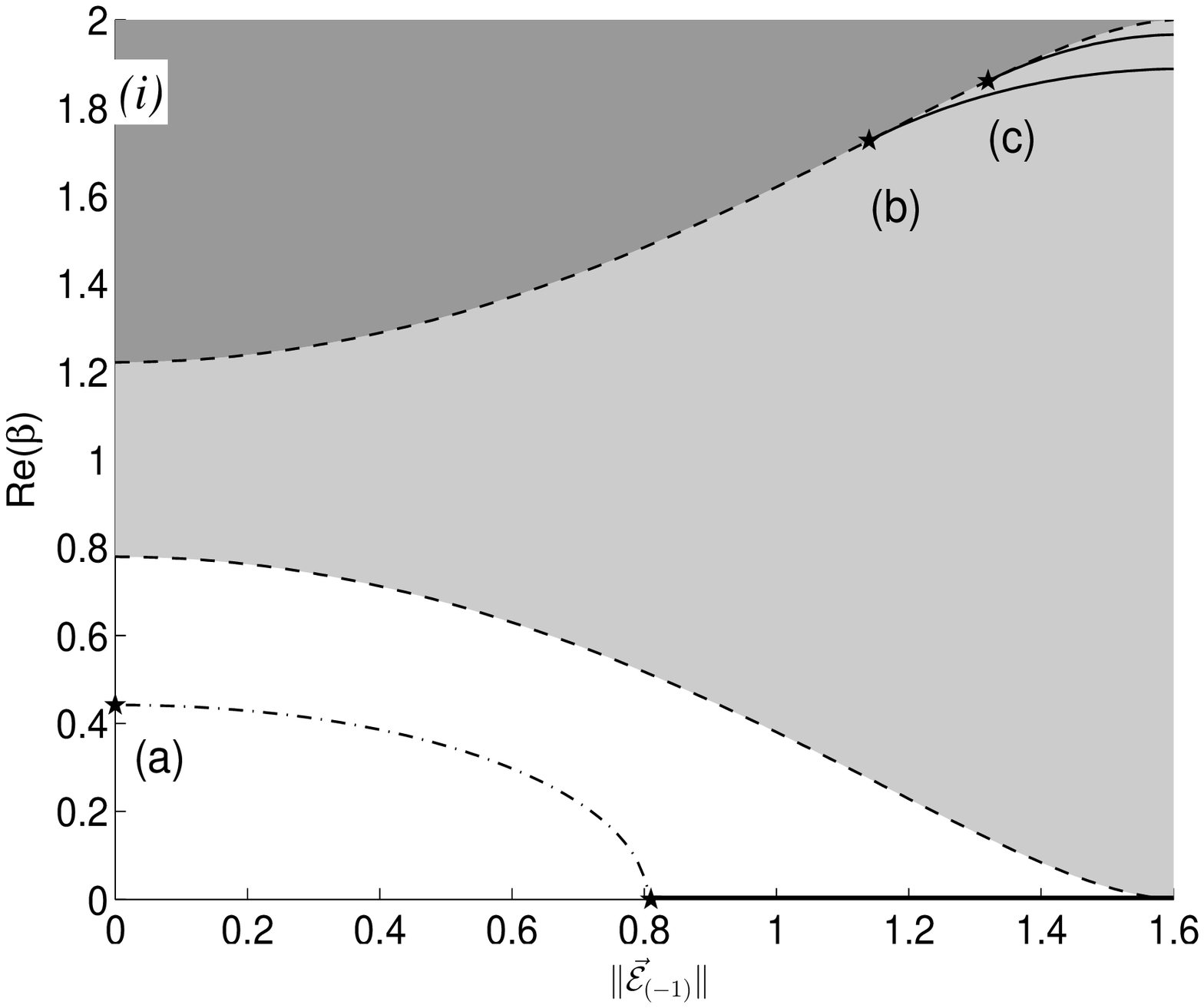}
\includegraphics[width=.45\textwidth]{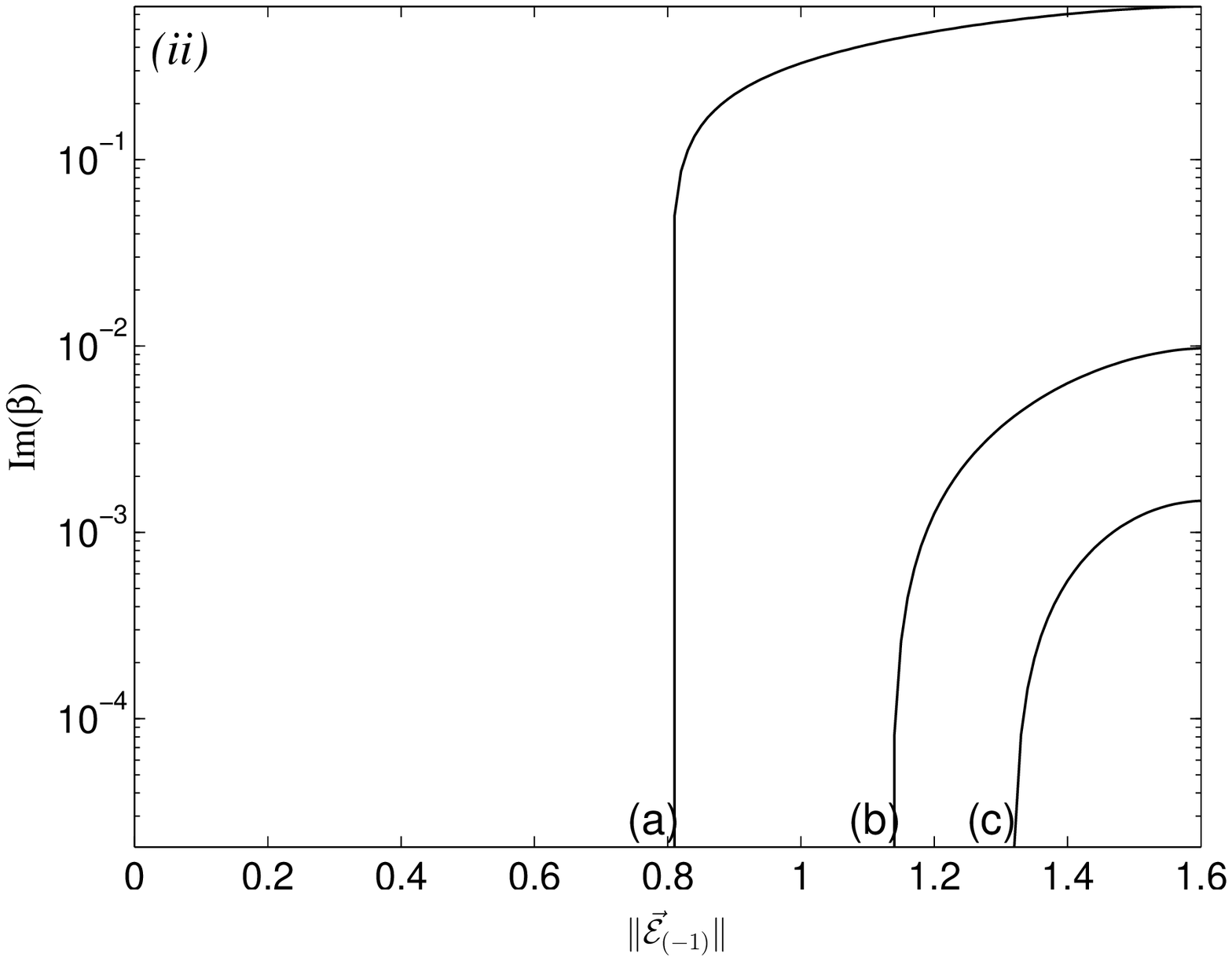}
\caption{Experiment 2.2, annotated as in figure~\ref{fig:111}: The frequencies \emph{(i)} and growth rates \emph{(ii)} for the linearization about the left nonlinear defect mode with $b=3$.}
\label{fig:Sol3_left}
\end{center}
\end{figure}

For the linear defect defined by~\eqref{eq:kappa} with $b=3$ and $k=1$, the linear frequencies decrease to $\w_{\pm1*} =\pm 0.221$ which puts the frequencies of the linearization inside the gap, as discussed in Example~1; see figure~\ref{fig:amp_freq}\emph{(iii)}.
The one discrete eigenfrequency (and its mirror image) present in the limit of zero amplitude is now in the gap (cf.\ the eigenvalues marked by `$+$' signs in figure~\ref{fig:spec_linearized}), and Hamiltonian symmetries prevent it from developing a nonzero imaginary part without colliding with another frequency or band edge.  We first look at the linearization about the mode $\vec{\cE}_{(-1)}$ with negative eigenvalue; see figure~\ref{fig:Sol3_left}. In this particular example the defect mode loses stability in a collision; frequencies labeled (a) at $\beta=\pm 0.442$ move toward zero and collide at amplitude $0.81$, and, as described by scenario 4, split into a pair of complex conjugate pure imaginary eigenvalues, leading to an instability which reaches a maximum growth rate of about $0.66$. This is the most significant instability seen in the three example defects explored.  Two further instabilities, labelled (b) and (c), arise from (scenario 3) edge bifurcations from the secondary band edge.

\begin{figure}
\begin{center}
\includegraphics[width=.45\textwidth]{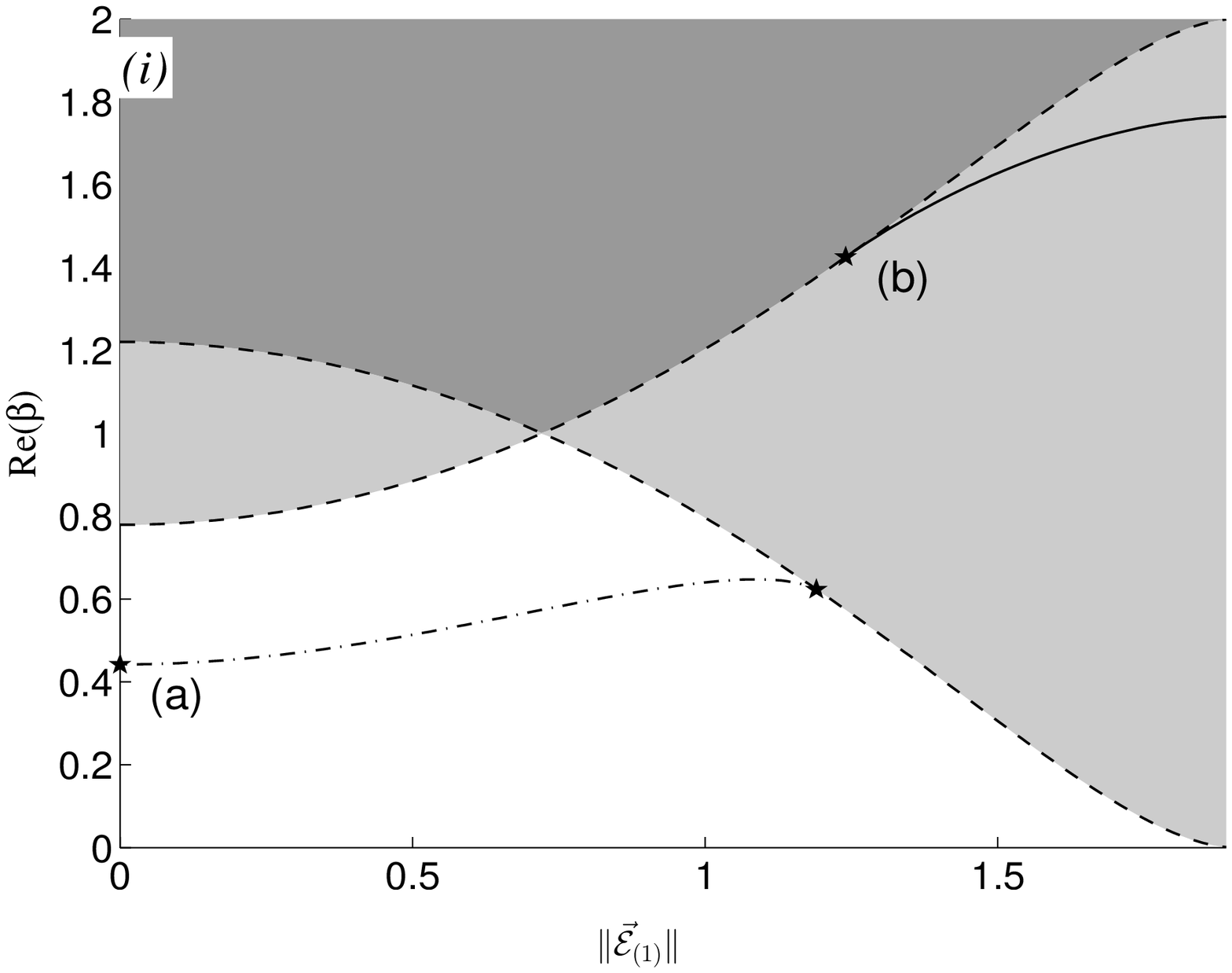}
\includegraphics[width=.45\textwidth]{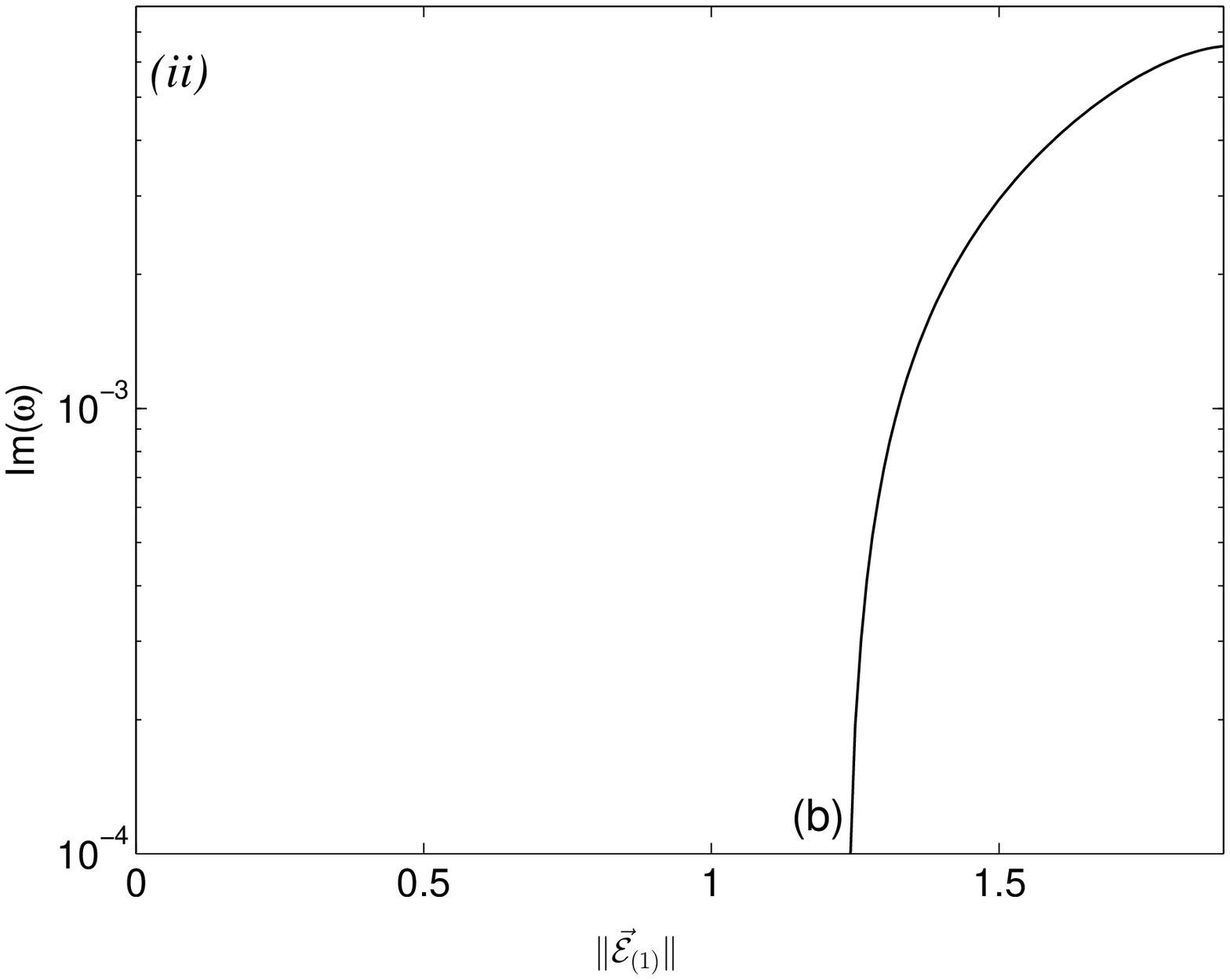}
\caption{Experiment 2.2, annotated as in figure~\ref{fig:111}: The frequencies  \emph{(i)} and growth rates  \emph{(ii)} for the linearization about the right nonlinear defect mode with $b=3$.}
\label{fig:Sol3_right}
\end{center}
\end{figure}

Finally, we look in figure~\ref{fig:Sol3_right} at the stability of the mode whose frequency in the linear limit is $+0.221$.  In this case the eigenvalue (a) in the gap moves toward the band edge, and eventually disappears in an edge bifurcation, moving to another ``sheet'' and becoming a \emph{resonance frequency} with negative imaginary part and corresponding resonance mode (non-$L^2$) with exponential decay in $t$.  Another instability (b) arises due to an edge bifurcation when the nonlinear mode has norm about 1.25 and reaches a maximum growth rate of about 0.0065.

\subsection*{Experiment 3: A defect supporting three bound states}
In the previous subsection, we were able to verify that the growth rate of nonlinear modes bifurcating from embedded frequencies scales as the fourth power of the $L^2$ norm, as was found analytically for scenario 1. In all the examples cited above the constants in front of the $\Norm{\vec \cE}^4$ term was small, meaning that at small amplitudes, energy does not move  quickly from one mode to another.  Here we show an example where the constant of proportionality is significantly larger.

\begin{figure}
\begin{center}
\includegraphics[width=.45\textwidth]{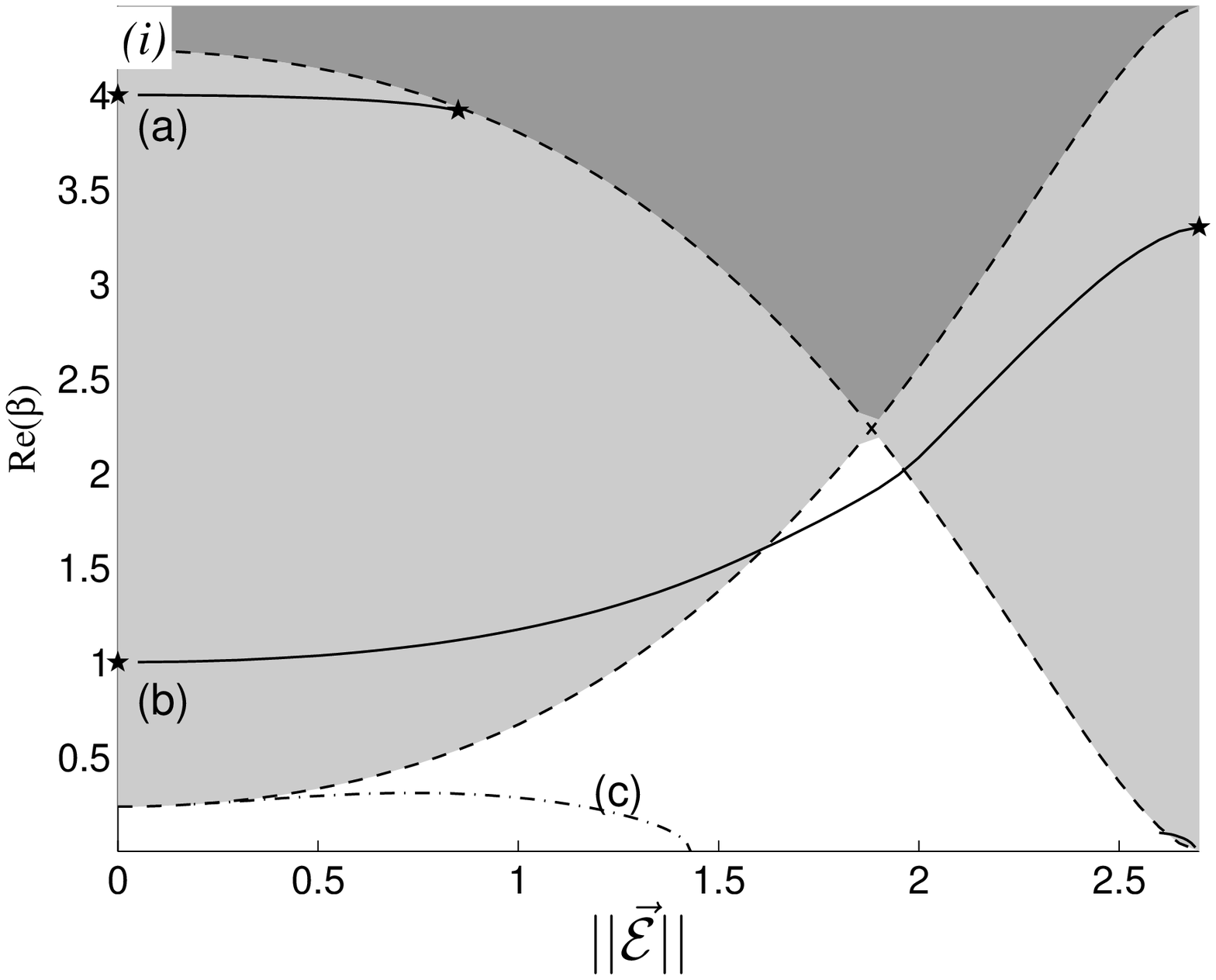}
\includegraphics[width=.45\textwidth]{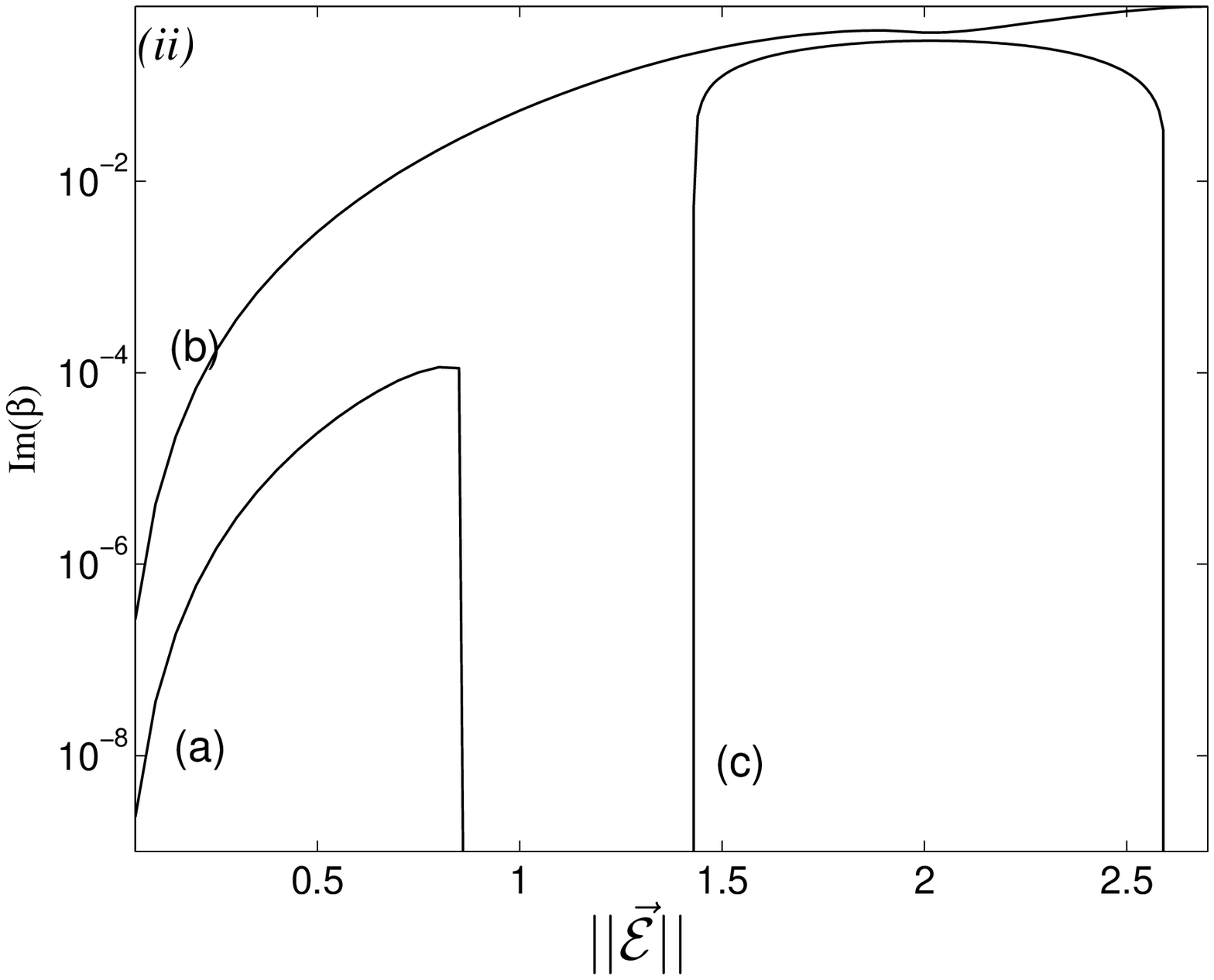}
\caption{Experiment 3, annotated as in figure~\ref{fig:111}: The frequency (left) and growth rate (right) of discrete modes of the linearization about the mode $\vec{\cE}_1$ of~\eqref{eq:stationary}.}
\label{fig:3mode_p1_instability}
\end{center}
\end{figure}

We consider the odd defect of the form~\eqref{eq:potential} with $\w=1$, $k=1$, $n=2$.  The linear problem has discrete eigenfrequencies $\w_{0*}=1$, $\w_{1*} = 2$, and $\w_{-1*}=-2$.  The nonlinear modes for this defect are described by figure~\ref{fig:all}  The linearizations about all three modes at zero amplitude possess embedded frequencies, so all three modes are unstable.  We will consider the instability of the mode $\vec \cE_1$, that with frequency $\w_{1*}=2$ in the linear limit (refer to figure~\ref{fig:all}), whose roots are plotted in figure~\ref{fig:3mode_p1_instability}.    The mode labeled (a) bifurcates from $\beta_{1,-1}^+=\w_{1*}-\w_{-1*}=2-(-2)=4$ moves to the other sheet in an edge bifurcation at $L^2$-norm about 0.84. The mode labeled (b) bifurcates from $\beta_{1,0}^+=\w_{1*}-\w_{0*}=2-1=1$ and develops a large growth rate with a maximum value of about 0.67, the one example we have found of an embedded mode leading to a large instability.  The mode labeled (c) appears in an edge bifurcation from the primary band edge and, as in scenario four, collides with its mirror image at the origin when $\Norm{\vec \cE}\approx 1.43$, becoming a pair of pure imaginary eigenvalues, which collide and become real again at amplitude 2.5 (not shown).  This instability is of the same order of magnitude as branch (b).

\section{Dynamics of competing discrete modes: Time-dependent numerical simulations}
\label{sec:timedependent}
In the previous two sections, by studying the spectral problem~\eqref{eq:eigen}, we provided analytical estimates and numerical approximations to the rate at which perturbations to given nonlinear defect modes grow.  Since the total intensity (squared $L^2$ norm) of a solution is preserved, growth of perturbations can be expected to lead to decay of the defect mode. 
We expect to observe a variant of the ground state selection scenario of~\cite{SofWei:03}, outlined in section~\ref{sec:NLS}; energy of an unstable mode is transferred to a preferred or selected discrete mode and to radiation modes. This scenario for NLCME is supported by simulations in~\cite{GSW}, reprinted in figure~\ref{fig:all}.  We now present further numerical evidence for this scenario.

At zero amplitude, the linearization about the mode $\vec{\cE}_{(j)*}$ has an eigenvalue $\beta_{j,k}^+= \pm(\w_{j*} - \w_{k*})$  whose eigenvector corresponds to a perturbation of the mode $\vec{\cE}_{(j)*}$ in the direction of the mode $\vec{\cE}_{(k)*}$; see section~\ref{sec:linearization}.  If this frequency is embedded in the continuous spectrum of $\Sigma_3H$ --condition~\eqref{eq:embed_cond}--and thus perturbs to one with positive imaginary part, then by equation~\eqref{A-final}, we expect the projection of the solution on the mode $\vec{\cE}_{(k)*}$ to grow.  Most of our simulations will begin with an initial profile described by a linear combination of linear defect modes. We plot, as time series, the projections of the numerical solution onto these modes.  In section~\ref{sec:embedded}, it was shown that embedded frequencies in the linearization generically perturb to isolated frequencies having nonzero imaginary parts. In particular, equation~\eqref{A-final} shows that the amplitude of the perturbation grows at the exponential rate~$\pi\ \abs{\left(\psi_{\tilde\lambda_0},W\psi_0\right)}^2/\abs{\left(\Sigma_3 \psi_0,\psi_0\right)}$.  We note that since the operator $W$ scales as $\Norm{\vec\cE}^2$, the growth rate is proportional to~$\Norm{\vec\cE}^4$, with an undetermined constant of proportionality. We have no a priori knowledge of the inner product $\left(\psi_{\tilde\lambda_0},W\psi_0\right)$ (its calculation requires the continuum (non-$L^2$) eigenfunction at frequency $\tilde\lambda_0$)  and the numerical calculations of section~\ref{sec:evans} demonstrate that the growth rate may in fact be quite small.

With that in mind, we now report on the results of direct numerical simulations of the dynamics of the competition between defect modes. We verify the analytical and numerical predictions of sections~\ref{sec:linearization} and~\ref{sec:evans}, and use the results of these sections to explain the results of~\cite{GSW}, summarized in figure~\ref{fig:JOSAB}.  

The PDE is discretized using a sixth-order upwind finite-difference scheme for the spatial derivatives and the fourth-order Runge-Kutta method is used for time-stepping.  The solutions lose energy to radiation through the endpoints of the computational domain, so outgoing boundary conditions are required, and we implement them using the method of perfectly matched layers (PML), as introduced by Dohnal and Hagstrom for coupled-mode equations~\cite{DohHag:07}.  For large-amplitude solutions, which shed a large quantity of radiation, the use of PML was absolutely vital to ensuring that the computations were not polluted by numerical artifacts.

\subsection*{Experiment 1: Dynamics in a NLCME defect supporting one bound state}

In experiment 1, the defect supports only one defect mode; there is no second mode to which the solution may transfer energy. This is analogous to a situation studied in detail for NLS / GP 
\cite{SW-Multichannel:90,SW-Multichannel:92,Pillet-Wayne:97,GNT:04}.  Our computations with the Evans function show the sole nonlinear bound state loses stability at an amplitude of just above 1.2 via a collision of discrete real eigenvalues with spectral gap edges (figure~\ref{fig:111}, scenario two).

\begin{figure}
\begin{center}
\includegraphics[width=2.5in]{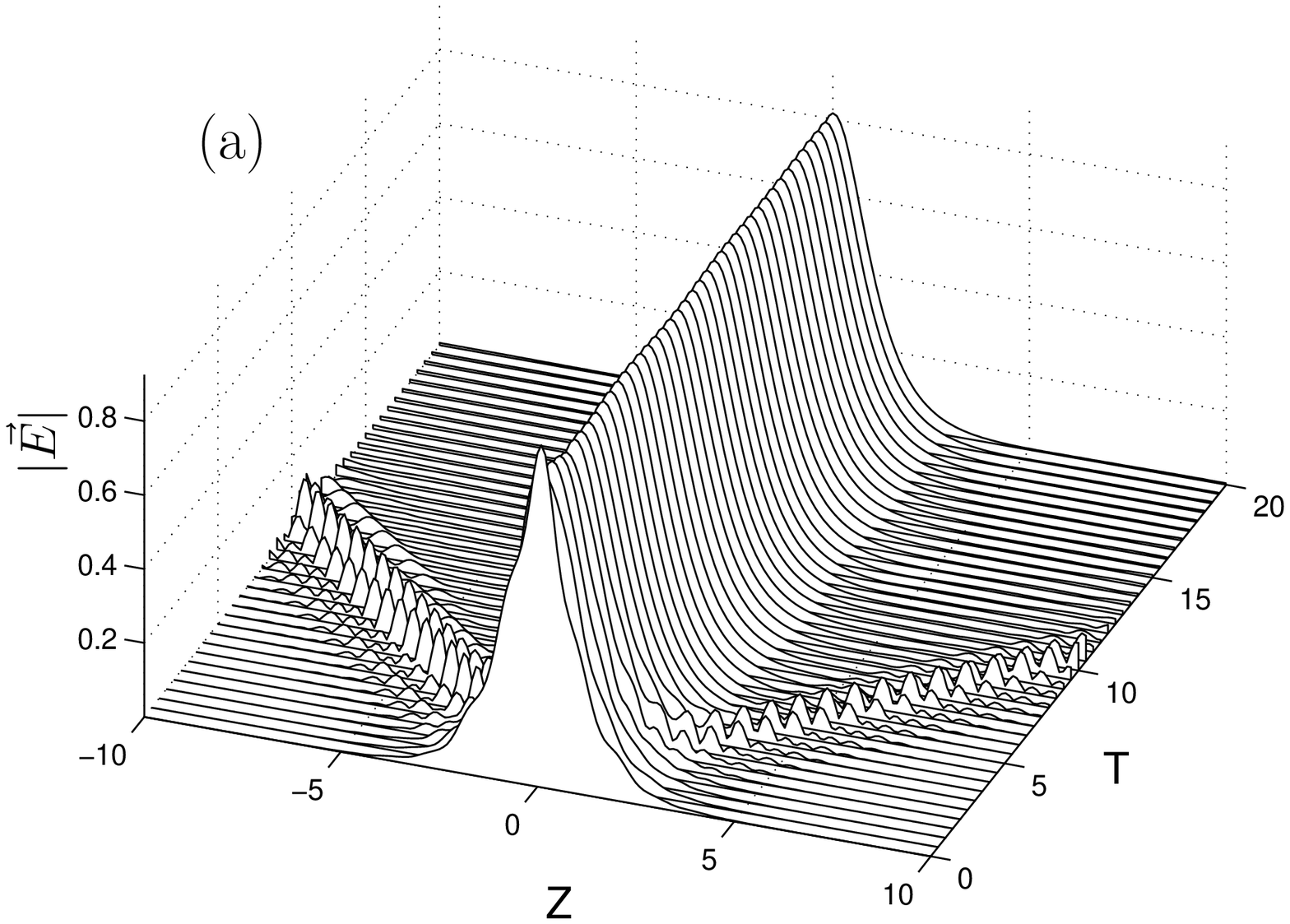}
\includegraphics[width=2.5in]{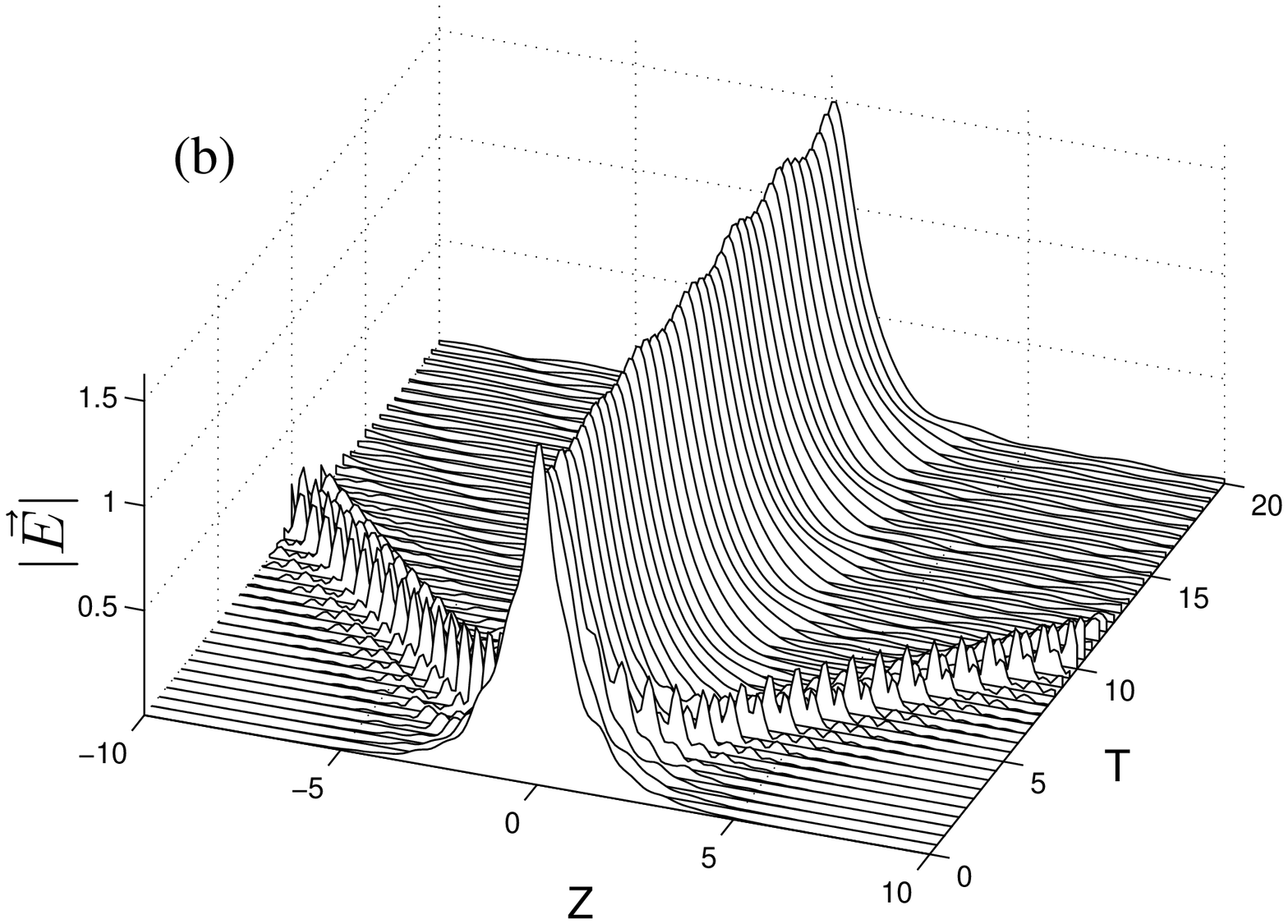}
\caption{Experiment 1: In the defect supporting only one bound state, \textbf{(a)} a perturbed nonlinear defect mode with amplitude 1 is stable, while \textbf{(b)} a perturbed nonlinear defect mode with amplitude 1.6 is unstable.  Note the continued shedding of radiation as well as the ``breathing'' of the pulse shape.}
\label{fig:experiment1}
\end{center}
\end{figure}

To test this prediction in a time-dependent numerical simulation, we begin with initial conditions consisting of a nonlinear defect mode (obtained by the methods described in appendix~\ref{sec:nl_numerics}) and a localized perturbation of about 10\% of its magnitude.  We show the results of two such experiments, the first with a nonlinear mode of amplitude ($L^2$-norm) 1.0 and the second of a nonlinear mode of amplitude 1.6 in figure~\ref{fig:experiment1}.  In the first case, the system quickly sheds a small amount of radiation and then settles down into a stable spatial profile (with an oscillating phase not shown). In the second case, the initial shedding of radiation is more dramatic and a very pronounced oscillation in the shape and amplitude of the trapped mode is seen throughout the length of the simulation, along with a continual shedding of radiation.  The perturbation oscillates as it grows, consistent with the quartet of complex eigenvalues shown in scenario two.

\subsection*{Experiment 2: Dynamics of NLCME defects supporting two bound states}
In Evans function Experiment 2, we looked at a family of defects that support exactly two defect modes.  In part 1, the linearization possesses an embedded frequency, while in part 2 all the frequencies of the linearized system are in the gap.  
\subsubsection*{Experiment 2.1: Dynamics of two bound states producing an embedded frequency}
\begin{figure}
\begin{center}
\includegraphics[width=.3\textwidth]{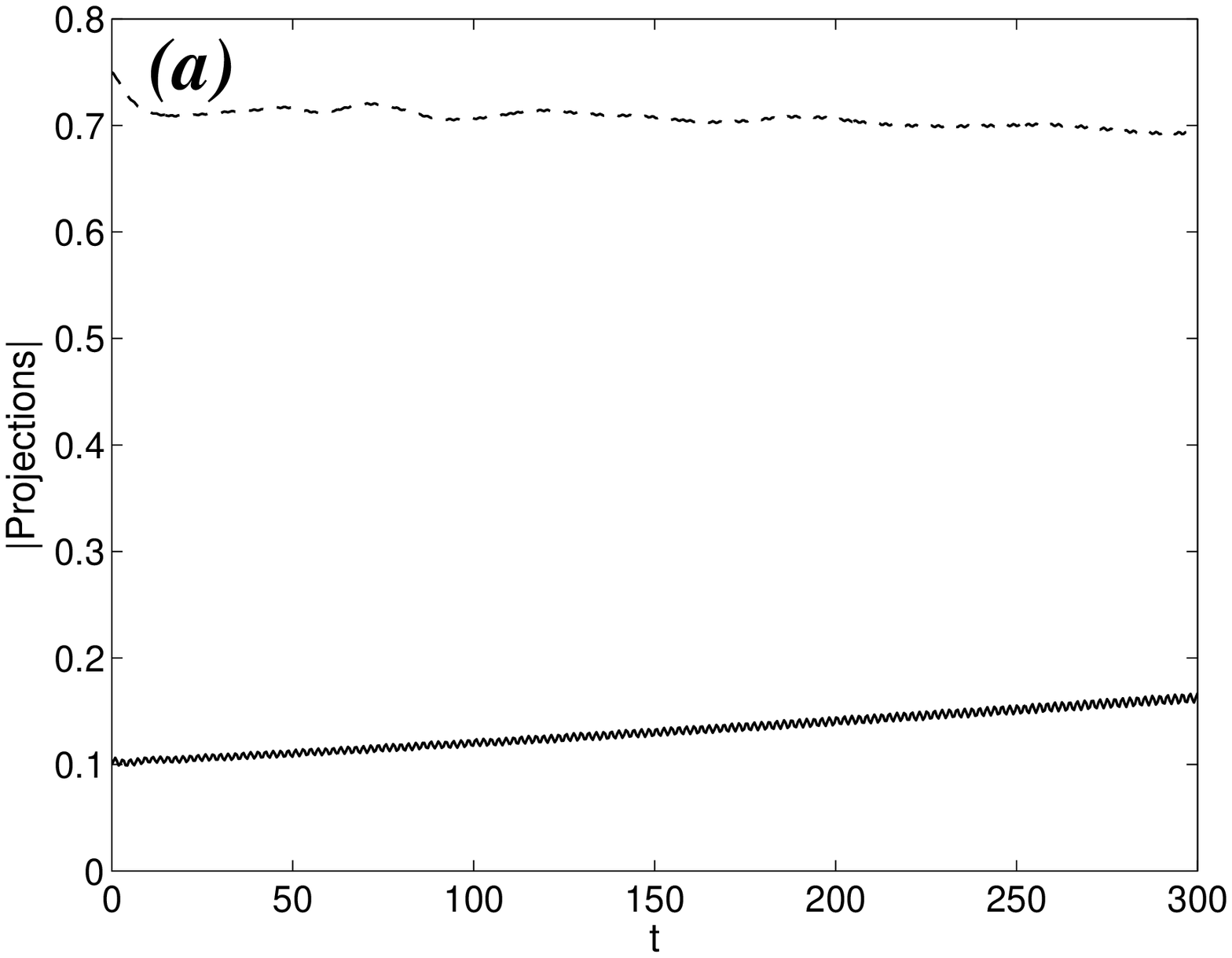}
\includegraphics[width=.3\textwidth]{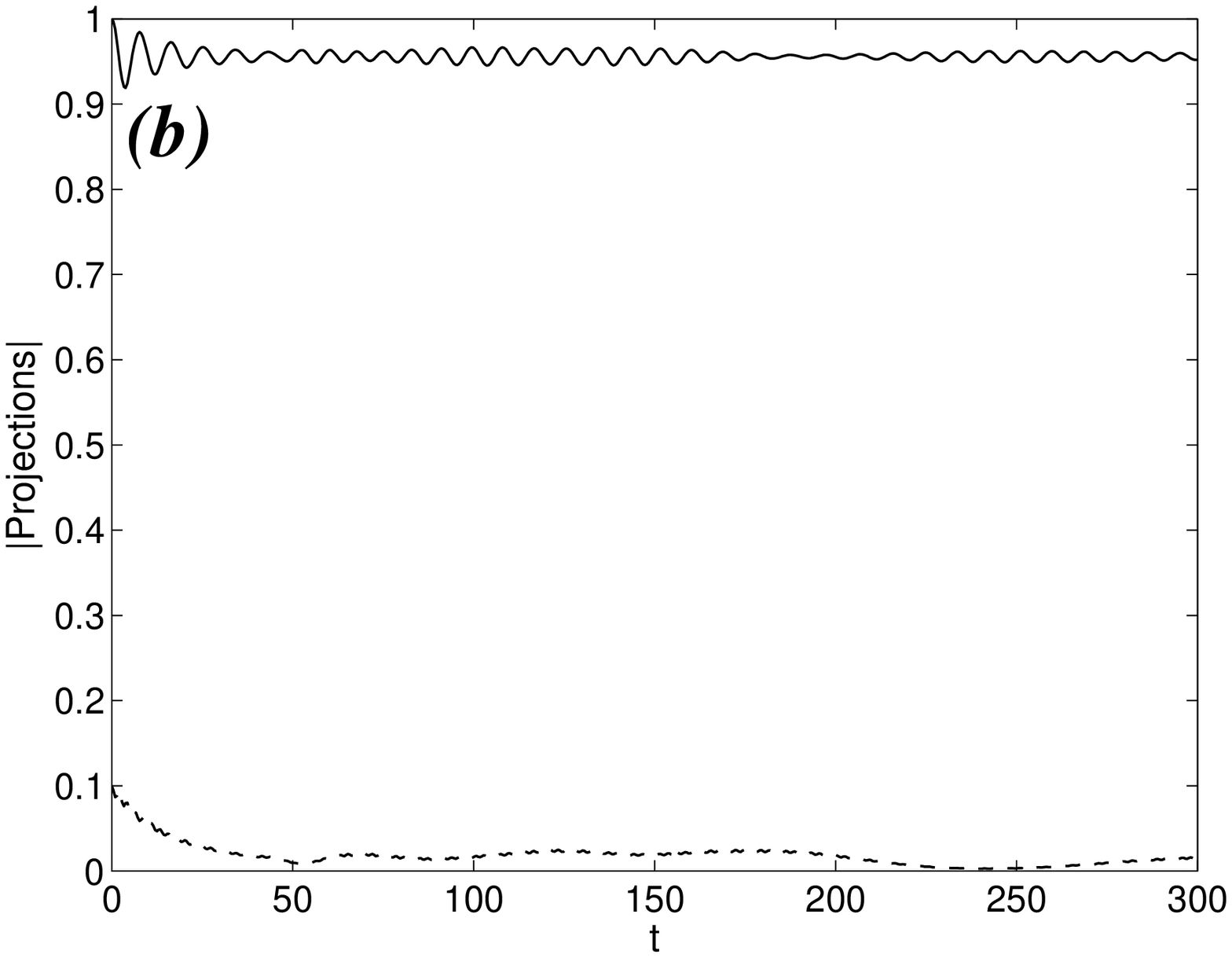}
\includegraphics[width=.3\textwidth]{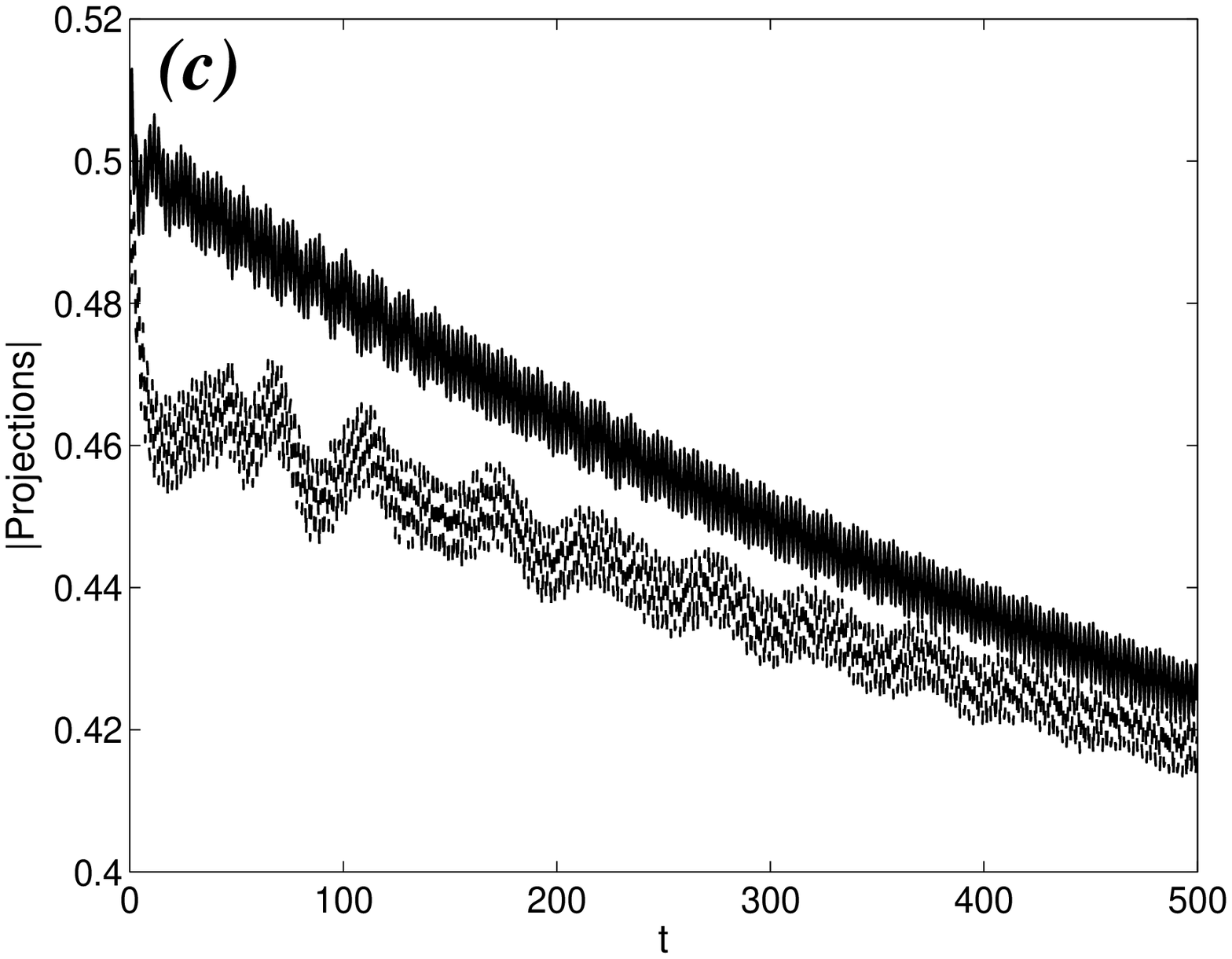}
\caption{Dynamics experiment 2.1:  \textbf{\emph{(a)}} energy initially mainly in the mode $\vec\cE_{(-1)}$ (case of figure~\ref{fig:Sol5_left}).  \textbf{\emph{(b)}} : energy mainly in mode $\vec\cE_{(+1)}$ (figure~\ref{fig:Sol5_right}); \textbf{\emph{(c)}}: equal energy initially in both modes simulated for a longer time. Key: in each figure, the solid line is the amplitude of $\vec\cE_{(+1)}$ and the dashed line is the amplitude of $\vec\cE_{(-1)}$}
\label{fig:Sol5}
\end{center}
\end{figure}

In section~\ref{sec:linearization}, we show that frequencies embedded in the continuous spectrum cause nonlinear defect modes to lose stability, but the numerical calculations summarized in figures~\ref{fig:Sol5_left}--\ref{fig:Sol5_right} show that this instability may be very weak.  We expect to see very slow energy transfer in the dynamics of this problem.  Using the defect defined by formula~\eqref{eq:kappa} with $k=3$ and $b=1$ gives embedded frequencies in the linearized problem.  Figure~\ref{fig:Sol5_left} shows that when $\Norm{\vec\cE_{(-1)}}=0.75$, perturbations should grow at a rate on the order of $10^{-4}$.  Figure~\ref{fig:Sol5}a shows that the growth of the perturbation (in the $\vec\cE_{(1*)}$ direction) is indeed quite small.
The calculations of the spectrum of the linearization about $\vec\cE_{(1)}$ summarized in figure~\ref{fig:Sol5_right} predicts that when $\Norm{\vec\cE_{(1)}}=1$, a perturbation in the direction of $\vec\cE_{(-1*)}$ should not grow at all (there are no non-zero growth rates at this point.) Figure~\ref{fig:Sol5}b shows that in this case the perturbation does not grow, and instead decays to zero rather quickly.  In figure~\ref{fig:Sol5}c, the two modes are initialized with equal nonzero amplitude, and both lose energy, with neither appearing to grow at the other's expense.

\subsubsection*{Experiment 2.2: Dynamics of two bound states with no embedded frequency}
Here we verify the scenario four instability described in Experiment 2.2. For that defect , which supports two bound states, neither of which has an embedded frequency in its linearization.  In the linearization about the bound state $\vec \cE_{(+1)}$ with positive frequency, figure~\ref{fig:Sol3_right} shows that no instabilities exist with significant growth rates.  Our numerical simulations we have confirmed this.  The results of these simulations are qualitatively very similar to those provided in figure~\ref{fig:Sol5}.

\begin{figure}
\begin{center}
\includegraphics[width=.3\textwidth]{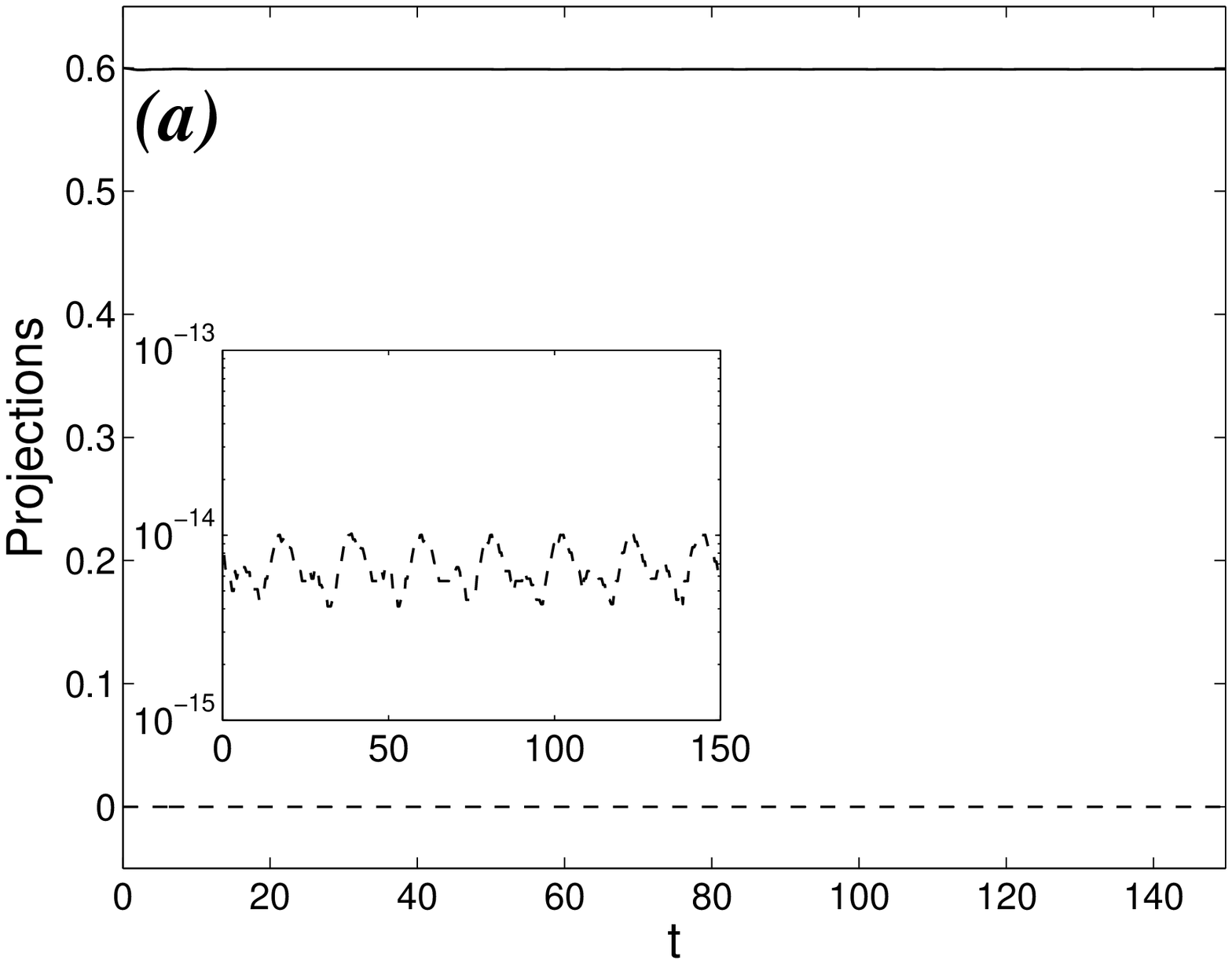}
\includegraphics[width=.3\textwidth]{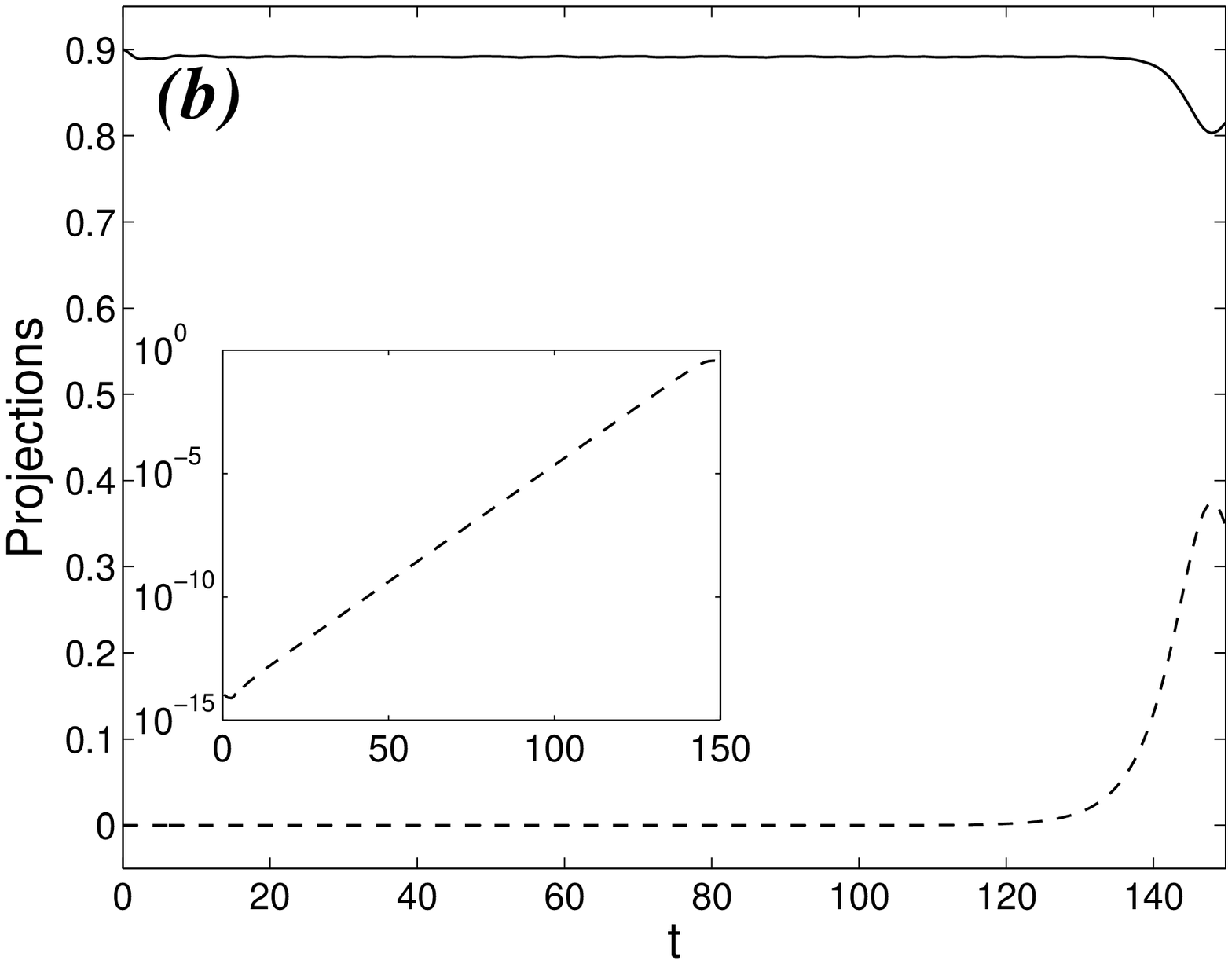}
\includegraphics[width=.3\textwidth]{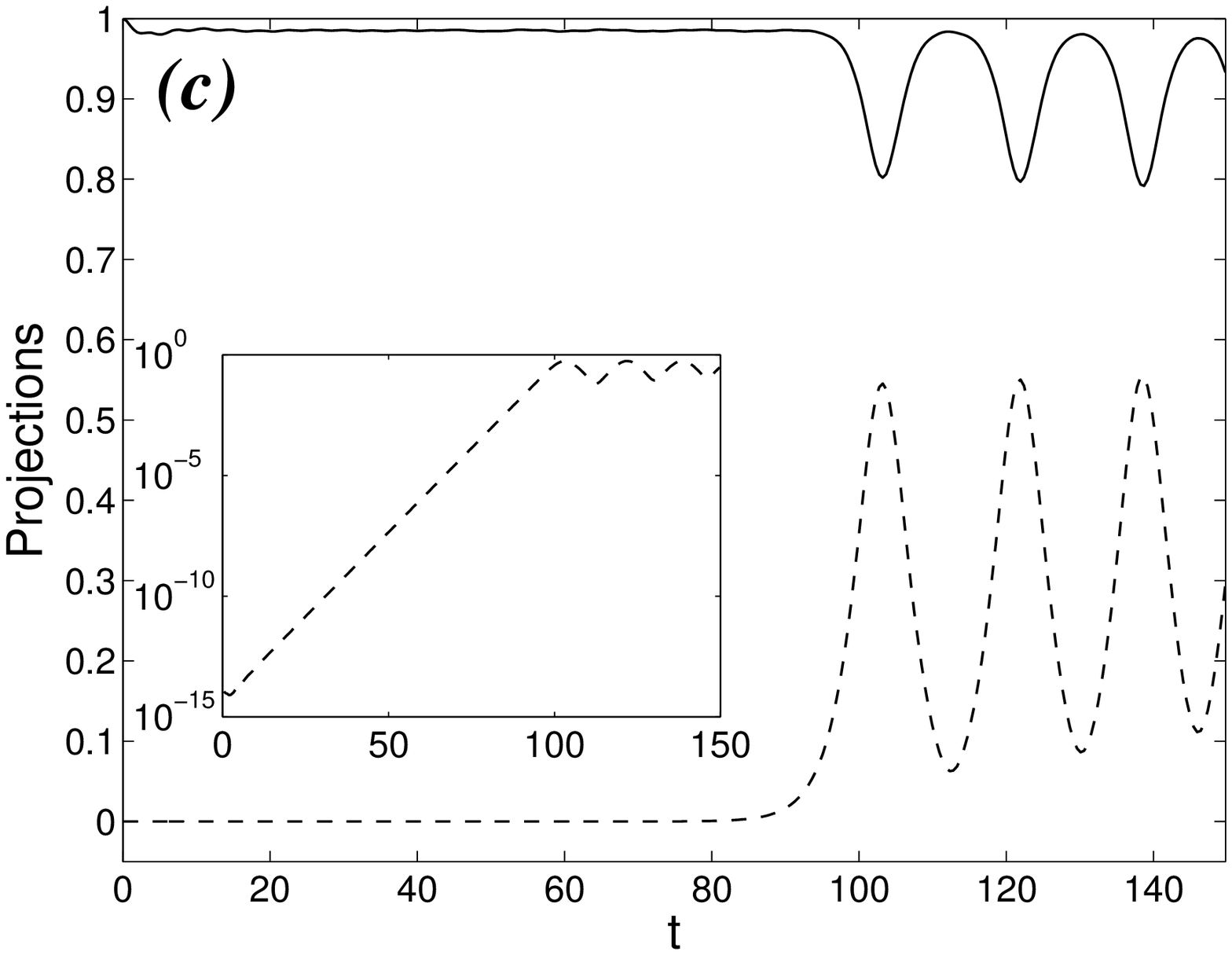}
\caption{Dynamical simulations illustrating the bifurcation shown in figure~\ref{fig:Sol3_left}.  \textbf{\emph{(a)}}: a stable solution (amplitude .6) below the merger of two real frequencies creates a pair of pure imaginary frequencies and instability.  \textbf{\emph{(b)}} and \textbf{\emph{(c)}}: the evolution of an initial condition with $L^2$-norm 0.9 and 1.0, respectively, above the instability threshhold. Here solid lines give the amplitude of $\vec \cE_{(-1)}$ and dashed lines give $\vec \cE_{(+1)}$. The insets present the same information with a logarithmic scale on the $y$-axis, demonstrating absolutely no growth of the second mode in (a) and exponential growth in~\textbf{\emph{(b)}} and~\textbf{\emph{(c)}}.}
\label{fig:Sol3}
\end{center}
\end{figure}

For the defect mode $\vec \cE_{(-1)}$ with negative frequency described by figure~\ref{fig:Sol3_left}, all frequencies in the linearization are real and in the gap for amplitudes below about $0.8$. At this point, two real frequencies collide at the origin and develop nonzero growth rates (scenario four). Figure~\ref{fig:Sol3} shows the evolution of the initial condition $0.6$ times the normalized linear eigenstate (below instability threshhold) and 0.9 and 1.0 times the normalized linear eigenstate (above the threshhold).  It is clear that energy flows from $\vec{\cE}_{(-1)}$ to $\vec{\cE}_{(1)}$ only when $\vec{\cE}_{(-1)}$ is unstable as predicted by experiment 2.2 (figure~\ref{fig:Sol5_left}) and that the rate of energy transfer increases as the amplitude is further increased.

\subsection*{Experiment 3: A defect supporting three bound states}
The spectrum of the linearization about the mode $\vec{\cE}_{(1)}$ of a defect supporting three bound states was discussed in figure~\ref{fig:3mode_p1_instability}.  This mode has frequency $\w_1=2$.  The other two frequencies of the linear problem are $\w_0=1$ and $\w_{-1}=-2$.  Our arguments show that the eigenmode of the linearization with frequency $\beta_{1,0}^+=1=\w_{1*}-\w_{0*}$ corresponds to perturbations in the direction of $\vec{\cE}_{(0)}$, while those with frequency $\beta_{1,-1}^+=4=\w_{1*}-\w_{-1*}$ correspond to perturbations in the direction of mode $\vec{\cE}_{(-1)}$.  Figure~\ref{fig:3mode_p1_instability} shows that the perturbation in the direction of $\vec{\cE}_{(0)}$, labeled \emph{(b)} in the figure grows at a much faster rate than the pertubation in the direction of  $\vec{\cE}_{(-1)}$, labeled \emph{(a)}.  In this last time-dependent experiment, we aim to test this prediction numerically.

\begin{figure}
\begin{center}
\includegraphics[width=.3\textwidth]{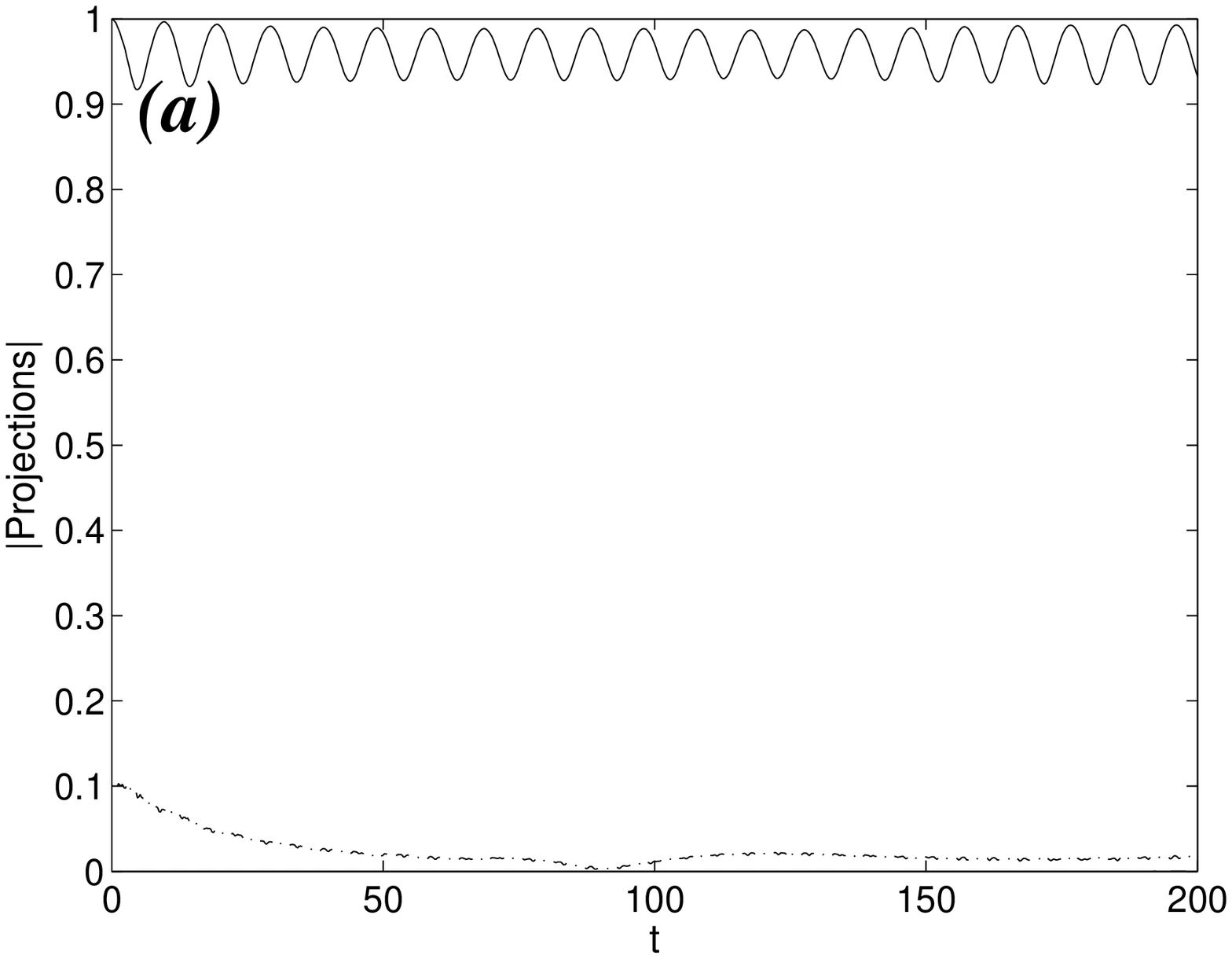}
\includegraphics[width=.3\textwidth]{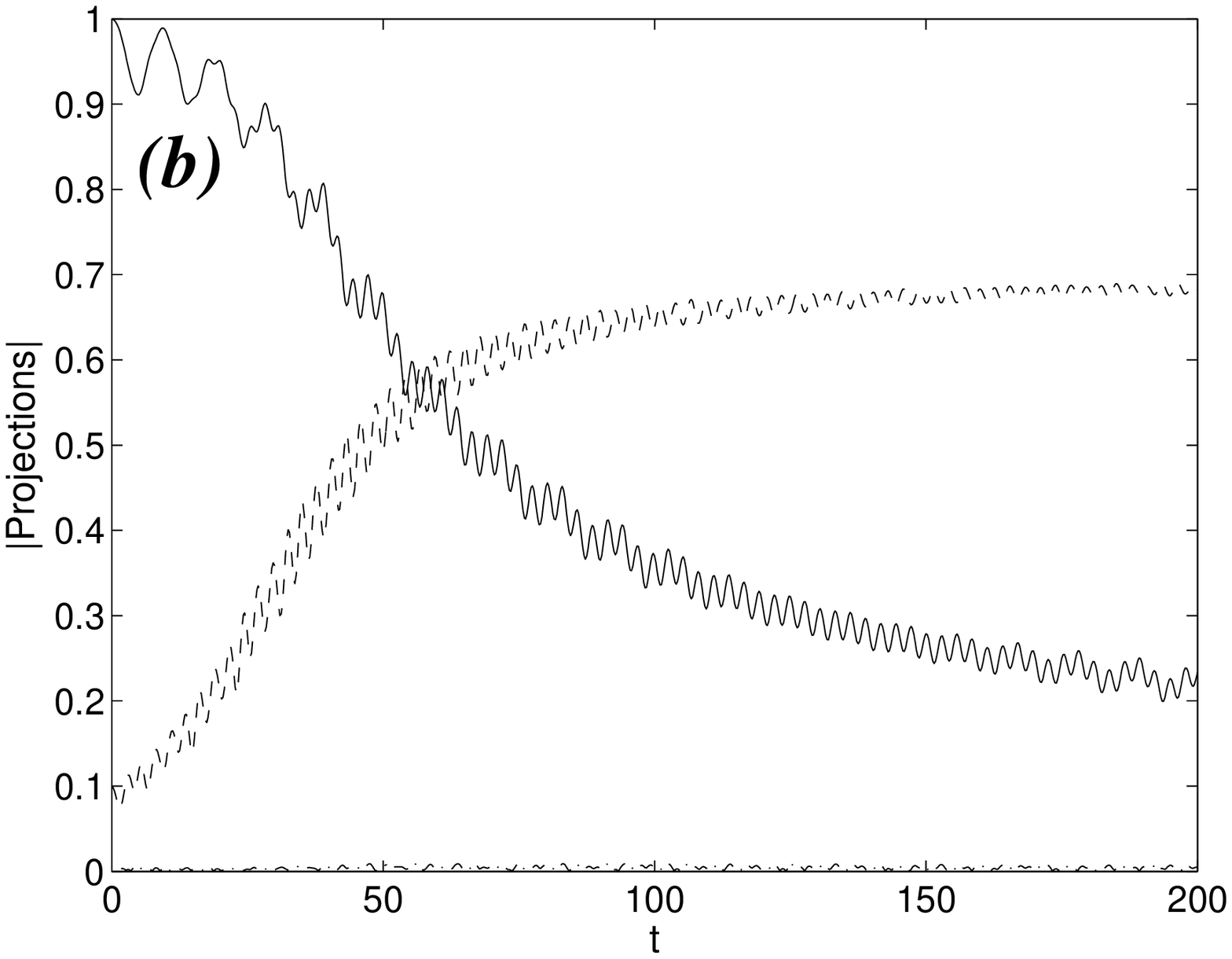}
\caption{\textbf{\emph{(a)}}:  Initially, almost all energy is in the mode $\vec\cE_{(1)}.$  In the left image, a small perturbation is added in the direction of mode $\vec\cE_{(-1)}$ with little transfer of energy.  \textbf{\emph{(b)}}: A perturbation is added in the direction of mode~$\vec\cE_{(0)}$, which grows in time.  This is predicted by figure~\ref{fig:3mode_p1_instability}.}
\label{fig:3mode_dyn}
\end{center}
\end{figure}

We have performed two simulations.  In the first, figure~\ref{fig:3mode_dyn}a, a perturbation of the mode $\vec{\cE}_{(1)}$  in the $\vec{\cE}_{(-1)}$ direction quickly decays, while in the second simulation, figure~\ref{fig:3mode_dyn}b, a perturbation of the mode $\vec{\cE}_{(1)}$  in the $\vec{\cE}_{(0)}$ direction grows, eventually overtaking the magnitude of the $\vec{\cE}_{(1)}$ mode. This is consistent with scenario one, as well as with the relative sizes of the growth rates found numerically in Experiment 3.
 
\section{Summary and discussion}
\label{sec:conclusion}
 
Wave propagation through a periodic one-dimensional structure at Bragg resonance is governed by NLCME. The introduction of a defect leads to a generalization of NLCME having spatially varying coefficients (potentials), which are determined by the defect. The linear coupled mode equations with defect potentials may have localized eigenstates. For a nonlinear medium, these linear states persist as nonlinear defect modes, which bifurcate from the zero state and the linear (defect) eigenfrequencies. Such nonlinear defect modes correspond to  optical pulse states, pinned at the defect sites. Such periodic structures and the  nonlinear defect states are candidates for optical storage schemes~\cite{GSW,patent}. Only stable nonlinear defect states could play this role.

Therefore, we have considered by analytical methods and numerical simulation the stability and instability of such nonlinear defect modes. We first classify the various scenarios for onset of instability and give examples which arise for  a large class of structures.
  
We investigate in particular detail the situation of a family of bifurcating states, whose linearization has an embedded eigenvalue in the continuous spectrum, in the limit of zero amplitude, {\it i.e.} at the bifurcation point. Here we find (generically)  that for arbitrarily small amplitudes such states are exponentially unstable with a  exponential rate of growth of order the fourth power of the nonlinear bound state. Although exponentially unstable, we find this growth rate in practice to be quite slow in many examples.

States for which the linearization, at the bifurcation point, exhibits no embedded eigenvalues appear to be stable, at least on long time intervals, for a range of amplitudes but exhibit instability via a collision of discrete eigenvalues of the linearization or the collision of a discrete eigenvalue with the endpoint of the continuous spectrum. For a given defect with multiple nonlinear bound states, different states may follow different scenarios and instability may result in the transfer of energy from one mode to another and to radiation.

It is instructive to compare the dynamics we have found with those of the NLS/GP equation.  In the small-amplitude limit, the behavior is slightly different because NLCME's second branch of continuous spectrum makes it harder to find defects which will produce linearizations without a resonance.  At higher amplitudes, we have found a large number of instabilities that arise due to collisions of discrete eigenvalues with the endpoint of the continuous spectrum. Instabilities due to collisions between pairs of eigenvalues  lead to the strongest instabilities (largest growth rates)  observed.  A related problem, also with two branches of continuous spectrum, is a coupled NLS system describing light propagation in birefringent optical fibers, studied by Li and Promislow in~\cite{LiPro:00}.  Using somewhat different methods than those employed here, they also show a solitary wave is unstable due to bifurcations involving embedded eigenvalues.

Another interesting behavior for the NLS/GP system has been seen by Kirr et al.~\cite{KirKevShl:07} and some of the numerical experiments discussed above suggest the same phenomenon exists in NLCME. They consider NLS/GP system  with a defect equal to the superposition of two potentials placed a fixed distance apart and discover a symmetry-breaking bifurcation. At small amplitudes, the stable ground state solution is given to leading order by the sum of two ground states of the individual potentials, in-phase and of equal amplitude, a highly symmetric configuration. At higher amplitudes, this configuration loses stability and the stable solution is concentrated almost entirely at one or the other defect location. This is what is known in dynamics as a supercritical pitchfork bifurcation.  Bifurcation scenario four is precisely the mechanism seen in a generic pitchfork bifurcation, so it is of interest to numerically locate the stable states that should exist at amplitudes above the critical amplitudes at which bifurcations are seen in experiments 2.2 and 3.

The symmetry breaking bifurcation could be understood by appeal to a variational principle. In cases where the {\it ground state} can be characterized as a minimum of $\cH[U]$, the NLS-GP Hamiltonian energy, subject to fixed $\| U\|_2^2=\cN$, it can be shown \cite{AFGST:02} 
that for small $\cN$ the ground state is symmetric, with peaks situated
over the local minima of each well, while for $\cN$ sufficiently large
the ground state is strongly localized in one well or the other. Symmetry breaking, if it occurs in NLCME, would be quite interesting and could be studied by the methods of~\cite{KirKevShl:07}.   Note however that in contrast to NLS-GP, nonlinear bound states of NLCME are critical points of infinite index, since the linearized energy has continuous spectrum which is unbounded above and below.
 
Finally, it would be instructive to extend~\eqref{A-final}---which describes the growth or decay of a single defect mode due to its interaction with the continuum---to a coupled system of equations describing the interactions of multiple defect modes with each other and with the continuum.  Such a derivation has been carried out rigorously by Soffer and Weinstein for the NLS/GP system with two defect modes.  They show that in the NLS/GP at low amplitude, the ground state is nonlinearly stable and derive a Fermi golden rule that predicts that half the energy of the excited state will be transferred to the ground state as $t\nearrow\infty$. Immediate extensions of this are to NLS/GP with 3 or more defect modes \cite{Tsai-multi:03} and to NLCME, especially in the case of 3 or more modes.  Such models were able to predict the symmetry-breaking bifurcation in~\cite{KirKevShl:07} and may be able to illuminate the Hamiltonian bifurcation of our experiment 2.2 in section~\ref{sec:evans}.  We note that Aceves and Dohnal have derived finite-dimensional models in~\cite{AceDoh:06}, although these models do not incorporate the coupling to the continuous spectrum seen to be important in section~\ref{sec:linearization}.

\section*{Acknowledgments} 
We thank Russell Jackson and Georg Gottwald for early conversations on the numerical computation of Evans functions.  RHG acknowledges support from NSF grants  DMS-0204881 and DMS-0506495; MIW from NSF grants DMS-0412305 and DMS-0707850.

 \appendix
\section{List of Symbols}
\noindent \textbf{Complex variables:}
$\Re z$ and $\Im z$\ denote the real and  imaginary parts of a  complex number $z$. Its complex conjugate is $z^*$.

\noindent \textbf{A projection operator:}
$ P_{c*}$\ -\ projection onto the continous spectral part of the self-adjoint 
 operator~$H$. 

\noindent \textbf{Inner Product Convention:} $\langle f,g\rangle = \int \bar{f}(Z)\ g(Z)\ dZ$

\noindent \textbf{Pauli matrices:}
$$\sigma_1 = \begin{pmatrix} 0 & 1 \\1 & 0\end{pmatrix};
 \sigma_3 = \begin{pmatrix} 1 & 0 \\0 & -1\end{pmatrix};
 \Sigma_3 = \begin{pmatrix} I & 0 \\ 0 & I \end{pmatrix}, \text{ where } I =  
 \begin{pmatrix} 1 & 0 \\0 & 1\end{pmatrix}
$$

\noindent \textbf{Linear and nonlinear solutions:} $\vec \cE = \binom{\cE_+}{\cE_-}$ is a solution to the nonlinear eigenvalue problem~\eqref{eq:stationary} while solutions with subscript aterisks, i.e.\ $\vec \cE_*$ are solutions to the linear problem~\eqref{eq:linear}.  Associated to these are frequencies $\omega$ (eigenfrequencies $\omega_*$ of $\Sigma_3 H$ in the linear limit.)

\noindent \textbf{Plemelj identities}
\begin{equation} (x\mp i0)^{-1}\ = \ 
 {\rm P.V.}\ x^{-1}\ \pm\ i\pi\ \delta(x)
\label{eq:Plemelj}\end{equation}
 
\section{Sketch of derivation of NLCME with potentials}\label{sec:nlcme-sketch}
Consider light propagation in a fiber with linear polarization and assume that the signal propagates with a constant (given) transverse profile.  The dielectric susceptibility is assumed to possess Kerr (cubic) nonlinearity.  Under these assumptions, Maxwell's equations for the scalar electric field $E(z,t)$ simplify to
\begin{equation}
\label{eq:maxwell}
\partial_z^2 E(z,t) -c^{-2}n^2(z,E) \partial_t^2 E(z,t) = 0.
\end{equation}

If the dielectric function is assumed to be constant $n(z,E)=\bar n$, this is the linear wave equation.  Seeking plane waves of the form $e^{i(kz-\w t)}$, leads to the dispersion relation 
\begin{equation}
\label{eq:dispersion}
\w^2 - c^2 k^2 / \bar n^2=0.
\end{equation}
For any pair $(k,\w)$ satisfying~\eqref{eq:dispersion}, the solution is a Fourier sum of backward and forward propagating plane wave solutions: 
\begin{equation}
E(z,t)= \Eplus e^{i(kz-\w t)} +\Eminus e^{-i(kz+\w t)} +{\rm c.c.}
\label{eq:planewave}
\end{equation}
where $\Eplus$ and $\Eminus$ are the complex amplitudes of the forward and backward moving Fourier components at wavenumber $k$.

Nonlinearity and periodic longitudinal variations in the linear dielectric constant will result in coupling between the forward and backward plane waves, and slow changes to the periodic profile will result in the creation of a potential. The dielectric constant consists of three parts $n^2 = \tilde n^2 + n^2_{\rm grating}(z) + n^2_{\rm NL}$
where
\begin{equation}
\label{eq:dielectric}
\begin{split}
\tilde n^2 &= \bar n^2 + \eps W(\eps z); \\
n^2_{\rm grating}(z) &= 2 \eps \chi^{(1)}(\eps z) \cos{(2 k_0 z + 2 \phi(\eps z))};\\
n^2_{\rm NL} & = \chi^{(3)} E^2(z,t).
\end{split}\end{equation}
The dimensionless parameter $\eps$ is taken to be small and the remaining parameters to be order one. 
The form of the dielectric constant~\ref{eq:dielectric} implies a specific relation between (a) the amplitude of refractive index variations, (b) 
the spatial scale over which the uniform grating is modulated (by apodization, introduction of defects etc.), and (c) the nonlinearity. This relation is chosen so that all these effects enter at the same order in the multiple scale analysis~\cite{KC}, i.e. the maximal balance. This yields an ansatz of the form:
\begin{equation}
\label{eq:expansion}
\begin{split}
E &= \sqrt{\eps} E_0 + O(\eps^{3/2}) \text{ where} \\
E_0 &= \tEplus(z_1,t_1) e^{i(k_0z-\w_0t +\phi(z_1))} 
+ \tEminus(z_1,t_1) e^{-i(k_0z+\w_0t +\phi(z_1))}.
\end{split}
\end{equation}
Here $z_1 = \eps z$ and $t_1= \eps t$.  Notice the approximate solution is chosen with twice he wavelength of the underlying grating---this is the Bragg resonance condition.  Inserting this ansatz into~\eqref{eq:maxwell}, separating terms by magnitude in~$\eps$, and eliminating secular terms at order $\eps^{3/2}$ in the expansion leads to ``slow equations'' for the evolution of the amplitudes $\tEplus$ and $\tEminus$:
\begin{align*}
i \partial_{z_1} \tEplus + i \frac{\bar n}{c} \partial_{t_1} \tEplus + \left[\frac{k_0 W(z_1) }{2 \bar n^2}- \phi'(z_1) \right] \tEplus + \frac{\w_0 \chi^{(1)}(z_1)}{2 \bar n c} \tEminus +
\frac{3 k_0\chi^{(3)}}{\bar n^2}(2\abs{\tEminus}^2 + \abs{\tEplus}^2) \tEplus & = 0; \\
-i \partial_{z_1} \tEminus + i \frac{\bar n}{c} \partial_{t_1} \tEminus + \left[\frac{k_0 W(z_1) }{2 \bar n^2}- \phi'(z_1) \right] \tEminus + \frac{\w_0 \chi^{(1)}(z_1)}{2 \bar n c} \tEplus +
\frac{3 k_0\chi^{(3)}}{\bar n^2}(2\abs{\tEplus}^2 + \abs{\tEminus}^2) \tEminus & = 0 .
\end{align*}
This multiple scale procedure~\cite{KC} is rigorously justified for a close analog of these equations in~\cite{GWH}; see also Martel~\cite{Mar:05,MarCas:07} for higher-order corrections that can effect the stability.

Introduce dimensionless distance  $Z = z_1/L$, time $T = \bar n Z/c$, and electric field $ \Epm = \tEpm/\mathbb{E}$ variables for some characteristic length scale $L$ and electric field strength $\mathbb{E}$ and scaled functions
\begin{equation}
W_L(Z) = W(L Z), \thickspace 
\phi_L(Z) = \phi(L Z), \text { and } 
\chi^{(1)}_L(Z)= \chi^{(1)}(LZ).
\label{eq:scales}
\end{equation}
A sketch of the above scalings and the Bragg resonance condition is given in figure~\ref{fig:schematic}; for a more detailed discussion of the physical length and time scales see~\cite{GSW}.
The above equations become
\begin{align*}
i\pT\Eplus + i \pZ\Eplus
+ \k(Z) \Eminus + V(Z)\Eplus
+ (\abs{\Eplus}^2 + 2\abs{\Eminus}^2)\Eplus
&=0 \\
i\pT\Eminus - i \pZ\Eminus
+ \k(Z) \Eplus + V(Z) \Eminus
+ (\abs{\Eminus}^2 + 2\abs{\Eplus}^2)\Eminus
&=0,
\end{align*}
where we have defined 
$$
\k(Z) = \frac{\w_0 L}{2 \bar n c} \chi^{(1)}_L(Z)  = \frac{L k_0}{2\bar n^2} \chi^{(1)}_L(Z),\,
V(Z) = \frac{L k_0}{2 \bar n^2} W_L(Z) - \phi'_L(Z),\, 
\text{ and } \mathbb{E}= \frac{\bar n}{\sqrt{3 k_0 L \chi^{(3)}}}.
$$
which is precisely~\ref{eq:NLCME}.  Note that the coefficient $\k(Z)$ arises entirely due to modulations in the depth of the grating, while $V(Z)$ may arise entirely due to slow changes to the average susceptibility or else due to local modulations in the wavelength of the grating.

 \section{Numerical Computation of Nonlinear Defect Modes}
\label{sec:nl_numerics}
Equation~\eqref{eq:stationary} is discretized  on the $Z$-interval $[-L,L]$,
setting $h = \frac{2L}{N}$ and $Z_j = -L + j\  h$,  for $j=0,\ldots,N-1$.
Approximating $\Epm(Z_j) \equiv e_{\pm,j}$, and defining the $N$-dimensional
vectors of unknowns $\vec{e}_+$ and $\vec{e}_-$, we arrive at the
system 
\begin{subequations}
\label{eq:discrete}
\begin{gather}
\w e_{+,j}  + i (\cD  {\vec e}_+)_j
+ \k(Z_j) e_{-,j} + V(Z_j)e_{+,j} 
+ (\abs{e_{+,j}}^2 + 2\abs{e_{-,j}}^2) e_{+,j}
=0  \label{eq:discrete1} \\
\w e_{-,j}  - i (\cD {\vec e}_-)_j
+ \k(Z_j) e_{+,j} + V(Z_j)e_{-,j} 
+ (\abs{e_{-,j}}^2 + 2\abs{e_{+,j}}^2) e_{-,j}
=0, \label{eq:discrete2}\\
h \cdot \sum_{j=1}^N \bigl( \abs{e_{+,j}}^2 + \abs{e_{-,j}}^2\bigr)  = I.
\label{eq:discrete3}
\end{gather}
\end{subequations}
with $2N$ complex unknowns $e_{\pm,j}$ and one real unknown $\w$.  $\cD$
is the discretized derivative operator, which will be discussed further below.
Below we use symmetries to reduce the number of real unknowns from $4N+1$ to $N+1$.

Assuming $\k(Z)$ and $V(Z)$ are even functions, the linear standing-wave solutions to~\eqref{eq:linear} all satisfy 
$$
\cE_{(j)-} = \fsign \cE^*_{(j)+} 
$$
where $\fsign = \pm 1$ and the overbar denotes complex conjugation.  The nonlinear defect modes are the continuations of the linear defect modes and preserve these symmetries, although it is possible that these solutions lose stability in symmetry-breaking bifurcations at higher amplitudes..  Incorporating this into the scheme~\eqref{eq:discrete} eliminates $2N$ complex unknowns.

By phase invariance, $\Eplus(0)$ may be assumed to be real and positive.  Then $\Re{(\Eplus)}$ is even  and $\Im{(\Eplus)}$ is odd.
Using this symmetry, we need solve only for $R_j=\Re{(e_{+,j})}$ for
$j=0,\ldots N/2$, and $I_j=\Im{(e_{+,j})}$ for $j=1,\ldots N/2 -1$.  The
remaining values may be found by symmetry, leaving $N+1$ real unknowns.  
Accounting for the symmetries eliminates a one-dimensional invariant set of fixed points and 
speeds the root-finder's convergence. 

The eigenvalue problem is posed on an infinite domain.  Solution of a  problem on
an infinite domain involves balancing two competing sources of error: truncation and discretization.  To eliminate error due to the truncating the computation to an interval $[-L,L]$, $L$, and thus $dZ$ should be taken large, but to
eliminate discretization error $dZ$ should be taken small.
To balance these two effects, a non-uniform discretization is imposed.  
The finite interval $[-\pi,\pi)$ is mapped to a much larger interval by 
\begin{equation}
\z = \sinh {\lambda Z},
\label{eq:stretch}
\end{equation}
as proposed by Weidemann and Cloot~\cite{WeiClo:90} and summarized by
Boyd~\cite{Boy:01}.  The uniform grid $Z_j = -\pi + j \Delta Z$ is then mapped
to a nonuniform grid $\z_j$, with points that are closely and almost
uniformly spaced near $\z=0$, and sparsely spaced for large $\z$.  Boyd
provides guidelines for the choice of the stretching parameter $\lambda$~\cite{Boy:94}. 
Because
the Evans function requires an integration from conditions at $\pm \infty$, it
is important to resolve the tails well at this stage.  We  found we could obtain 
 more accurate numerical defect modes using 256 points on a
stretched grid than using 2048 points on a uniform grid, where both grids were
of the same width, in addition to the obvious savings in computation time. 

On the periodic domain $[-\pi,\pi)$, Fourier collocation is used to
derive the discretized derivative operator $\cD$ in~\eqref{eq:discrete}, which
is then multiplied by a weighting factor $\cW_j = \frac{1}{\lambda
  \sqrt{1+\z_j^2}}$ to account for the stretching  
transformation~\eqref{eq:stretch}.  As long as the solutions are sufficiently
small at $\z_{\rm max} = \sinh{\lambda \pi}$, the error introduced into the
equations by the assumption of periodicity is small, and the numerical
solutions converge quickly as $n$, the number of grid points, is increased.
What is important is that the exponential decay rate in the tails is computed
accurately and that the  numerical solution is computed accurately far enough
into the tail so that $\abs{\Epm}\sim 10^{-8}$, then since the linearized
equations will have coefficients depending on $\abs{\Epm}^2\sim 10^{-16}$, any
error due to the truncation will be on the order of that due to using finite-precision arithmetic. 
 
System~\eqref{eq:discrete} was solved numerically using programs from the
MINPACK suite~\cite{minpack}.  For the smallest chosen value of $I$, the
initial guess for the nonlinear solver was a suitably scaled linear defect
mode.  For larger values of $I$, the initial guess was extrapolated from
previously computed solutions.  The program continued producing new iterates
until it reached the left edge of the band gap, and was able to find
reasonable solutions quite close to the band edge.

\section{Technical details from section~\ref{sec:embed_cond}}
\label{sec:details}
We explain the manner in which the limit $\veps\to0$ is taken in equation~\eqref{long}. 
\begin{align}
 &\int_{\sigma_c^\#}\ e^{i(\lambda-\tilde\lambda_0)t}\  \int_0^t\ ds\ 
 \left(\
e^{-i(\lambda-\tilde\lambda_0)s}\psi_\lambda,  \ W\psi_0\ \right) \psi_\lambda\ d\lambda\ \ A(s)\ ds\nn\\
 &=\ \lim_{\veps\to0}\ \int_{\sigma_c^\#}\ e^{i(\lambda-\tilde\lambda_0+i\veps)t}\ \psi_\lambda\ d\lambda\ 
  \int_0^t\ ds\ A(s)\ \left(\
e^{-i(\lambda-\tilde\lambda_0+i\veps)s}\psi_\lambda,  \ W\psi_0\ \right)  \nn\\
&=\ iA(t)\ \int_{\sigma_c^\#}\  \frac{1}{\lambda-\tilde\lambda_0+i0}\ \psi_\lambda\ d\lambda\ \left(\ \psi_\lambda,W\psi_0\ \right)\ -\ iA_0\   \int_{\sigma_c^\#}\  \frac{e^{i(\lambda-\tilde\lambda_0)t}}{\lambda-\tilde\lambda_0+i0}\ \psi_\lambda\ d\lambda\ \left(\ 
 \psi_\lambda,W\psi_0\ \right)\nn\\
&\ \ \ +\ i\int_0^t\  \int_{\sigma_c^\#}\ \frac{e^{i(\lambda-\tilde\lambda_0)(t-s)}}{\lambda-\tilde\lambda_0+i0}\ \psi_\lambda\ d\lambda\  \left(\ \psi_\lambda\ ,\ W\psi_0\ \right)\ \D_sA(s)\ ds\label{error-est}
\end{align}
The first term on the right hand side of~\eqref{error-est} contributes to~\eqref{A-final}. We now show that the remaining terms decay. In particular, it suffices  to show that for any smooth and decaying function, $f$, the term
\begin{equation}
 \int_{\sigma_c^\#}\ \frac{e^{i(\lambda-\tilde\lambda_0)t}}{\lambda-\tilde\lambda_0+i0}\ \psi_\lambda\ d\lambda\ \left(\ 
 \psi_\lambda\ , f\ \right)
 \nn\end{equation}
 decays as $t\uparrow+\infty$. Let $\chi(\lambda)$ be a smooth function, which is equal to one on $\sigma_c^\#$, vanishes outside a neighborhood of $\sigma_c^\#$ and vanishes near thresholds and infinity. For any $k\ge1$
 \begin{equation}
 \begin{split}
 \int_{\sigma_c^\#}\  \frac{e^{i(\lambda-\tilde\lambda_0)t}}{\lambda-\tilde\lambda_0+i0}\ \psi_\lambda\  
 \left(\ \psi_\lambda\ , f\ \right)\ d\lambda  
 &=\ \lim_{\veps\to0}\  \int_{\sigma_c^\#}\  \frac{e^{i(\lambda-\tilde\lambda_0 +i\veps)t}}{\lambda-\tilde\lambda_0+i\veps} \left(\  \psi_\lambda\ , f\ \right )
\ \psi_\lambda\ d\lambda \\
 &=\ \lim_{\veps\to0}\ -i\int_t^\infty\  \int_{\sigma_c^\#}\ 
  \left(\  \psi_\lambda\ , f\ \right)\  e^{i(\lambda-\tilde\lambda_0 +i\veps)s}\ \psi_\lambda\ d\lambda \ ds
 \\
 &=\ \lim_{\veps\to0}\ -i\int_t^\infty\  \int 
  \left(\  \psi_\lambda\ , f\ \right)\  \psi_\lambda\ \chi(\lambda)\ e^{i(\lambda-\tilde\lambda_0 +i\veps)s}\  d\lambda \ ds
 \\
 &=\  \lim_{\veps\to0}\ -i\int_t^\infty\  \int 
  \left(\  \psi_\lambda\ , f\ \right)\  \psi_\lambda\ \chi(\lambda)\  
 \left(\ (it)^{-1}\D_\lambda\right)^k\ e^{i(\lambda-\tilde\lambda_0 +i\veps)s}\  d\lambda \ ds\\
 &=\ \lim_{\veps\to0}\ -i(-it)^{-k}\int_t^\infty\  \int 
  \D_\lambda^k\left[\ \left(\  \psi_\lambda\ , f\ \right)\  \psi_\lambda\ \chi(\lambda)\right]\ 
 e^{i(\lambda-\tilde\lambda_0 +i\veps)s}\  d\lambda \ ds\\
 &\ =\ \cO(t^{-k}).
 \end{split}
 \end{equation} 
 \section{Numerical computations with Evans Functions}
\label{sec:numer_evans} 
Before describing the numerical method used to compute the Evans function, we
first comment on the sources of error in this calculation.  The largest source
of error, is a loss-of-significance in the calculation of the
Wronskian~\eqref{eq:wronskian}.  As each column in this determinant is the
image of a solution that grows exponentially as $Z$ moves from $\pm\infty$,
the norm of each column vector may be quite large.  Near zeros
of the Evans function $D(Z)$, there must be large cancellations that arise in
evaluating the determinant that defines it.  To mitigate this loss-of-significance, the solutions to the
differential equations~\eqref{eq:eigen_general} must be extremely accurate.
First that means that the defect modes around which we linearized must be very
accurate, and must be well-computed deep into the exponentially decaying
tails, to allow the shooting to start in a region where the linearized problem
is very nearly the linearized differential equation at $Z=\pm \infty$.
Additional considerations that arise are more technical and are discussed
below. 
 
\subsection{Numerical Computation on Exterior Product Spaces}
The numerical calculation the Evans function is extremely delicate, and there
exists a wide literature discussing how to accurately implement the
calculation~\cite{BD:99a,Bri:01,Swinton:92}.  We will summarize briefly and
point to the appropriate 
references.  The most straightforward calculation would be to first find the
$n$ vectors $\vetap_j(0)$ andn $\vetam_j(0)$ and calculate their
Wronskian~\eqref{eq:wronskian} directly.  That is, for each of the $k$
positive eigenvalues, choose $L$ large enough that $A(\pm L) \approx A_\infty$
and numerically integrate  
\begin{equation}
\begin{split}
\label{eq:fromleft}
    \diff{\vetap_j}{Z}& =A(Z) \vetap_j  \\
    \vetap_j\rvert_{Z=-L}&  = \vv_j e^{-\lambda_j L}
\end{split}
\end{equation}
from $Z=-L$ to $Z=0$ for $j=1\ldots k$ and
\begin{equation}
\begin{split}
\label{eq:fromright}
    \diff{\vetam_j}{Z}& =A(Z) \vetam_j  \\
    \vetam_j\rvert_{Z=L}&  = \vv_j e^{\lambda_j L}
\end{split}
\end{equation}
from $Z=L$ (backwards) to $Z=0$ for $j=k+1\ldots n$.  

When the stable and unstable subspaces of $A_\infty$ are one-dimensional, this
strategy works well, but when they are of higher dimension, this works poorly.
Consider the equation  
\begin{equation}
\diff{\vec y}{Z} = A(Z) \vec y
\label{eq:generic}
\end{equation}
$\vec y \in \bbC^n$, such that the two largest eigenvalues of $A$ are positive with
$\mu_2>\mu_1>0$ and corresponding eigenvectors $\vec{y}_2$ and $\vec{y}_1$.
In attempting to compute the solution $\vec y \sim \vec{y}_1 e^{\mu_1 Z}$,
rounding errors will inevitably introduce a small error in the direction of
$\vec{y}_2$ which will grow at the faster exponential rate $\mu_2$, which, for
the necessarily large values of $L$, may cause significant errors.  One might
hope to correct this error by performing a re-orthogonalization at each step,
but this, too, is problematic, as discussed in~\cite{Bri:01,DerGot:05}.   An algorithm that overcomes these difficulties without resorting to a compound matrix formalism has recently been constructed by Humpherys and Zumbrun~\cite{HumZum:06}.  This is important for computations in ${\mathbb C}^n$ with $n\gg1$ since the dimension of the exterior product space grows very rapidly with $n$.

A method was proposed by Brin~\cite{Bri:01} and significantly refined
and made mathematically precise by Bridges and collaborators~\cite{BD:99a,BD:03,BDG:02}, working in various combinations.  We
briefly introduce the basic features of the method, and refer the reader to
the cited papers.  Let $\ve_j$ be a set of orthonormal basis vectors for
$\bbC^m$, then the exterior product (the wedge product) is defined on the
$\ve_j$ by the properties of bilinearity: 
\begin{align*}
    \ve_i \wedge (a\ve_j+ \ve_k) &= a \ve_i\wedge\ve_j + \ve_i\wedge\ve_k   \\
(a\ve_i +\ve_j) \wedge \ve_k &= a \ve_i\wedge\ve_j + \ve_j\wedge\ve_k 
\end{align*}
and antisymmetry:
$$
\ve_i \wedge \ve_j = - \ve_j \wedge \ve_i,
$$
which implies that $\ve_i \wedge\ve_i = 0$.  The exterior product is extended
to $\bbC^m$ using (bi)linearity and antisymmetry: if $\vec u = \sum_{j=1}^n
u_j \ve_j$ and $\vv = \sum_{j=1}^n v_j \ve_j$, then 
$$
\vec u \wedge \vv = \sum_{i=1}^{n-1} \sum_{j=i+1}^{n} (u_i v_j - v_i u_j)
\ve_i \wedge \ve_j. 
$$
One may easily see that the set $\{\ve_i \wedge\ve_j\}$ forms the basis for the vector
space $\bbC^{\binom{m}{2}}$ of 2-forms over $\bbC^m$. 
The 2-form $\vv_i \wedge \vv_j$ gives the complex area of the parallelogram
with sides $\vv_i$ and $\vv_j$.  Given an $i$-form and a $j$-form with $i+j
\le m$, we may define their wedge product as an $(i+j)$ form using the
anticommutativity and bilinearity.  This is a binary operation of the form 
$$ 
\wedge: \bbC^{\binom{m}{i}} \times \bbC^{\binom{m}{j}} \to \bbC^{\binom{m}{i+j}}.
$$
Note now that the Evans function can be written in exterior product notation as
\begin{equation}
\begin{split}
\label{eq:wedge_evans}
D(\beta) &= 
      \vetap_1(0) \wedge \ldots \wedge \vetap_k(0) \wedge \vetam_{k+1}(0) 
      \wedge \ldots \wedge  \vetam_n(0) \\
      &= \cW^+(0) \wedge \cW^-(0),
\end{split}	
\end{equation}
where $\cW^+(Z)$ is the $k$-form given by the wedge product of the
$\vetap_j(Z)$ and $\cW^-(Z)$ is the $(n-k)$-form given by the wedge product of
the $\vetam_j(Z)$.  In the present case, ${\ve_j : 1\le j\le 4}$ is the
standard basis of $\bbC^4$, and the space of 2-forms $\wedge^2(\bbC^4)$ is
six-dimensional with orthonormal basis vectors: 
\begin{equation}
\vf_1 = \ve_1 \wedge \ve_2, \;
\vf_2 = \ve_1 \wedge \ve_3, \;
\vf_3 = \ve_1 \wedge \ve_4, \;
\vf_4 = \ve_2 \wedge \ve_3, \;
\vf_5 = \ve_2 \wedge \ve_4, \;
\vf_6 = \ve_3 \wedge \ve_4.
\label{eq:basis}
\end{equation}

The essential idea of all the so-called compound matrix methods is to
numerically evolve $\cW^{\pm}(Z)$ from $Z=\mp L$ to $Z=0$, rather than
integrating the individual solutions whose wedge product forms $\cW^{\pm}$.
If $\vec y^1$ through $\vec y^k$ satisfy evolution equation~\eqref{eq:generic}
then $Y = \vec y^1 \wedge \ldots \wedge \vec y^k$ satisfies the linear
evolution equation: 
\begin{align*}
\diff{Y}{Z} &= 
\sum_{j=1}^{k}  \vec y^1 \wedge \ldots \wedge \vec y^{(j-1)} \wedge  
\diff{}{Z} \vec y^j \wedge \vec y^{(j+1)} \wedge \ldots \wedge \vec y^k \\
&= \sum_{j=1}^{k}  \vec y^1 \wedge \ldots \wedge \vec y^{(j-1)} \wedge  
\left(A(Z) \vec y^j\right) \wedge \vec y^{(j+1)} \wedge \ldots \wedge \vec y^k.
\end{align*}
Using basis vectors for the wedge space $\wedge^k (\bbC^{n})$, this equation
may be rewritten in matrix form as  
\begin{equation}
\label{eq:wedge_evolution}
\diff{Y}{Z} = \cA^{(k)}(Z) Y
\end{equation}
where $\cA^{(k)}(Z)$ is a $\binom{n}{k} \times \binom{n}{k}$ matrix. In the
present situation, $n=4$ and $k=2$, then this may be written out, using
basis~\eqref{eq:basis}: 
\begin{equation}
\label{compoundA}
\cAA (Z)= \begin{pmatrix}
      a_{11} + a_{22} & a_{23} & a_{24} & -a_{13} & -a_{14} & 0  \\
      a_{32} & a_{11} + a_{33} & a_{34} & a_{12} & 0 & -a_{14}   \\
      a_{42} & a_{43} & a_{11} + a_{44} & 0 & a_{12} & a_{13}   \\
      -a_{31} & a_{21} & 0 & a_{22} + a_{33} & a_{34} & -a_{24}   \\
      -a_{41} & 0 & a_{21} & a_{43} & a_{22}  + a_{44} & a_{23}   \\
      0 & -a_{41} & a_{31} & -a_{42} & a_{32} & a_{33} + a_{44}
\end{pmatrix}.
\end{equation}
If $\mu_i$ and $\mu_j$ are eigenvalues of $A$ with eigenvectors $\vv_i$ and
$\vv_j$.  Then $\vv_i \wedge \vv_j$ is an eigenvector of $\mathcal A^{(2)}$
with eigenvalue $\mu = \mu_i + \mu_j$.    Thus only the solution with fastest
growth need be (stably) computed.

\subsection{Integration scheme}
\label{sec:glrk}
A vector $\vv \in \wedge^k(\bbC^n)$ is called decomposable if it can be
written as the single wedge product of $k$ elements of $\bbC^n$, and not all such
vectors are decomposable.  A vector $\vv = \sum_{j=1}^{6} v_j \vf_j$ in
$\wedge^2(\bbC^4)$ is decomposable if and only if the 4-form $\vv \wedge \vv =
0$, which gives the algebraic condition 
\begin{equation}
\label{eq:onGrassmannian}
v_1 v_6 - v_2 v_5 + v_3 v_4 =0.
\end{equation}
This quantity is the Grassmannian invariant, and the set on which it vanishes
is called the Grassmannian manifold.  The $k$-form $\cW^+$ is by its
definition decomposable, and it is important for numerical accuracy that the
computed approximation to be so as well.  Therefore we require a numerical
integration routine that conserves the Grassmannian invariant.  One such
family of methods is the fully implicit Runge-Kutta schemes using the
Gauss-Legendre points~\cite{BDG:02,Ise:96}.  We use the three-step
sixth-order scheme of this type: 
\begin{equation}
\label{eq:glrk}
\vy^{n+1} = \vy^n + \Delta Z \sum_{i=1}^{3} b_j \vec{K}_i
\end{equation}
where $\Delta Z$ is the stepsize, and the superscript $n$ denotes points in
the spatial discretization. The increments $\vec{K}_i$ are implicitly defined
by 
\begin{equation}
\label{eq:Kj}
\vec{K}_i  = A(Z_n + c_i \Delta Z)\cdot(\vy^n+ \sum_{j=1}^{3}  a_{ij}
\vec{K}_j) 
\end{equation}
and the coefficients are given by:
$$a = \begin{pmatrix}
\frac{5}{36} & \frac{2}{9}-\frac{\sqrt{15}}{15} & \frac{5}{36} -
\frac{\sqrt{15}}{30} \\ 
\frac{5}{36}+\frac{\sqrt{15}}{24}& \frac{2}{9} & \frac{5}{36} -
\frac{\sqrt{15}}{24} \\ 
\frac{5}{36}+\frac{\sqrt{15}}{30} & \frac{2}{9}+\frac{\sqrt{15}}{15} & \frac{5}{36}
\end{pmatrix},\,
 \vec{b} = \begin{pmatrix}
\frac{1}{2}- \frac{\sqrt{15}}{30}  \\ \frac{1}{2} \\ \frac{1}{2}+
\frac{\sqrt{15}}{30}  
\end{pmatrix}, \,
\vec{c} = \begin{pmatrix}
 \frac{5}{18} \\ \frac{4}{9}  \\ \frac{5}{18}
\end{pmatrix}.
$$
Since the matrix $a$ has no nonzero elements, all the substeps $K_i$ of the
Runge-Kutta algorithm must be calculated simultaneously by
solving~\eqref{eq:Kj}.  Thus, although equation~\eqref{eq:evansform}, used to
construct the Evans function has four dependent variables, the number of
unknowns at each step is increased to six by the wedge product formulation, and
then tripled to eighteen by the numerical scheme.  Since
equation~\eqref{eq:wedge_evolution} is linear, this is still a small amount of
work at each step. 

\subsection{Additional digit-preserving numerical measures}
For large values of $Z$, the matrix in equation~\eqref{eq:evansform}, and thus
in~\eqref{compoundA}  is nearly constant, the solution is not far from the product of  the 
exponential of this matrix and the initial vector, which is
especially simple given that the initial condition is an eigenvalue of $\cAA$.
For most values of $\beta$ we are interested in, we find that  the contribution of the constant part of $\cA$ dominates the computed solution so it is useful to solve
this part exactly--the growth or rapid oscillation due to the largest
eigenvalue.  

If $\vv$ is an eigenvector of a matrix $M$, then it is also an eigenvector of
$M-c I$ for any $c$, so we modify equation~\eqref{eq:fromleft} to use  
$$\tilde A = A(Z)-\frac{\lambda_1+\lambda_2}{2}$$
instead of A.  Thus the compound matrix $\cAA$ is replaced in
equation~\eqref{eq:wedge_evolution} by $\tcAA=\tilde A \wedge \tilde A
=\cAA-\lambda_1 -\lambda_2$, and the Runge-Kutta algorithm is only used to
compute the non-constant part of the evolution.  The solution at $Z=0$ can
then be multiplied by $e^{(\lambda_1 + \lambda_2)z}$ to find the solution.  
While the integration scheme converges at sixth order in $\Delta Z$ with or
without this modification, we found empirically that the error found using the
modified method to be much smaller for each fixed value
of $\Delta Z$, even for values of  $\beta$ near zero. 

An additional source of error comes from solving~\eqref{eq:wedge_evolution} from $\pm L$ to 0, rather than from $\pm \infty$.  We take L large by working on the same nonuniform grid as was used in the solution to~\eqref{eq:discrete}.  This is especially important if the nonlinear mode is only known numerically. Another solution to the tail problem is to use a rapidly convergent series form of $D(\beta)$ to obtain the solution of equations~\eqref{eq:fromleft} and~\eqref{eq:fromright} at some finite value $z=\pm z_0$, and then solve the odes numerically from $\pm z_0$ to 0~\cite{LiPro:00}.

\begin{figure}
\begin{center}
\includegraphics[width=3in]{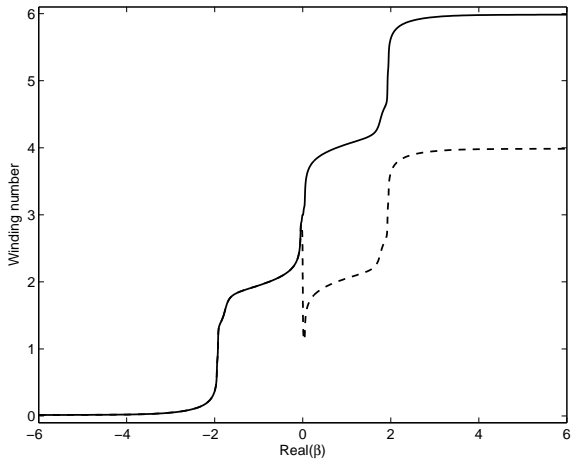}
\caption{The winding number ($\arg{(D(\beta)})/2\pi$) arising from the
  linearization about a gap soliton with $\rho=0.1$, $v=0.9$ and
  $\d=0.9\pi$, with $\Im{\beta}=0.005$.  The solid line corresponds to a
  uniform grid of 2048 points, and the dashed line to a variable grid with 512
  points.  For both computations, the ODE~\eqref{eq:wedge_evolution} is solved
  beginning from $Z_{\rm max}=\pm 50.05$, chosen so that $\abs{\Epm(Z_{\rm
  max})}^2<10^{-16}$ which makes the error due to tail truncation on the same
  order of magnitude as the roundoff error. The uniform grid calculation
  overcounts the roots of $D(\beta)$ by two because of numerical errors near
  $\beta=0$ .} 
\label{fig:compare_grids}
\end{center}
\end{figure}

The code was tested by confirming the results of Derks and Gottwald~\cite{DerGot:05} on the  stability of gap solitons in NLCME without defect. 
We found that using the non-uniform grid for
integrating the ordinary differential equation~\eqref{eq:wedge_evolution} in the
shooting method greatly improved the accuracy of $D(\beta)$, especially in a
neighborhood of $\beta=0$, where a zero of high multiplicity decreases the accuracy of the computation.  In figure~\ref{fig:compare_grids}, we show that the computation on the nonuniform grid with 512 points gives a more accurate count of the zeros than a uniform grid calculation with 2048 points which predicts two spurious unstable eigenvalues.  The tests also showed that the interval of computation $[-L,L]$ must be taken quite large to properly resolve the tail's effect on the computed $D(\beta)$.  Our numerical solutions of~\eqref{eq:discrete} were not able to resolve the solution for $Z$ this large without using the stretched grid.  The value of $L$ used in the calculations section~\ref{sec:numer_stability} had to be taken fairly large before the computation converged.

\bibliographystyle{amsplain}
\bibliography{defectmodes}
\end{document}